\pdfoutput=1
\documentclass[%
 reprint,
 amsmath,amssymb,
 aps,
 pra,
]{revtex4-2}

\usepackage{hyperref}
\hypersetup{
    colorlinks,
    citecolor=blue,
    filecolor=black,
    linkcolor=blue,
    urlcolor=black,
    pdftitle={State Preparation on Quantum Computers via Quantum Steering},
    pdfauthor={Daniel Volya},
    pdfkeywords={Quantum Steering, State Preperation, Quantum Computing}
}

\usepackage{tikz}

\usepackage{physics}
\usepackage{dcolumn}
\usepackage{bm}
\usepackage{quantikz}
\usepackage{adjustbox}
\usepackage{float}
\usepackage[caption=false]{subfig}
\usepackage[linesnumbered,ruled,vlined]{algorithm2e}
\usepackage{orcidlink}

\usepackage{booktabs,multirow}

\newcommand{\bigzero}{\mbox{\normalfont\Large\bfseries 0}}

\usepackage[caption=false]{subfig} 
\usepackage{ragged2e} 
\DeclareCaptionJustification{justified}{\justifying}

\begin{document}

\preprint{APS/123-QED}

\title{State Preparation on Quantum Computers via Quantum Steering}

\author{Daniel Volya\,\orcidlink{0000-0001-5026-5646}}
\author{Prabhat Mishra\,\orcidlink{0000-0003-3653-6221}}%
\affiliation{%
  University of Florida, Gainesville, Florida, USA
}%

\date{\today}


\begin{abstract}
  One of the major components for realizing quantum computers is the ability to initialize the computer to a known fiducial state, also known as state preparation. 
  We demonstrate a state preparation method via measurement-induced steering on contemporary, digital quantum computers.
  By delegating ancilla qubits and systems qubits, the system state is prepared by repeatedly performing the following steps:
  (1) executing a designated system-ancilla entangling circuit,
  (2) measuring the ancilla qubits, and
  (3) re-initializing ancilla qubits to known states through active reset.
  While the ancilla qubits are measured and reinitialized to known states, the system qubits are steered from arbitrary initial states to desired final states.
  We show results of the method by preparing arbitrary qubit states and qutrit (three-level) states.
  We also demonstrate that the state convergence can be accelerated by utilizing the readouts of the ancilla qubits to guide the protocol in an active manner.
  This protocol serves as a nontrivial example that incorporates and characterizes essential operations such as  qubit reuse (qubit reset), entangling circuits, and measurement. 
  These operations are not only vital for near-term noisy intermediate-scale quantum (NISQ) applications but are also crucial for realizing future error-correcting codes.
\end{abstract}
\maketitle

\vspace{-0.2in}
\section{Introduction}
\vspace{-0.2in}

\begin{figure}[htb]
  \centering
  \includegraphics[width=1\linewidth]{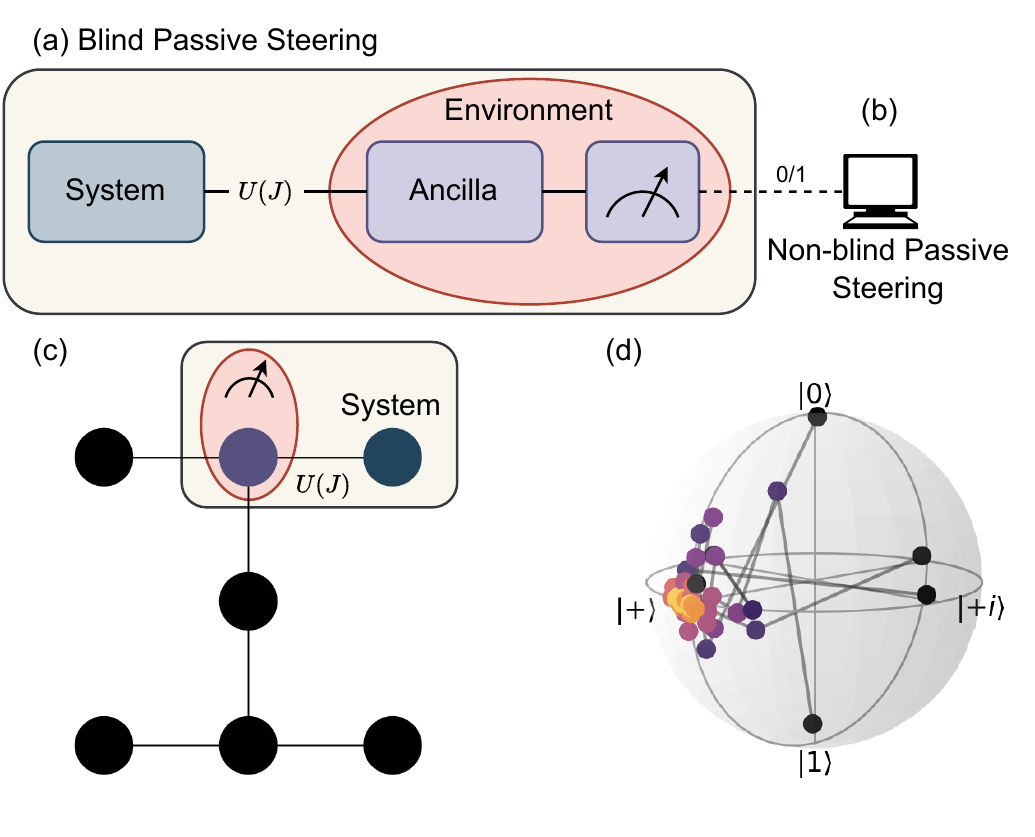}
\vspace{-0.4in}
  \caption{The measurement-induced steering protocol conceptually consists of (a)  passively steering a system (qubit or qutrit) to an arbitrary state via coupling to an ancilla qubit that is exposed to an environment for measurement and simple state reset (i.e., to $\ket{0}$). 
    A specifically chosen unitary operator $U(J)$, parameterized by an arbitrary coupling strength $J$, acts upon the system-ancilla. By repeatedly applying the unitary and measuring the ancilla, a back-action is induced on the system whereby the average of all readout outcomes steer the system to a desired state.
    Instead of averaging the measurement readouts (b) non-blind passive steering processes the readouts on a classical computer to accelerate the convergence of the system state.
    We experimentally realize the protocol on IBM's superconducting quantum computers, such as \textit{ibm\_perth} with the device connectivity graph shown in (c).
    To select our system qubit (qutrit), we choose the transmon with the best measurement discrimination between the computational states. The ancilla qubit is then selected as nearest neighbor given by the device connectivity.
    The Bloch sphere (d) shows the results of passive steering a system qubit on \textit{ibm\_perth} to prepare an equal superposition state (shown as yellow dots) where the initial states are arbitrary (shown as black dots).}
    \label{fig:overview}
\end{figure}

One of the primary requirements in quantum computing is the ability to prepare an arbitrary quantum state \cite{kakInitializationProblemQuantum1999,divincenzoPhysicalImplementationQuantum2000a}. This requirement is fulfilled by: (1) initializing the quantum computer to a known fiducial state ($\ket{0}^{\otimes n}$) of $n$-qubits, and (2) applying a series of discrete quantum gates to the known state to obtain a desired final state ($\ket{\psi_\oplus} = \mathcal{U}\ket{0}^{\otimes n}$) \cite{kitaevQuantumComputationsAlgorithms1997}.
Initialization of the quantum computer is commonly achieved by waiting for the system to thermalize to the ground state (\textit{passive reset}) -- with the waiting time roughly correlated to $T_1$ coherence times~\cite{rigettiSuperconductingQubitWaveguide2012a, hartyHighFidelityPreparationGates2014a}.
Although the waiting time for qubits to thermalize is feasible for today's contemporary quantum computers, as technology improves and the coherence times of large collection of qubits increases, the waiting time will dominate in comparison to the program duration. Moreover, the passive reset is not applicable for scenarios when we need to initialize to an arbitrary (non-fiducial) state. Alternatively, recent efforts investigate \textit{active reset} such as through projective measurements~\cite{basilewitschFundamentalBoundsQubit2021a, tornowMinimumQuantumRunTime2022a}.
In reality, a desired state may not be an eigenstate of a measurement operator and thus  leads to probabilistic outcomes.
Therefore, when the state of the qubits is collapsed via measurement, single-qubit rotations are applied to correct the state based on the readout outcomes \cite{tornowMinimumQuantumRunTime2022a}.
However, such an approach faces two main challenges: first, measurement itself can be a long and error-prone operation depending on the underlying technology~\cite{johnsonHeraldedStatePreparation2012, risteInitializationMeasurementSuperconducting2012}; and secondly, arbitrary state preparation requires carefully calibrating the necessary quantum gates, as well as extreme fine-tuning on large quantum computers to guarantee an appropriate fidelity.

Alternative strategies for initializing fiducial states have been proposed via engineered dissipative dynamics \cite{krausPreparationEntangledStates2008,verstraeteQuantumComputationQuantumstate2009,weimerRydbergQuantumSimulator2010}, reversible \cite{boykinAlgorithmicCoolingScalable2002,fernandezAlgorithmicCoolingSpins2004} or irreversible \cite{brassardProspectsLimitationsAlgorithmic2014, rodriguez-brionesHeatbathAlgorithmicCooling2017} algorithmic cooling \cite{parkHeatBathAlgorithmic2015}.
The methods assume Markovian dynamics, whereby a system state is driven to a pure steady state.
However, for real world open quantum systems undergoing non-Markovian dynamics \cite{breuerColloquiumNonMarkovianDynamics2016, whiteDemonstrationNonMarkovianProcess2020a}, a successful state reset implies not only purification, but also erasure of initial correlations between qubits and the environment \cite{reedFastResetSuppressing2010, geerlingsDemonstratingDrivenReset2013, basilewitschBeatingLimitsInitial2017}.
Furthermore, passive reset, active reset, and strategies based on dissipative dynamics and algorithmic cooling are not applicable for scenarios when we need to initialize to an arbitrary (non-fiducial) state.
To further prepare an arbitrary state from the initial fiducial state requires careful calibration of quantum gates, as well as extreme fine-tuning on large quantum computers to guarantee an appropriate fidelity.

In additional to careful calibration of unitary quantum gates, near-term and far-term quantum computers must also effectively perform mid-circuit measurement and qubit reuse. Recent demonstrations of NISQ algorithms necessitate both of these operations, such as in circuit cutting \cite{pengSimulatingLargeQuantum2020,loweFastQuantumCircuit2023} and error mitigation techniques \cite{temmeErrorMitigationShortDepth2017a, liEfficientVariationalQuantum2017a}. Furthermore, quantum error-correcting codes involve repeated measurements of ancilla qubits for error detection, a key part for realizing fault-tolerant quantum computer \cite{satzingerRealizingTopologicallyOrdered2021a, acharyaSuppressingQuantumErrors2023}. However, despite the importance of these operations, efficient and effective characterization of these operations remains an open problem.
In this paper, we explore an alternative approach to state preparation, incorporating unitary gates, mid-circuit measurement, and qubit reuse. We are able to characterize the holistic performance of these operations via state fidelity.  In principle, the approach reduces the number of qubits that undergo active resets, lowers the classical processing involved during quantum computation for correction, and can prepare arbitrary states $\ket{\psi_\oplus}$ without having to first prepare a known initial state.


\vspace{-0.2in}
\subsection{Quantum Steering}
\vspace{-0.1in}

Historically, an important yet perplexing feature of quantum mechanics is the apparent non-local correlations (or entanglement) between distant particles.
Schr\"odinger famously introduced the term \textit{quantum steering} in concern of the ability to remotely steer a particle's state through measurements on another particle that is entangled with it \cite{schroedingerErfassungQuantengesetzeDurch1929, schrodingerDiscussionProbabilityRelations1935}. 
Recently, quantum steering has been creatively exploited in developing a protocol for preparing arbitrary quantum states irrespective of their initial (mixed) state \cite{royMeasurementinducedSteeringQuantum2020a}. The protocol consists of a repetition of simple steps:
\begin{enumerate}
  \item a fixed unitary operation $U$ couples ancilla qubits and system qubits;
  \item a measurement is conducted on the ancilla qubits, decoupling it from the system qubits;
  \item the ancilla qubits are reinitialized to known simple states.
\end{enumerate}
As the ancilla qubits are measured, the back-action on the system qubits steers them from arbitrary (unknown) mixed states to a desired final state.
The protocol has been theoretically analyzed in preparing a two-qubit system to arbitrary (mixed or pure) states \cite{kumarEngineeringTwoqubitMixed2020}, noting the strength of the entangling operator $U$ for preparing classical, discorded, or entangled target states.
Subsequently, the protocol's rate of convergence was studied in preparing an arbitrary qubit state where it is remarked that significant speedup can be achieved with slight compromise to the fidelity of the final target state \cite{kumarOptimizedSteeringQuantum2022}.
Extensions to the protocol have been proposed where instead of ignoring the results of measurement, the ancilla readouts are utilized to perform online decisions in navigating the Hilbert space \cite{herasymenkoMeasurementdrivenNavigationManybody2022a}. By utilizing the readouts, the protocol's convergence may be improved and the entangling operation $U$ may be changed via a feedback mechanism without collapsing the system state.
The variations of the protocol can be summarized as follows: (a) \emph{blind passive steering} where readouts are ignored and $U$ remains fixed, (b) \emph{non-blind passive steering} where readouts are utilized and $U$ remains fixed, (c) \emph{blind active steering} where readouts are ignored and $U$ is changed with each iteration, and (d) \emph{non-blind active steering} where readouts are utilized and $U$ is changed with each iteration.

In this paper, we demonstrate the (non-)blind passive steering protocol on contemporary cloud-accessible quantum computers by delegating ancilla and system qubits (qutrits) that undergo $N$-repetitions of an operation $U$ implemented as a digital quantum circuit.
After repeating the protocol steps, we show that the state of the system approaches a desired state. In summary, we make the following major contributions.
\begin{itemize}
    \item We realize measurement-induced steering \cite{royMeasurementinducedSteeringQuantum2020a} for arbitrary state preparation on physical quantum computers.
    \item We develop quantum circuits to implement the steering protocol with primary focus on a qubit-qubit coupled system (an ancilla qubit to steer a qubit) and a qubit-qutrit coupled system (an ancilla qubit to steer a qutrit.)
    \item We also investigate the \textit{non-blind} approach, where instead of disregarding the measurement results, we take advantage of the measurement readouts to accelerate the convergence.
    \item Furthermore, we show that the quantum steering operator can be divided into local and non-local operations using Cartan decomposition~\cite{dalessandroDecompositionsUnitaryEvolutions2006,dalessandroIntroductionQuantumControl2021}. The non-local operations convey the strength of the entanglement necessary to perform quantum steering. Furthermore, this decomposition can be viewed as a graphical representation for a qubit-qubit coupled system, providing visualization for non-local operations. 
\end{itemize}

Figure~\ref{fig:overview} conceptually summarizes the quantum steering protocol and shows an overview of mapping the protocol to a cloud-accessible quantum computer.

\vspace{-0.2in}
\section{Digital Implementation of MIQS}
\vspace{-0.1in}

The goal of the measurement-induced quantum steering (MIQS) protocol is to prepare a desired target state $\ket{\psi_\oplus}$, irrespective of the initial state.
This is achieved by exploiting the back-action caused by measuring part of an entangled system, steering our system to the target state.
In this section, we first provide a review of the formal specification of the MIQS protocol.
Next, we describe the circuit implementation of the MIQS protocol, focusing on steering a qubit and a qutrit, providing quantum circuits that satisfy the steering conditions.
Finally, we explore the properties of the generated circuits.



\vspace{-0.2in}
\subsection{Formulation of MIQS Protocol}
\vspace{-0.1in}
Suppose we have a system of ancilla qubits initialized to the state $\ket{\psi_A}$ (density matrix $\rho_A$) and system qubits in an arbitrary state $\rho_S$.
The general MIQS protocol involves the following steps:

\begin{enumerate}
  \item Couple the ancilla qubits and system qubits with a composite unitary operator $U$. The state of the ancilla-system after the $n$-th application of the unitary evolution is $\rho_{A-S}^{n+1} = U \rho_A \otimes \rho_S^{n} U^\dagger$.
  \item {The ancilla qubits are then decoupled from the system, giving the density state of the system as:
        \vspace{-0.05in}
        \begin{equation}\label{eq:meas-steer}
          \rho_S^{n+1} = \mathrm{Tr}_A \left[\rho_{A-S}^{n+1} \right] = \mathrm{Tr}_A \left[ U \rho_A \otimes \rho_S^{n} U^\dagger\right]
        \vspace{-0.05in}
        \end{equation}
        }
  \item The ancilla qubits are reinitialized to their initial states and the steps are repeated.
\end{enumerate}
The goal is to steer the system state to a desired target state $\ket{\psi_{\oplus}}$ ($\rho_{\oplus})$. The dynamics of $U$ should be chosen such that the following steering inequality is satisfied:
\vspace{-0.05in}
\begin{equation}\label{eq:steer-ineq}
  \bra{\psi_{S\oplus}} \rho_S^{n+1} \ket{\psi_{\oplus}} \ge \bra{\psi_{\oplus}} \rho_S^n \ket{\psi_{\oplus}}.
  \vspace{-0.05in}
\end{equation}
\begin{figure}[tp]
  \subfloat[Starting from an unknown initial state $\rho_S$, the system qubit $S$ is steered via a repeated application of an ancilla-system entanglement operation $U_{A-S}$, followed by measurements and active resets of the ancilla qubit $A$. After $N$ applications, the system qubit arrives to a target state $\ket{\psi_{S\oplus}}$. \label{fig:sqrt-swap-circuit}]{%
  \includegraphics[width=1\linewidth]{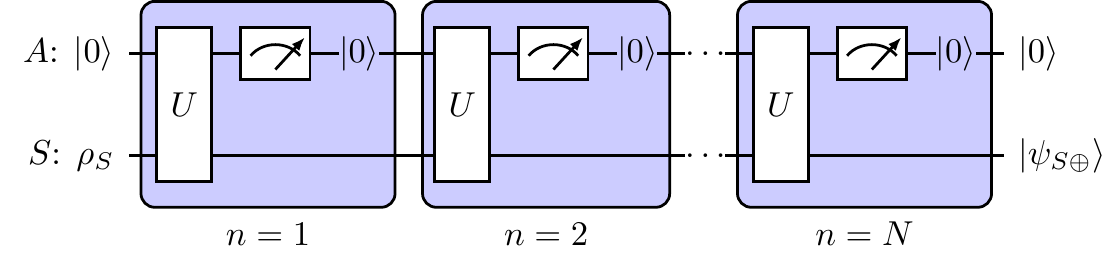}%
  }\hfill
  \subfloat[A Pauli-based quantum circuit representation of the ancilla-system entangling operator $U$ for the target state $\ket{0}\otimes\ket{+} = \ket{0} \otimes \frac{1}{\sqrt{2}}(\ket{0} + \ket{1})$.\label{fig:two-qubit-decomposition}]{%
    \includegraphics[width=1\linewidth]{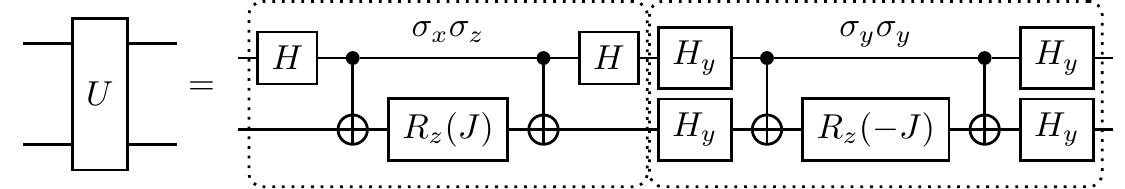}
  }\hfill
  \vspace{-0.1in}
  \caption{An overview of the quantum steering protocol.}
  \vspace{-0.1in}
\end{figure}
In other words, with each repetition of the steps in the MIQS protocol, the state of our system should get closer to our desired pure target state $\ket{\psi_\oplus}$.
The general theory under which Equation~\ref{eq:steer-ineq} will be satisfied is derived in \cite{royMeasurementinducedSteeringQuantum2020a}.
In brevity, if the quantum dynamics is given as the time evolution $U = \exp(-i H \delta t)$ of a Hamiltonian $H$, then for $H$ to satisfy Equation~\ref{eq:steer-ineq} it has the following form
\vspace{-0.05in}
\begin{equation}\label{eq:steer-H}
  H = \sum_n \left( O_A^{(n)}\ket{\psi_A}\bra{\psi_A}\right) \otimes \Omega_S^{(n)} + \mathrm{h.c.,}
\vspace{-0.05in}
\end{equation}
where $n$ labels the ancilla qubits.
The Hamiltonian consists of direct product of operators $O_A^{(n)}$ that rotate the ancillas from their initial state to an orthogonal subspace and operators $\Omega_S^{(n)}$ that rotate the system to an orthogonal subspace.


Rather than physically engineering and realizing a system with the satisfactory Hamiltonian (Equation \ref{eq:steer-H}), we instead \textit{simulate} the Hamiltonian on a quantum computer through the application of discrete unitary operators \cite{lowHamiltonianSimulationQubitization2019a, kokcuFixedDepthHamiltonian2022}.

\vspace{-0.2in}
\subsection{Implementation of Qubit-Qubit MIQS Protocol}\label{sec:qubit-qubit-miqs}
\vspace{-0.1in}
In this section, we derive the unitary operator $U$ that steers a qubit to a desired state.
An arbitrary target state of a qubit (excluding global phase) has the form
\vspace{-0.05in}
\begin{equation}
\ket{\psi_\oplus} = \cos(\theta/2) \ket{0} + e^{i\phi}\sin(\theta/2)\ket{1},
\vspace{-0.05in}
\end{equation}
with $0 \leq \theta \leq \pi$ and $0 \leq \phi < 2\pi$.
A Hamiltonian that satisfies Equation~\ref{eq:steer-ineq} is 
\vspace{-0.05in}
\begin{multline}\label{eq:qubit-H}
  H_{A-S} = \frac{J}{2}\left( -\cos(\phi)\cos(\theta)\sigma_A^x\sigma_S^x - \cos(\phi)\sigma_A^y\sigma_S^y
   \right.\\
  \left. + \sin(\phi)\sigma_A^y\sigma_S^x
  +\sin(\theta)\sigma_A^x\sigma_S^z
  -\sin(\phi)\cos(\theta)\sigma_A^x\sigma_S^y \right)
\end{multline}
where $J$ is an arbitrary coupling constant, and $\sigma_u^{\{x,y,z\}}$ are the standard Pauli matrices acting on the individual subsystem $u$.
Assuming the standard computational basis, the matrix corresponds to
\vspace{-0.05in}
\begin{equation}
  H = \frac{J}{2}\begin{bmatrix}
    0 & 0 & \alpha & -\beta_-^*\\
    0 & 0 & -\beta_+ & -\alpha \\
    \alpha & -\beta_+^* & 0 & 0\\
    -\beta_- & -\alpha & 0 & 0\\
  \end{bmatrix}
  \vspace{-0.05in}
\end{equation}
with $\alpha = \sin\theta$ and $\beta_\pm = e^{i\theta}(\cos\theta \pm 1)$.
A quantum circuit that reproduces the unitary operator
\vspace{-0.05in}
\begin{equation}
  U = \exp(-iH)
  \vspace{-0.05in}
\end{equation}
will essentially swap the ancilla-qubit space with the system-qubit space.

In Section~\ref{sec:geometry} we provide the optimal quantum circuits that implements the operator with single qubit rotations and CNOT gates.
However, for the remainder of this section we provide an illustrative example with a simple circuit construction. 




\noindent \textbf{Example: } A systematic method to construct the quantum circuit is to consider each Pauli string in the Hamiltonian~$H$.
As an example, consider the case when $\phi=0$, then Equation~\ref{eq:qubit-H} simplifies to
\begin{equation}\label{eq:H_steer}
  \hat{H}_{A-S}=\frac{J}{2} (-\cos(\theta) \underbrace{\sigma_A^x\sigma_S^x}_{H_{XX}} + \sin(\theta)\underbrace{\sigma_A^x\sigma_S^z}_{H_{XZ}} - \underbrace{\sigma_A^y\sigma_S^y}_{H_{YY}} ).
\end{equation}
Therefore, the unitary evolution operator is given as
\begin{equation}
  U_{A-S} = \exp(-i\hat{H}_{A-S}) = U_{XX+XZ} \circ U_{YY} \label{eq:U}; 
\end{equation}
with two commuting terms
\begin{align}
U_{XX+XZ} & = \exp(i\alpha H_{XX} -i \beta H_{XZ}),  \label{eq:U_XX_XZ}\\
U_{YY} &= \exp(i\frac{J}{2}H_{YY}), \label{eq:U_YY}
\end{align}
where $\alpha = \frac{J\cos(\theta)}{2}$ and $\beta = \frac{J\sin(\theta)}{2}$. The circuit decomposition is done in two main steps. First, the non-commuting terms in Equation~\ref{eq:U_XX_XZ} are decomposed  using an approximation. Next, all the Pauli Hamiltonians, $H_{XX}$, $H_{XZ}$, and $H_{YY}$, are decomposed to their circuit representations.

A nice simplification occurs when either $\sin(\theta) =0$ or $\cos(\theta)=0$, leaving either $U_{XX}$ or $U_{XZ}$ terms in combination with $U_{YY}$. This specifically occurs when the target state $\ket{\psi_\oplus} = \ket{+} = \frac{1}{\sqrt{2}}\left(\ket{0} + \ket{1}\right)$. With $\theta = \pi/2$, the Hamiltonian in Equation~\ref{eq:H_steer} simplifies to
\begin{equation}
  \hat{H} = \frac{J}{2}\left( \sigma_A^x\sigma_S^z - \sigma_A^y\sigma_S^y\right).
\end{equation}
Since the Pauli operators $H_{XZ}$ and $H_{YY}$ commute, we can express the evolution operator as
\begin{equation}\label{eq:U_plus}
  U_{A,S} = \exp(-i\frac{J}{2}\sigma_A^x\sigma_S^z)\circ \exp(i\frac{J}{2}\sigma_A^y\sigma_S^y)
\end{equation}
and obtain the quantum circuit as shown in Figure~\ref{fig:two-qubit-decomposition}. The $\ket{+}$ state is particularly interesting due to its prevalence in quantum algorithms, primarily in preparing entangled Bell states by applying a subsequent CNOT operation.
Appendix~\ref{sec:eg-single-qubit} provides an analytical analysis of steering to the $\ket{+}$ state.

\vspace{-0.2in}

\subsection{Implementation of Qubit-Qutrit MIQS Protocol}
\vspace{-0.1in}
In the previous section, we show a derivation of the quantum circuit to steer a qubit to a desired state.
In this section, we derive a quantum circuit to prepare an arbitrary qutrit state.
Control of qutrits is typically harder to do via conventional means compared to qubits, therefore, there is additional benefit to using the MIQS protocol.

An arbitrary qutrit state (excluding global phase) can be written in terms of four parameters as
\begin{align}
  \ket{\psi_\oplus} &= \sin(\xi/2)\cos(\theta/2)\ket{0} \nonumber \\
                    &+ e^{i\phi_{01}}\sin(\xi/2)\sin(\theta/2)\ket{1} \nonumber \\
                    &+ e^{i\phi_{02}}\cos(\xi/2)\ket{2},
\end{align}
where $0 \leq \theta,\xi \leq \pi$ quantify the magnitude of the components of $\ket{\psi_\oplus}$ while $0 \leq \phi_{01},\phi_{02} \leq 2\pi$ describe the phases of $\ket{0}$ relative to $\ket{1}$ and $\ket{2}$, respectively.

A Hamiltonian that steers the qutrit will have the following form
\vspace{-0.1in}
\begin{equation}
  H = \sigma^+ \otimes \ket{\psi_\oplus}\bra{\psi_\oplus}_1^\perp + \sigma^+ \otimes \ket{\psi_\oplus}\bra{\psi_\oplus}_2^\perp + \mathrm{h.c.}
  \vspace{-0.1in}
\end{equation}
where $\sigma^+$ is the raising operator and $\ket{\psi_\oplus}_i^\perp$ are orthogonal states to our desired state. We note that we may rewrite the Hamiltonian in terms of $\sigma_x$ and $\sigma_y$ Pauli-matrices and $\lambda_j$ Gell-Mann matrices, with some coupling $\alpha_{i,j}$ between them. Similar to the previous section, we may take the strings consisting of Pauli and Gell-Mann terms and map them to simple building blocks for our quantum circuits.    

For our experimental realization of a qutrit state, we will focus on one particular state: an equal superposition as defined by
\vspace{-0.2in}
\begin{equation}
  \ket{\psi_\oplus} = \frac{1}{\sqrt{3}}\left(\ket{0} + \ket{1} + \ket{2} \right).
  \vspace{-0.05in}
\end{equation}
We may express the orthogonal subspace as being spanned by two vectors
\vspace{-0.05in}
\begin{align}
  \ket{\psi_\oplus}^\perp_1 &= \frac{1}{\sqrt{3}}\left(\ket{0} + \nu\ket{1} + \nu^*\ket{2} \right),\\
  \ket{\psi_\oplus}^\perp_2 &= \frac{1}{\sqrt{3}}\left(\ket{0} + \nu^*\ket{1} + \nu\ket{2} \right)
\vspace{-0.05in}
\end{align}
where $\nu = \exp(i 2\pi/3)$. Thus, a Hamiltonian that will steer the overall qutrit state to the desired target $\ket{\psi_\oplus}$ has the following matrix form
\begin{equation}\label{eq:qutrit-hamiltonian}
H_{A-S} = \frac{1}{3}\left(\begin{array}{@{}c|c@{}}
  \bigzero_{3 \times 3} & 
  \begin{matrix}
  2 & 2 & 2 \\
  -1 & -1 & -1 \\
  -1 & -1 & -1
  \end{matrix}
  \\
\hline
  \begin{matrix}
  2 & -1 & -1 \\
  2 & -1 & -1 \\
  2 & -1 & -1
  \end{matrix} &
  \bigzero_{3\times 3} 
\end{array}\right)
\end{equation}
again showing that overall operation moves both subsystems to their orthogonal subspace.

\vspace{-0.2in}
\subsection{Geometrical Considerations}\label{sec:geometry}
\vspace{-0.1in}

We have derived the quantum circuits that steer qubit and qutrit states to their respective desired states.
The quantum circuits specifically entangle the ancilla and systems states such that they satisfy target state convergence given by Equation~\ref{eq:steer-ineq}.
This section presents the quantum circuits from a geometrical point of view, offering insight to the \textit{kinds} of entanglement necessary.

\newtheorem{definition}{Definition}[section]

The machinery for providing our insight is based on the Cartan decomposition of the $\mathfrak{su}(d_1 d_2)$ Lie algebra, where $d_1 = 2$ and $d_2 = 2,3$ for the qubit or qutrit case, respectively \cite{dalessandroDecompositionsUnitaryEvolutions2006, dalessandroIntroductionQuantumControl2021}.

\begin{definition}\label{def:cartan}
  A \textbf{Cartan decomposition} of a Lie algebra $\mathfrak{g}$ is defined as an orthogonal split $\mathfrak{g} = \mathfrak{k} \oplus \mathfrak{m}$ satisfying
  \vspace{-0.1in}
  \begin{equation}
    [\mathfrak{k}, \mathfrak{k}] \subset \mathfrak{k}, \quad [\mathfrak{m}, \mathfrak{m}] \subset \mathfrak{k}, \quad [\mathfrak{k}, \mathfrak{m}] = \mathfrak{m}.
    \vspace{-0.05in}
  \end{equation}
  A \textbf{Cartan subalgebra} denoted by $\mathfrak{a}$ refers to a maximal Abelian algebra within $\mathfrak{m}.$
\end{definition}

Picking basis elements one by one, and finding a Cartan decomposition directly through Definition~\ref{def:cartan} is difficult in practice.
Instead, partitioning the Lie algebra into $\mathfrak{k}$ and $\mathfrak{m}$ is done by an involution: a Lie algebra homomorphism $\theta:\mathfrak{g}\to\mathfrak{g}$, such that $\theta(\theta(g)) = g$ for any $g\in\mathfrak{g}$ and preserves all commutators.
The involution is then used to split the Lie algebra by defining subspaces via $\theta(\mathfrak{k}) = \mathfrak{k}$ and $\theta(\mathfrak{m}) = -\mathfrak{m}$.
Cartan's classification revealed that there are only three types of decomposition for $su(n)$. However, we utilize the decomposition given by the corresponding involution $\theta(g) = -g^\mathrm{T}$ for all $g\in\mathfrak{g}$ (referred in literature as an \textbf{AI} type decomposition).
The result of the Cartan decomposition is the ability to write any unitary operator $U$ as
\vspace{-0.05in}
\begin{equation}
  U = K_1 A K_2
  \vspace{-0.05in}
\end{equation}
where $K_1$ and $K_2$ are elements of $e^{i\mathfrak{k}}$ and $A \in e^{i\mathfrak{a}}$ are elements defined by the Cartan subalgebra.

It is well-known that an arbitrary operator acting on two-qubits $U \in U(4)$ can be decomposed as product of a gate $U\in SU(4)$ and a global phase shift $e^{i\theta}$.
Since the global phase does not impact the underlying quantum mechanics, we focus specifically on the $SU(4)$.
We are particularly interested in the operations that are \textit{non-local}, giving insight to the necessary entanglement.
Such operations are then given as elements in $SU(4)\backslash SU(2)\otimes SU(2)$.
The Cartan decomposition of $\mathfrak{su(4)}$, any two-qubit operation can be written as
\vspace{-0.05in}
\begin{equation}
  U = k_1 A k_2
  \vspace{-0.05in}
\end{equation}
where $k_1,k_2 \in SU(2) \otimes SU(2)$ and the non-local part $A = exp(i/2 (c_1 \sigma_x\sigma_x + c_2 \sigma_y\sigma_y + c_3 \sigma_z\sigma_z))$.
This representation allows separation of steering operator into local ($K_1$, $K_2$) and nonlocal ($A$) parts. 
The coefficients $c_k \in [0, \pi]$ are the non-local coordinates, and contain a geometrical structure \cite{zhangGeometricTheoryNonlocal2003}. 
The coefficients for any possible ancilla-qubit steering operator $U(J)$ is given by
\begin{equation}
  c = [J, J, 0]
\end{equation}

\begin{figure}[tp]
  \centering
  \vspace{-0.1in}
  \includegraphics[width=1\linewidth]{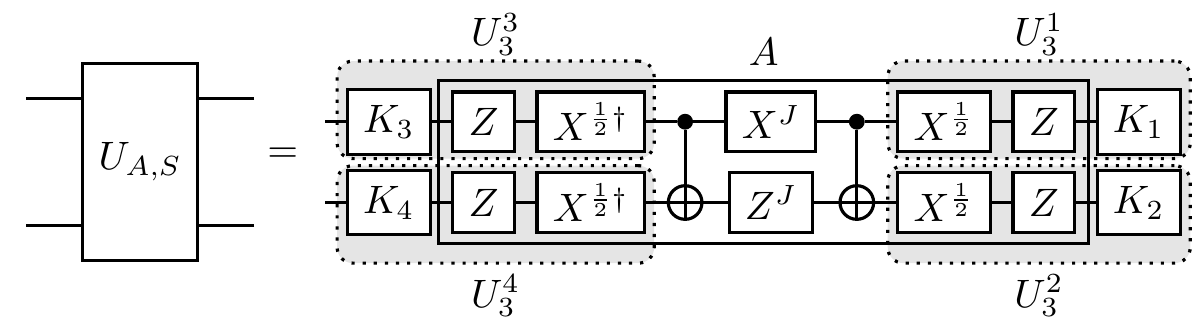}
  \vspace{-0.1in}
  \caption{The optimized decomposition of the qubit-qubit steering operator. $K_i$ gates are single-qubit rotations produced by the Cartan decomposition and are parameterized by $\theta$ and $\phi$ of a desired state. The non-local operator $A$ is decomposed using two CNOT gates and local qubit rotations along X and Z axis. The circuit is further simplified by combining possible local rotations into a single qubit rotation $U_3$ -- a native arbitrary rotation gate on IBM Quantum computers.  The $X^{(J/2)}$ and $Z^{(J/2)}$ gates are defined as $e^{i\frac{\pi}{4}J} R_x(\pi J/2)$ and $e^{i\frac{\pi}{4}J} R_z(\pi J/2)$ respectively. $X^{(1/2)}$ gate is then defined as $R_x(\pi/2)$. \label{fig:two-qubit-kak}}
\end{figure}

\begin{figure}[tp]
  \centering
  \includegraphics[width=1\linewidth]{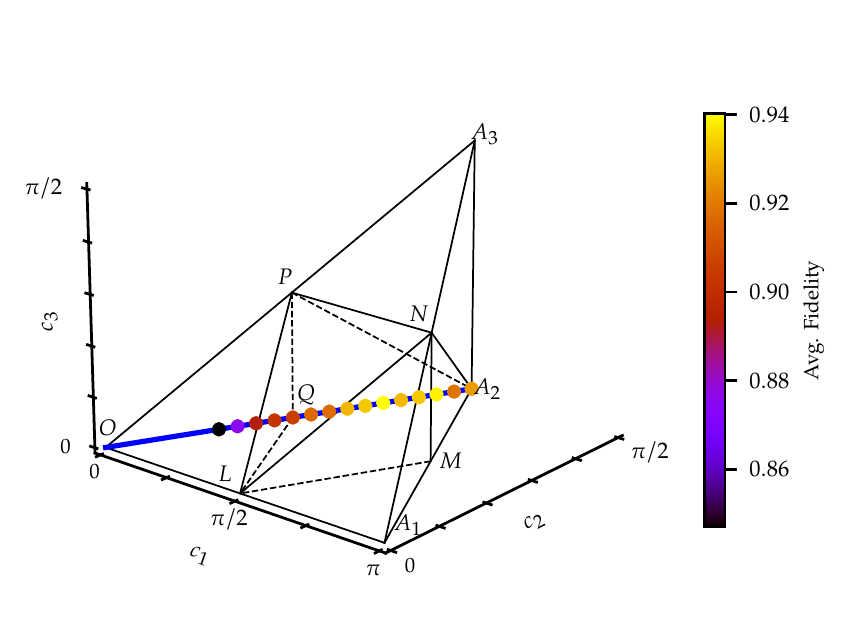}%
  \vspace{-0.4in}
  \caption{The Weyl Chamber representing coordinates of non-local two-qubit unitaries.
  All possible two-qubit steering operators $U$ are represented by the blue line.
  The coordinates are given by the coupling parameter $J$, namely $[J, J, 0]$.
  Maximum entanglement is achieved when $J=\pi/2$, corresponding to the point $A_2$ in the chamber.
  Individual points correspond to the maximum fidelity achieved when executing the steering protocol with a steering operator given by a choice of $J$.  \label{fig:weyl}}
\end{figure}

Figure~\ref{fig:weyl} displays these parameters for any ancilla-qubit steering operator $U$ on the Weyl chamber -- which is the symmetry-reduced version of a cube. 
The point $L$ corresponds to the gate CNOT and all gates that are locally equivalent, including the CPHASE gate. 
As shown, CNOT and CPHASE gates are not locally equivalent to the steering operator $U$.
Thus, despite being characterized as perfect entanglers, the CNOT and CPHASE gates do not satisfy the steering conditions and in fact are unital operators on the qubit. 
Therefore, capability of the steering operator to create entanglement between qubit and ancilla is a necessary but not sufficient condition to steer the qubit. 
Digital quantum computers, fortunately, allow for implementation of arbitrary unitary operations that satisfy the non-local criteria. Figure~\ref{fig:two-qubit-kak} is the optimal circuit given by the Cartan decomposition for the ancilla-qubit steering operator which we execute on digital quantum computers.



\vspace{-0.2in}
\subsection{Rapid Reset via Measurement Readouts}
\vspace{-0.1in}
In our current description of the protocol, the results of measuring the ancilla qubits are discarded -- i.e. blind passive steering.
Effectively, by averaging all possibilities of readout outcomes, the state of our system converges to a desired state.
This is advantageous as, in general, classical processing of data is not required avoiding additional overhead.
However, by utilizing the readout results of the ancilla qubits we can accelerate convergence of our system state.  
Contemporary quantum computers have the infrastructure to process readout results during the execution of a quantum circuit.
Hence, we take advantage of this capability to demonstrate preparation of a desired state by utilizing readout results via the non-blind passive steering protocol.

As a simple demonstration, note in Section~\ref{sec:qubit-qubit-miqs} that the steering operator swaps the detector and system spaces.
Therefore, if the ancilla qubit has swapped to its orthogonal state (a readout of ``1"), that means the system qubit has successfully swapped to the desired state. In general, the measurement of an ancilla qubit with a readout of ``1'' is given by the projection operator
\vspace{-0.05in}
\begin{equation}\label{eq:measurement-proj}
  \Pi_1 = \ket{1}_{A}\bra{1}_{A} \otimes \mathbb{I}_{S}.
\vspace{-0.05in}
\end{equation}
The ancilla-system state after applying the steering operator $U$ and measuring the ancilla state in ``1'' is
\vspace{-0.05in}
\begin{equation}
  \rho_{A-S}^{n+1} = \frac{\Pi_1 U \rho_{A-S}^{n} U ^\dagger \Pi_1}{p_1}
\vspace{-0.05in}
\end{equation} 
where $p_1 = \mathrm{Tr}\left[U\rho_{A-S}^nU^\dagger\Pi_1\right]$ is the probability of measuring a ``1''.
For further analysis and extensions of this idea, we refer to Reference \cite{herasymenkoMeasurementdrivenNavigationManybody2022a}.

\vspace{-0.2in}
\section{Experiments}
\vspace{-0.1in}
In this section, we describe the different steps followed to physically prepare states via measurement-induced quantum steering (MIQS) protocol with the superconducting transmon qubits and qutrits.


\vspace{-0.2in}
\subsection{Experimental Setup}
\vspace{-0.1in}


\begin{table*}[htbp]
  \scriptsize
  \hspace*{-1cm}
\centering
\begin{tabular}{@{}llc ccc c cc@{}}
  \toprule
  \multicolumn{3}{c}{Processor}		& \multicolumn{3}{c}{Mean fidelity} & \multicolumn{3}{c}{Blind Steering fidelity} 	\\
  \cmidrule(lr){1-3}
  \cmidrule(lr){4-6}
  \cmidrule(lr){7-9}
  Name	& Type	& QV	& 2Q gate	& 1Q gate	& SPAM & $U$ & $\ket{+}$ & Mean $\ket{\psi_\oplus}$	\\		
  \midrule[\heavyrulewidth]
  ibmq\_lima & Falcon r4T & 8 & 0.9898 & 0.9998 & 0.973 & 0.827 &     $\mathbf{0.772\pm0.034}$  &  --\\
  ibmq\_belem & Falcon r4T & 16 & 0.9874 & 0.9998 & 0.9775 & 0.877 &  $\mathbf{0.786\pm0.033}$  & --\\
  ibm\_perth & Falcon r5.11H & 32 & 0.9781 & 0.9997 & 0.987 & 0.846 & $\mathbf{0.954\pm0.026}$  & $\mathbf{0.925\pm0.012}$\\
  \bottomrule
\end{tabular}
\caption{Table of the processors used to demonstrate the steering protocol. Values in the Processor Quantum Volume (QV), as well as single-qubit 1Q gate, two-qubit 2Q gate, and State Preparation and Measurement (SPAM) fidelities are provided by the IBM Quantum team \cite{IBMQuantum}. 
The Steering operator $U$ fidelity is computed as an average of process tomography results. The maximum state fidelity of steering to $\ket{+}$ is shown for each processor. The average state fidelity of steering to stabilizer states was only performed on \textit{ibm\_perth}.   }
\label{table:NISQ_Devices}
\end{table*}

The experiments were performed using different IBM Quantum computers (accessed through IBM Cloud~\cite{IBMQuantum}): \textit{ibm\_lima}, \textit{ibm\_belem}, and \textit{ibm\_perth}.
The hardware commands are coded using Qiskit, utilizing the recent additions of mid-circuit measurements and active reset operations. Furthermore, we took advantage of Qiskit Pulse \cite{alexanderQiskitPulseProgramming2020a} -- a pulse-level programming model -- which allowed us to define, calibrate, and execute quantum circuits outside conventional definitions.
The low-level access to the underlying quantum hardware enables processing quantum information on qutrits (three-level system), extending the concept of quantum computation on two-level systems.
For most operations, we used gates calibrated by the IBM team.

\begin{figure}[htp]
  \centering
    \vspace{-0.2in}
  \includegraphics[width=0.8\linewidth]{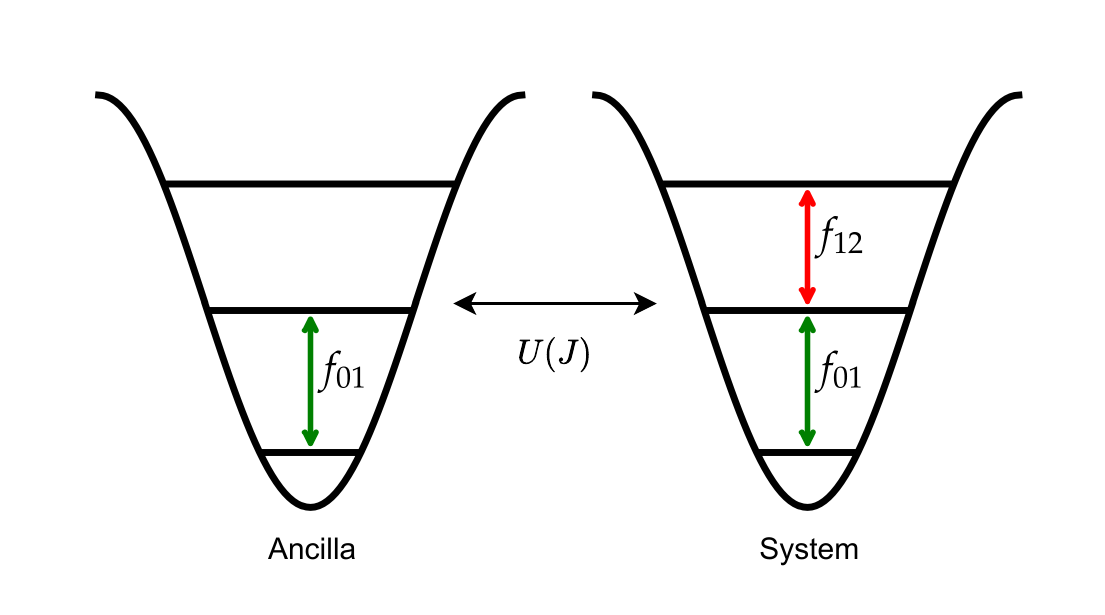}
  \vspace{-0.2in}
  \caption{The schematic of superconducting computers that realizes our qubit-qubit and qubit-qutrit coupling. \label{fig:superconducting-transmons}}
  \vspace{-0.1in}
\end{figure}

For each transmon, the local oscillator (LO) frequency is given by IBM's calibrated $\ket{0} \to \ket{1}$ frequency, which was kept fixed for the experiments.
Transitions between the $\ket{1}$ and $\ket{2}$ states are achieved by using amplitude-modulated microwave pulses via sinusoidal side-band at a frequency $f_{12} - f_{01}$.
This results in an effective shift of frequency for the pulses from $f_{01}$ to $f_{12}$ \cite{krantzQuantumEngineerGuide2019a}.
Appendix~\ref{sec:chip} shows the results of the calibration.
Figure~\ref{fig:superconducting-transmons} represents the energy levels of the superconducting transmons architecture.

The MIQS circuits are designed using a combination of:  default single-qubit gates, which operate in the $\{\ket{0}, \ket{1}\}$ subspace $(01)$; default entangling CNOT gate; and custom calibrated single-qutrit gates, which operate on the $\{\ket{1}, \ket{2}\}$ subspace $(12)$. The single-qutrits gates are defined by utilizing the amplitude of the $\pi_{1\to2}$ pulse -- which we obtained via a Rabi experiment.
We use the default implementation of the CNOT gate as defined by IBM Quantum. Extended to a qubit-qutrit system, it acts as a $SU(2\times3=6)$ gate with the truth table as shown in Table~\ref{tab:cnot}. For the control qubit in the (01) subspace, it acts as a standard qubit CNOT gate but with an additional phase of $\pi/2$ to the $\ket{2}$ state of the target qutrit \cite{galdaImplementingTernaryDecomposition2021, yurtalanImplementationWalshHadamardGate2020}.
IBM Quantum allows the reuse of qubits through mid-circuit measurements and conditional-reset.
The \emph{reset} is achieved by applying a not-gate conditioned on the measurement outcome of the qubit.
During the execution of the MIQS protocol, the ancilla qubit is measured and subsequently reset.

\begin{table}[htp]
  \centering
  \vspace{-0.1in}
      \begin{tabular}{c c c} 
       \hline
       Control & Target & Output \\ 
       \hline\hline
       $\ket{0}$ & $\ket{0}$ & $\ket{00}$\\ 
       $\ket{0}$ & $\ket{1}$ & $\ket{01}$ \\
       $\ket{0}$ & $\ket{2}$ & $\ket{02}$ \\
       $\ket{1}$ & $\ket{0}$ & $\ket{10}$ \\
       $\ket{1}$ & $\ket{1}$ & $\ket{11}$ \\
       $\ket{1}$ & $\ket{2}$ & $i\ket{12}$ \\
      \end{tabular}
      \vspace{-0.1in}
      \caption{Truth table for the default IBM CNOT gate where the control qubit acts on a target qutrit. The operation is implemented as two consecutive CNOT gates (more details can be found in Ref. \cite{galdaImplementingTernaryDecomposition2021}).\label{tab:cnot}}
      \vspace{-0.1in}
  \end{table}
  
For qubit readout, we used the  $0-1$ discriminator provided by IBM Quantum. However, this discriminator is unable to correctly identify excitations to the $\ket{2}$ state, misclassifying them as $\ket{1}$. Therefore, to read out the qutrits, we developed our own custom $0-1-2$ discriminator to classify in-phase and quadrature (IQ) points.

For a desired system state $\ket{\psi_\oplus}$, we construct a batch of MIQS circuits where the total iterations ($N$) of $U_{A,S}$ is incremented from $1$ to a maximum of $\mathcal{N}$. This enables us to estimate the state of the system as the number of  $U_{A,S}$ iterations varies, and reduces the overhead due to cloud access to hardware. For each iteration $N$, we conduct quantum state tomography on the system qubit.  The measurement results from the quantum computer are processed locally. The estimated state of the system qubit is taken as an unbiased average over all ancilla qubit outcomes (i.e., a projective measurement), and estimates of the mixed system state is computed using maximum likelihood, minimum effort method \cite{smolinEfficientMethodComputing2012}. Once we are content with the results, we fix $N = \mathcal{N}$ which provides one MIQS circuit that faithfully prepares the state $\ket{\psi_\oplus}$. We repeat this process for different coupling parameters $J$, noting the relationship between $J$, numbers of iterations $\mathcal{N}$, and the achieved state $\ket{\psi_\oplus}$ fidelity.

\begin{figure*}[tp]
  \includegraphics[width=0.9\linewidth]{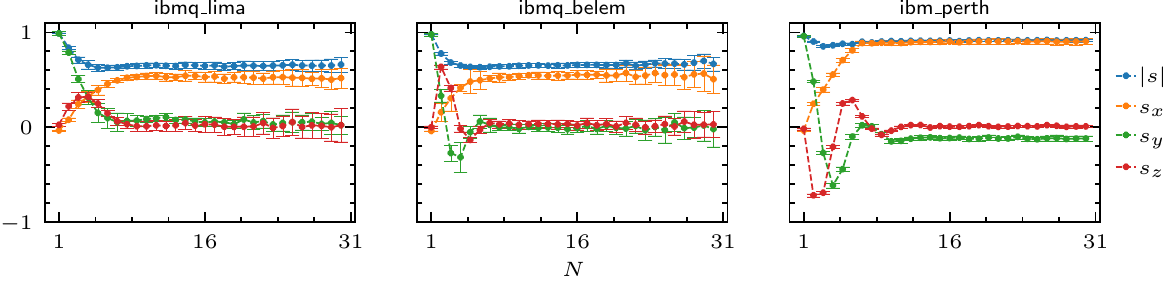}%
  \caption{Steering experiment on three IBM Quantum (IBMQ) machines. \label{fig:sqrt-swap-steer-ibmq}}
\end{figure*}

Before executing the MIQS protocol, we further verify the correctness of the steering operator $U_{A,S}$ through quantum process tomography (QPT). QPT is a procedure for experimentally reconstructing a complete description of a noisy quantum channel $\mathcal{E}$. This is done by preparing a set of input states $\{ \ket{a_i} \}$ and performing measurements on a set of operators $\{B_j\}$ to estimate probabilities $p_{ij} =\mathrm{Tr}[B_j^\dagger \mathcal{E}(\ket{a_i}\bra{a_i})].$ If the input states and measurement operators span the input and output spaces respectively, then the set $\{p_{ij}\}$ reconstructs the channel $\mathcal{E}$. For a $n$-qubit channel, the input space is constructed via tensor products of $\{\ket{0}, \ket{1}, \ket{+} = \frac{1}{\sqrt{2}}(\ket{0} + \ket{1}), \ket{+i} = \frac{1}{\sqrt{2}}(\ket{0} + \ket{1}) \}$, and the measurement space via tensor products of $\sigma_x$, $\sigma_y$, and $\sigma_z$. Thus a total of $4^n 3^n$ experiments are conducted to estimate $4^{2n}$ probabilities.

\begin{figure}[tp]
  \vspace{-0.1in}
  \includegraphics[width=1\linewidth]{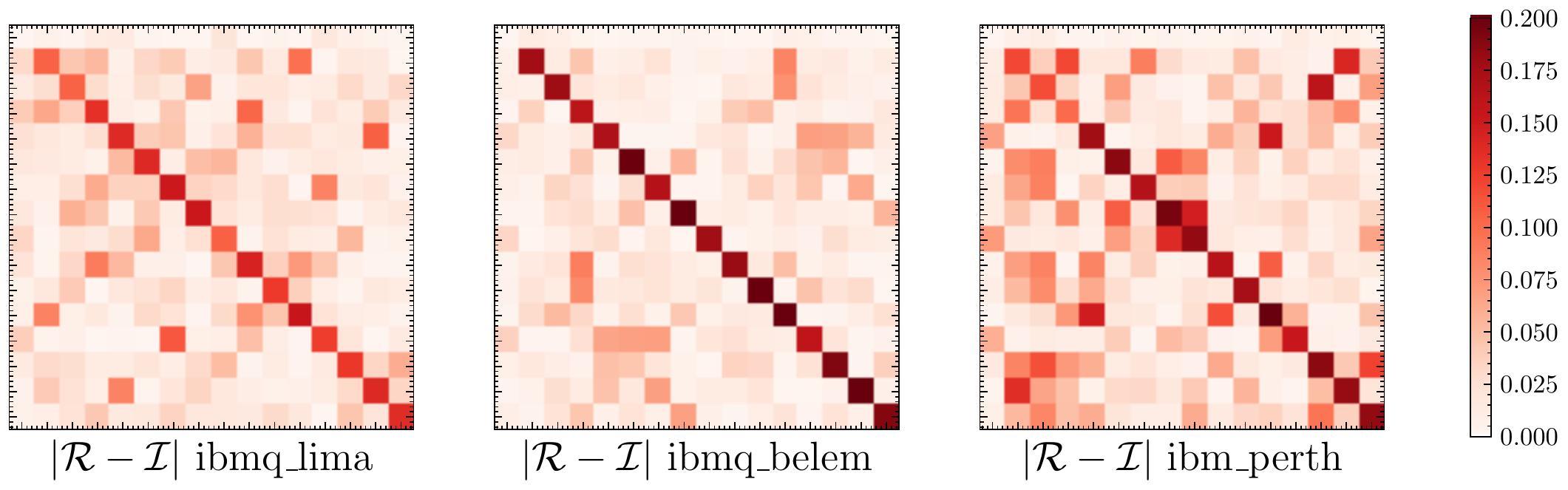}%
  \vspace{-0.1in}
  \caption{Process Tomography of the steering circuit to prepare $\ket{+}$ on IBM Quantum machines. 
  Both \textit{ibm\_lima} and \textit{ibm\_belem} are 5-qubit Falcon r4 (year 2020) processors with a quantum volume of 8 and 16, respectively. \textit{ibm\_perth} is a 7-qubit Falcon r.511H (year 2021) processor with a quantum volume of 32. As indicated by the quantum volume benchmark, \textit{ibm\_perth} qubits are expected to have higher stability and lifetime.} \label{fig:sqrt-swap-process-tomo}
    \vspace{-0.1in}
\end{figure}

After reconstructing the channel $\mathcal{U_{A-S}}$ through QPT, we extract the error channel by composing with the inverse of the ideal channel $\mathcal{E} = \mathcal{U} \circ \mathcal{U}_{\text{ideal}}^{-1}$. The error channel is converted to the Pauli-transfer matrix representation $\mathcal{R}$, which is strictly real. In the ideal case, $\mathcal{R} = I$, the identity matrix -- representing no errors. The absolute difference between the noisy reconstructed $\mathcal{R}$ and the ideal $|\mathcal{R} - I|$ is shown in Figure~\ref{fig:sqrt-swap-process-tomo}. The average gate fidelity of the reconstructed channels were $F=0.827$, $F=0.877$, and $F = 0.846$ for ibmq\_lima, ibmq\_belem, and ibm\_perth, respectively. While the average gate fidelities are comparable, we can see clear differences in matrix entries in Figure~\ref{fig:sqrt-swap-process-tomo}. Typically, two-qubit gates will have coherent errors due to imperfections in calibration from unwanted terms in the cross-resonance interaction Hamiltonian \cite{sheldonProcedureSystematicallyTuning2016, woodSpecialSessionNoise2020b}.

\vspace{-0.2in}
\subsection{Evaluation of Qubit-Qubit Protocol}
\vspace{-0.1in}

We employed the MIQS protocol to prepare 1-qubit stabilizer states.
The stabilizer states serve as a suitable unitary 3-design for the randomized benchmarking protocol.
Stabilizer states can also be defined as the states that are produced by gates from the Clifford group ($H$, $CNOT$, and $S$ gates) applied to $\ket{0}$ state. We express the system-qubit density state as
\vspace{-0.05in}
\begin{equation}
  \rho_S(n) = \frac{1}{2}(I + \vec{s}(n) \cdot \vec{\sigma})
  \vspace{-0.05in}
\end{equation}
where $\vec{s}(n)$ is a three-component vector that depends on the current iteration $n$ of the steering protocol, and $\vec{\sigma}$ is a vector of the Pauli matrices. The single qubit stabilizers, their vector coordinates $\vec{s}$, and the necessary steering operator $U_{A,S}$ are summarized in Table~\ref{tab:stabilizers}. 
\begin{table}[tp]
  \centering
      \begin{tabular}{c | c c | c |c} 
       \hline
       $\ket{\psi_\oplus}$ & $\theta$ & $\phi$ & $\vec{s}$ & $U_{A,S}$ \\ 
       \hline\hline
       $\ket{0}$ & $0$ & $0$                           & (0, 0, 1) & $\exp(-i\frac{J}{2}\left(\sigma_A^x\sigma_S^x + \sigma_A^y\sigma_S^y\right))$\\ 
       $\ket{1}$ & $\pi$ & $0$                         & (0, 0, -1) & $\exp(-i\frac{J}{2}\left(\sigma_A^x\sigma_S^x - \sigma_A^y\sigma_S^y\right))$\\ 
       $\ket{+}$ & $\frac{\pi}{2}$ & $0$               & (1, 0, 0) & $\exp(-i\frac{J}{2}\left(\sigma_A^x\sigma_S^z - \sigma_A^y\sigma_S^y\right))$\\ 
       $\ket{-}$ & $\frac{\pi}{2}$ & $\pi$             & (-1, 0, 0) & $\exp(-i\frac{J}{2}\left(\sigma_A^x\sigma_S^z + \sigma_A^y\sigma_S^y\right))$\\
       $\ket{i}$ & $\frac{\pi}{2}$ & $\frac{\pi}{2}$   & (0, 1, 0) & $\exp(-i\frac{J}{2}\left(\sigma_A^y\sigma_S^x - \sigma_A^x\sigma_S^z\right))$ \\
       $\ket{-i}$ & $\frac{\pi}{2}$ & $\frac{3\pi}{2}$ & (0, -1, 0)  & $\exp(-i\frac{J}{2}\left(\sigma_A^x\sigma_S^z -\sigma_A^y\sigma_S^x\right) )$
      \end{tabular}
      \vspace{-0.05in}
      \caption{Single qubit stabilizers parameterized by angles $\theta$ and $\phi$ the steering operator $U_{A,S}$ for the MIQS protocol.\label{tab:stabilizers}}
  \end{table}
Following Section~\ref{sec:qubit-qubit-miqs}, we develop the quantum circuits for each desired stabilizer state. 
We ran the experiment $30$ times, with $1024$ shots each, using quantum process tomography to estimate the density state of the system at each step $n$ of the MIQS protocol.
Figure~\ref{fig:sqrt-swap-steer-ibmq} shows the average result, along with error bars, of running the circuit from Figure~\ref{fig:sqrt-swap-circuit} to prepare $\ket{\psi_\oplus} = \ket{+}$ for $n$ up to 30.
The error bars indicate the decoherence associated with the system qubit.
Namely, for increased $n$, we see an increase in uncertainty of the measured density state.
We then compute the fidelity for all stabilizer states, and find their average. 


\begin{figure}[tp]
  \vspace{-0.2in}
  \subfloat[Average state fidelity between $\rho^n$ and target state. \label{fig:avg-fidelity-plus}]{%
    \includegraphics[width=0.5\linewidth]{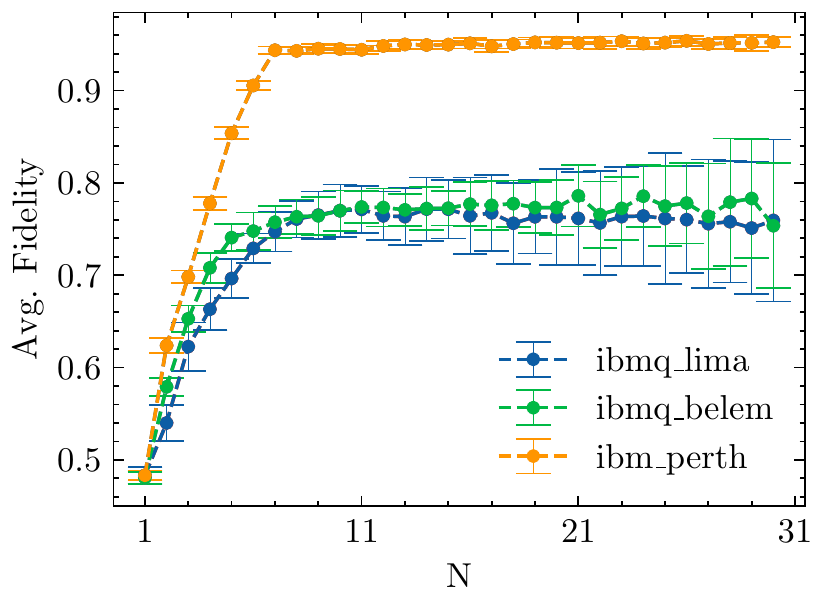}%
  }
  \subfloat[Steering inequaility. \label{fig:steering-inequality}]{%
    \includegraphics[width=0.5\linewidth]{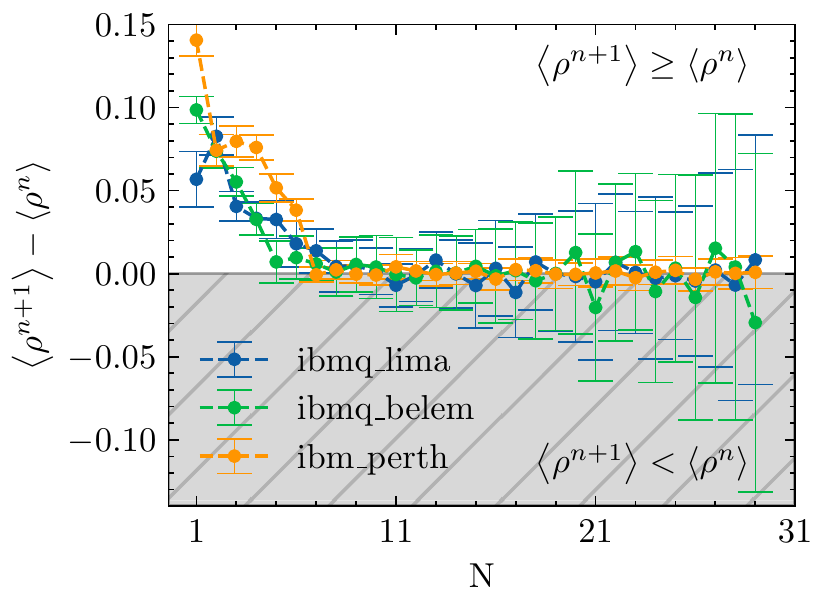}%
  }

  \vspace{-0.1in}
  \caption{Convergence of qubit fidelity throughout the execution of the steering protocol.
  (a) Depicts the estimated fidelity across three IBM quantum machines, with the best fidelity being achieved by \textit{ibm\_perth}.
  (b) Shows that the steering inequality given by Equation~\ref{eq:steer-ineq} is satisfied.}
\end{figure}
\begin{figure*}[tp]
  \centering
  \subfloat[Average fidelity of preparing stabilizer states versus the number of repetitions $N$ with different coupling strengths $J$. For certains values of $J$, the fidelity decreases at first before increasing.\label{fig:fid-v-N}]{%
  \includegraphics[width=0.35\linewidth]{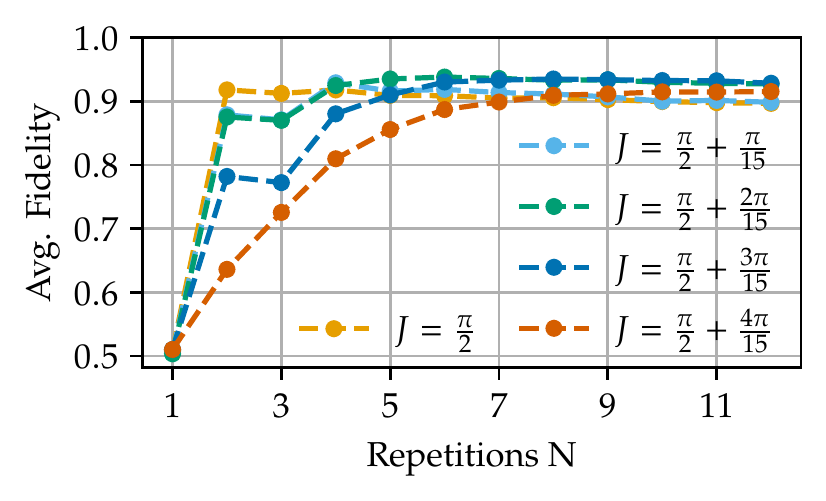}
  }
  \hspace{0.1in}
  \subfloat[Average fidelity of steering to all stabilizer states with different coupling strengths $J$. The number of repetitions of the protocol (vertical dots) is optimally chosen for each $J$. Maximum fidelity of $93 \pm 1\%$ is observed for $J = \pi/2 + \pi/8$. \label{fig:avg-stabilizer}]{%
    \includegraphics[width=0.35\linewidth]{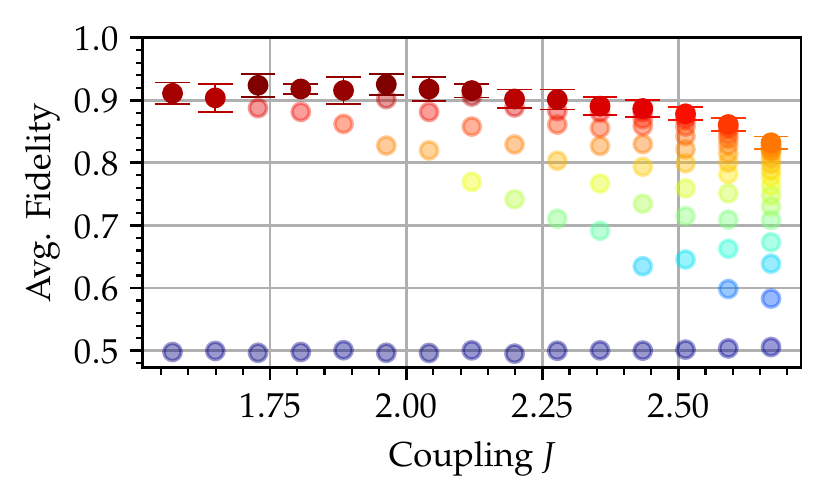}
  }
  \vspace{-0.1in}
  \caption{Preparation of qubit stabilizer states with various coupling parameter $J$. The fidelity is given as an average of all stabilizer states. All experiments are performed on \textit{ibm\_perth}.}
\end{figure*}
Figure~\ref{fig:avg-fidelity-plus} shows the average fidelity for all single-qubit stabilizer states.
Furthermore, Figure~\ref{fig:steering-inequality} confirms that the steering inequality (Equation~\ref{eq:steer-ineq}) is satisfied.
The quantum computer ibmq\_perth, achieved the highest overall fidelity and stability.

As noted in Section~\ref{sec:geometry}, the qubit-qubit operator $U_{A,S}$ can be characterized by the coupling parameter $J$.
In theory, the parameter is associated with the strength of entanglement necessary.
To experimentally analyze the role that $J$ plays, we prepare the stabilizer states with varying coupling $J$.
Figure~\ref{fig:fid-v-N} shows the fidelity of preparing the $\ket{+}$ state for varying $J$ on ibmq\_perth.
Although $J=\pi/2$ achieves the fastest convergence, it does not correspond to the highest fidelity.
Figure~\ref{fig:avg-stabilizer} shows the average of steering all the stabilizer states as computed by
\vspace{-0.2in}
\begin{equation}
  \mathcal{F} = \frac{1}{6}\sum_{i=1}^6\bra{\psi_i}\rho_i\ket{\psi_i}.
  \vspace{-0.1in}
\end{equation}
On average, the fidelity tends to decrease with smaller $J$ values.
Table~\ref{table:NISQ_Devices} summarizes the details and the published backend characteristics of the quantum computers.

\begin{figure}[tp]
  \centering
  \vspace{-0.1in}
  \includegraphics[width=0.9\linewidth]{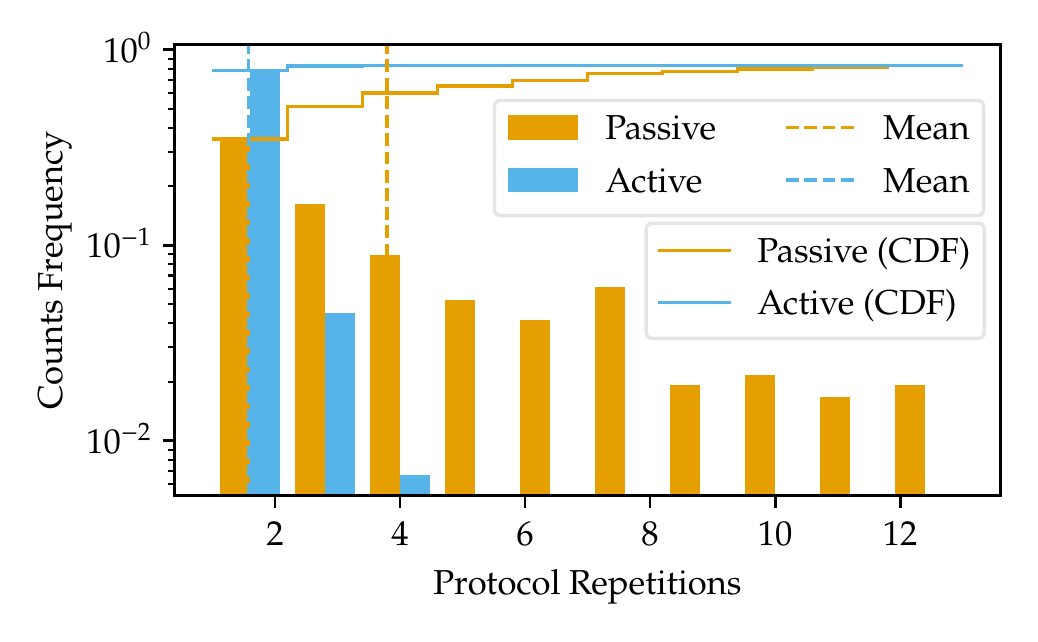}%
    \vspace{-0.2in}
  \caption{Histogram of protocol repetitions (effort) for preparing stabilizer states with varying steering operators determined by coupling strength $J$.
  Passive steering exhibits a Poissonian process \cite{herasymenkoMeasurementdrivenNavigationManybody2022a}, with an exponential decaying count frequency (log scale).
  The mean number of repetitions is $\mathcal{N}_{\mathrm{mean}}^{\mathrm{passive}} \approx 3.8$.
  The active approach has a $2.5$ times improvement compared to the passive approach with a mean repetition of $\mathcal{N}_{\mathrm{mean}}^{\mathrm{active}}\approx 1.6$.
  The cumulative distribution function (CDF) is also shown, further displaying the faster convergence of the active protocol.
   \label{fig:passive-vs-active-time}}
\end{figure}

Figure~\ref{fig:passive-vs-active-time} takes that average number of repetitions (application of ancilla-system entanglement operation in Figure~\ref{fig:sqrt-swap-circuit}) needed to obtain a fidelity $\mathcal{F} > 0.9$ and compares it against the active steering approach. Note that we end the protocol once the readout of the ancilla is a $1$. Each bar in the figure indicates what percentage of runs lead to the desired fidelity. For example, the leftmost bar shows that the passive quantum steering can reach the desired fidelity 10\% of the time (e.g., out of 100 runs) if we apply the entanglement operation only once ($n$=1 in Figure~\ref{fig:sqrt-swap-circuit}).  




\vspace{-0.2in}
\subsection{Evaluation of Qubit-Qutrit Protocol}
\vspace{-0.1in}

Quantum control beyond the two-level system has been exploited in superconducting quantum processors since the beginning of this technology.
Examples include utilizing the higher levels for qubit readout \cite{martinisRabiOscillationsLarge2002, cooperObservationQuantumOscillations2004, luceroHighFidelityGatesSingle2008}, faster qubit initialization \cite{valenzuelaMicrowaveInducedCoolingSuperconducting2006}, and spin-1 quantum simulation \cite{neeleyEmulationQuantumSpin2009}.
Steps towards ternary quantum computation with superconducting transmon devices have developed in the last 10 years \cite{bianchettiControlTomographyThree2010, abdumalikovElectromagneticallyInducedTransparency2010, abdumalikovjrExperimentalRealizationNonAbelian2013a, jergerContextualityNonlocalitySuperconducting2016, tanTopologicalMaxwellMetal2018, honigl-decrinisMixingCoherentWaves2018, vepsalainenSimulatingSpinChains2020, fedorovImplementationToffoliGate2012}. Recently, these efforts have led to
the implementation of high-fidelity single-qutrit gates
\cite{yurtalanImplementationWalshHadamardGate2020, morvanQutritRandomizedBenchmarking2021}. 


Many physical devices, such as superconducting transmons, naturally have higher-energy states which are often ignored to realize qubits. However, controlling the higher-energy states can be tricky, requiring additional techniques to produce a desired evolution. Our goal is to prepare a qutrit in an arbitrary state utilizing an ancilla qubit.
However, controlling qutrits can be a difficult task. There are various factors that need to be calibrated, such as frequency of the drive, amplitidue of the drive, leakage, etc. We believe MIQS can simplify initialization of a qutrit, by coupling it to a qubit. 

We demonstrate the protocol by the preparing an equal superposition qutrit state
  \vspace{-0.05in}
\begin{equation}
  \ket{\psi_\oplus} = \frac{1}{\sqrt{3}}\left(\ket{0} + \ket{1} + \ket{2}\right)
    \vspace{-0.05in}
\end{equation}
via a qubit-qutrit operator as defined by Equation~\ref{eq:qutrit-hamiltonian}.
The protocol is repeated $N$ times, where at each step $n$ we perform qutrit quantum state tomography (see Appendix~\ref{sec:qutrit-tomography}). 
Figure~\ref{fig:qutrit-fidelity} shows the estimated average fidelity at each step $n$ on ibmq\_perth.
In comparison with the qubit case, the qutrit fidelity has increased error as a result of:
(1) measurement error for classifying the $\ket{2}$ state,
(2) coherence time of the $\ket{2}$ state,
(3) heightened complexity of perform full qutrit state tomography.





\begin{figure}[htp]
  \includegraphics[width=0.8\linewidth]{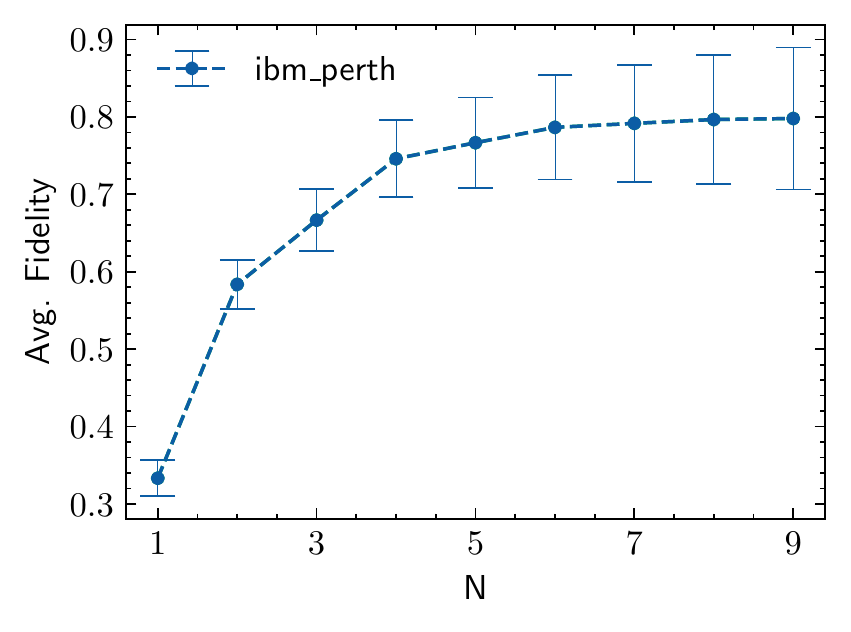}%
  \vspace{-0.2in}
  \caption{Average qutrit state fidelity between $\rho^n$ and the desired target state $\ket{\psi_\oplus} = \frac{1}{\sqrt{3}}\left(\ket{0} + \ket{1} + \ket{2}\right)$. The errors are primarily from inherent measurement error in discriminating the qutrit state, weaker $T_1$ coherence time of the $\ket{1}\to\ket{2}$ subspace, and increased overhead in performing qutrit state tomography. We obtained a state fidelity of $80\pm 9\%$.   \label{fig:qutrit-fidelity} }
\end{figure}



\vspace{-0.2in}
\section{Conclusions and Outlooks}
\vspace{-0.1in}

A major challenge in quantum computing is efficiently preparing an initial (arbitrary) state.
We experimentally demonstrate measurement-induced steering on contemporary superconducting quantum computer to prepare arbitrary qubit and qutrit states.
By applying a simple repetition of gates and ancilla measurements, we generate arbitrary qubit states with fidelity $93 \pm 1\%$ and arbitrary qutrit states with fidelity $80\pm 9\%$.
To achieve this, we generate optimal quantum circuits that implement the steering operator, and experimentally reconstruct the density states via quantum state tomography to obtain the fidelity.
We explored the dependence of a tunable parameter that relates fidelity convergence with the number of repetitions of the protocol. 
Additionally, we noted that by taking advantage of readout outcomes, we may accelerate the convergence. 
Furthermore, for qutrit functionality, we calibrate qutrit gates using the pulse-level programming model Qiskit Pulse via cloud access to IBM Quantum devices.

Traditionally, the fidelity of an initialized state and the fidelity of a quantum gate are considered independently. 
We demonstrate that by utilizing the programmability of a digital quantum processor, arbitrary quantum state can be prepared via a simple protocol of repeatedly executing the same small set of quantum gates. The success of the protocol -- achieving high state initialization fidelity -- depends primarily on the fidelity of the quantum gates and stability of qubits. Therefore, from a quantum engineers point of view, the task of state preparation may be considered a byproduct of achieving high gate fidelity.
Additionally, we demonstrate state preparation of a qutrit, escaping the conventional notation of a binary quantum system.
From a quantum technology point of view, the ability to access more quantum information in higher dimensions has direct advantages in quantum error-correcting codes, as well as asymptotic improvements in computation in comparison with binary computation.
Traditional control of a qutrit introduces further engineering overhead, such as careful calibration of drive frequency, drive amplitude, and phases.
From a device design standpoint, several compromises need to be made, including speed of readout versus the coherence of a qutrit.
However, for the task of qutrit state preparation via steering, a specific subclass of qutrit gates is needed to prepare an arbitrary state which lowers the engineering overhead. We demonstrated the necessary calibrations and executions of qutrit gates on superconducting transmons to prepare an equal-superposition qutrit state. We believe this research paves a path to reliably prepare higher-dimensional quantum states on experimental platforms. 

Future work in utilizing steering for state preparation on experimental quantum devices consists of several challenges and possible directions:

\textit{Entangled-state preparation:} highly-entangled states are crucial for implementing error-correcting codes and performing quantum information processing. However, preparing an arbitrary entangled state via steering requires appropriately coupling to measurement-capable ancilla qubits. Contemporary superconducting quantum devices have restrictive device connectivity between qubits, which introduces additional overhead to transfer quantum information (i.e. via SWAP). Trapped ion quantum computers may be better suited for this task due to all-to-all coupling between qubits. Unfortunately, compared to superconducting qubits, measurement operations on trapped ion qubits are more disruptive due to stray light \cite{gaeblerSuppressionMidcircuitMeasurement2021}. Assessing the feasibility of steering on various contemporary hardware platforms remains an open challenge. 

\textit{Device-specific measurement:} it is rarely the case that measurements are conducted on a qubit directly. Instead, measurement typically observes what effect a system $\ket{\psi}$ has on an environment. Generally, the system is coupled with an apparatus $\ket{\theta}$ to give an overall state $\ket{\Psi} = U\ket{\theta}\otimes\ket{\psi}$ after an entangling operation $U$. Then a measurement is conducted on the apparatus which disentangles it from the system. For example, superconducting transmon qubits are measured through a readout resonator which couples with the transmon. A frequency shift of the resonator is observed depending on the state of the transmon \cite{mcclureRapidDrivenReset2016}. Therefore, assuming an appropriate entanglement $U$, it is possible to utilize quantum steering to prepare arbitrary system quantum states by coupling and measuring an apparatus -- thereby reducing the overall use of expensive qubits to act as ancillas.

\textit{Parameterized quantum algorithms:} many near-term quantum algorithms utilize parameterized quantum circuits to prepare quantum states such that an expectation value is minimized \cite{peruzzoVariationalEigenvalueSolver2014a}. Unfortunately, parameterized circuits suffer from barren plateaus whereby a classical optimizer is unable to solve the high-dimensional non-convex optimization \cite{grantInitializationStrategyAddressing2019a}. Quantum steering provides theoretical guarantee to state initialization, and may overcome pitfalls in traditional parameterized quantum circuits. Namely, active steering provides a feedback mechanism whereby the optimization may be aided by conducting local decisions rather than finding a global optimal directly. 

\textit{Steering quantum gates:} certain systems contain a dark space that is spanned by several dark states.
A closed (non-)adiabatic trajectory can be used to induce a unitary operator in the dark space \cite{wilczekAppearanceGaugeStructure1984a, snizhkoNonAbelianGeometricDephasing2019a}.
In other words, the generalization of the Berry phase -- a non-abelian holonomy -- can be used to realize quantum gates \cite{zanardiHolonomicQuantumComputation1999b}.
An intriguing direction is to study the role that a steering protocol may play in realizing quantum gates via a holonomy.


\vspace{-0.2in}
\begin{acknowledgments}
\vspace{-0.1in}
The authors gratefully thank the IBM Quantum team and the services offered through the IBM Quantum Researchers Program.
The authors also acknowledge support from the National Science Foundation, Grant No. CCF-1908131.
\end{acknowledgments}

\appendix

\vspace{-0.2in}
\section{Operator-Sum Representation}
\vspace{-0.1in}


To analyze the recurrence relation given by Equation~(\ref{eq:meas-steer}),  it is helpful to utilize the theory of open quantum systems. Specifically, we may diagonalize the state of the ancilla qubits, $\rho_A = \sum_i p_i \ket{\psi_i}\bra{\psi_i}$, and evaluate the partial trace:
\begin{equation}
  \begin{split}
    \rho_S^{n+1} & = \mathrm{Tr}_A \left[ U \left( \sum_i p_i \ket{\psi_i}_A \bra{\psi_i}_A \otimes \rho_S^{n}\right) U^\dagger\right] \\
    & = \sum_k \bra{k}_A U \left( \sum_i p_i \ket{\psi_i}_A \bra{\psi_i}_A \otimes \rho_S^{n}\right) U^\dagger \ket{k}_A \\
    & = \sum_k \sum_i A_{k,i} \rho_S^n A_{k,i}^\dagger.
  \end{split}
\end{equation}
The Kraus operators $A_{k,i} = \sqrt{p_i} \bra{k}U\ket{\psi_i}$
express the evolution of the system $\rho_S$ assuming that it is initially separable from the ancilla. In our work, we prepare the ancillas to known states, such as $\ket{\psi_A}$ = $\{\ket{0}, \ket{1}\}$. Therefore, we have a fixed $\ket{\psi_i}$ and need only Kraus operators $A_k$. Hence, the evolution is
\vspace{-0.05in}
\begin{equation}\label{eq:steer-kraus}
  \rho_S^{n+1} = \sum_k A_k \rho_n A_k^\dagger
\vspace{-0.05in}
\end{equation}
where $k$ enumerates the possible measurement outcomes of the ancilla.

\vspace{-0.2in}
\section{Analysis of Steering to $\ket{+}$}\label{sec:eg-single-qubit}
\vspace{-0.1in}

The ancilla-system entanglement operator that drives the system to $\ket{+}$ is given by Equation~\ref{eq:U_plus}. The Kraus operators that govern system-qubit evolution are given by:
\vspace{-0.1in}
\begin{equation}
  \begin{split}
    A_0 = \bra{0_A}U\ket{1_A} = \frac{\cos J - 1}{2}\mathbb{I_S} + \frac{1-\cos J}{2}\sigma_S^x\\
    A_1 = \bra{1_A}U\ket{1_A} = \frac{\sin J}{2} \left(\sigma_S^y + i\sigma_S^z\right).
  \end{split}
\vspace{-0.05in}
\end{equation}
By solving the operator-sum evolution given by Equation~\ref{eq:steer-kraus}, we have the following recurrence relations for the components of $\vec{s}(n)$
  \begin{align}
    s_x(n) &= 1 - \cos^{2n}(J)s_x(0) + \cos^{2n}(J)s_x(0), \nonumber\\
    s_y(n) &= \cos^{n}(J)s_y(0), \nonumber \\
    s_z(n) &= \cos^{n}(J)s_z(0).
  \end{align}

This relation explicitly shows that
\begin{equation}
  \lim_{n \to \infty} s(n) = (1, 0, 0); \quad 0 < J < \pi.
\end{equation}
In other words, the system state converges exponentially to our desired $\ket{+}\bra{+} = (\mathbb{I} + \sigma_x)/2$ state with respect to the number of steps $n$ and does not depend on the initial conditions. Furthermore, the fastest convergence is achieved in one step with $J=\pi/2$.

\vspace{-0.2in}
\section{Chip Characterization}\label{sec:chip}
\vspace{-0.1in}

In this paper, we used the IBM Quantum Falcon Processors \emph{ibmq\_lima}, \emph{ibmq\_belem}, and \emph{ibm\_perth}. Qiskit Pulse was used to perform pulse-level control, particularly to establish qutrit operations.

\begin{figure}[h]
  \centering
  \includegraphics[width=0.7\linewidth]{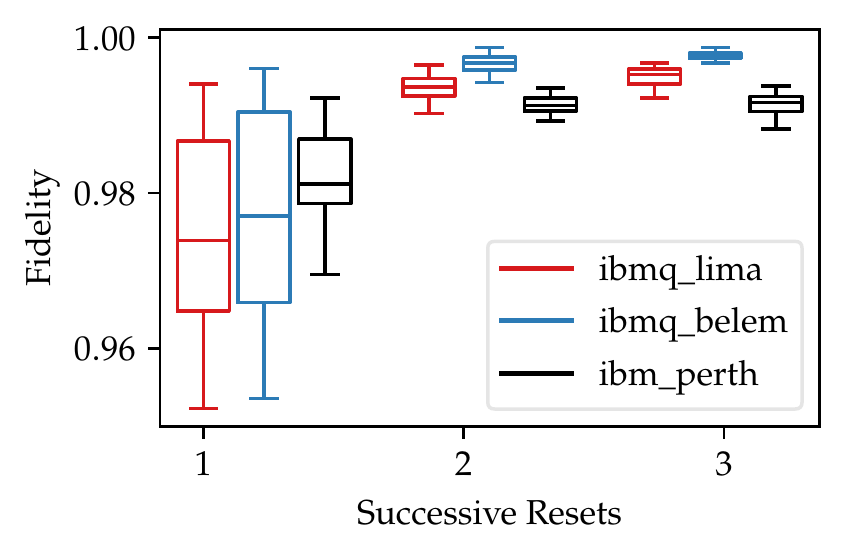}
  \caption{Spread of active reset fidelity in initializing $\ket{0}$ on different IBM Quantum computers. Active reset measures the qubit, classically checks the readout, and then rotates the qubit if necessary.\label{reset-fidelity}}
\end{figure}

In the steering protocol, the ancilla qubit must be reset to a known state. This is achieved on IBM Quantum Computers via a mid-circuit qubit active reset. We benchmark the probability of the measurement result $\ket{0}$ as shown in Figure~\ref{reset-fidelity}. Applying several consecutive active resets improves the fidelity of preparing $\ket{0}$.  

\begin{figure}[htp]
  \centering
  \subfloat[Frequency sweep of qutrit on \emph{ibm\_perth}. \label{fig:freq}]{%
    \centering
    \includegraphics[width=0.45\linewidth]{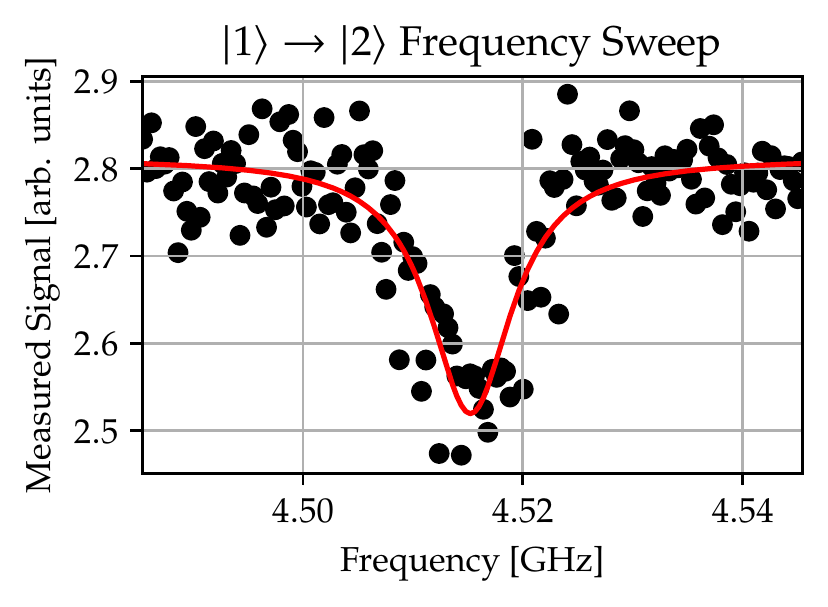}
  }\hfill
  \subfloat[Rabi $1\to 2$ experiments performed to calibrate the amplitude of the $\pi_{1\to 2}$ pulse on \emph{ibm\_perth}. \label{fig:rabi}]{%
    \centering
    \includegraphics[width=0.45\linewidth]{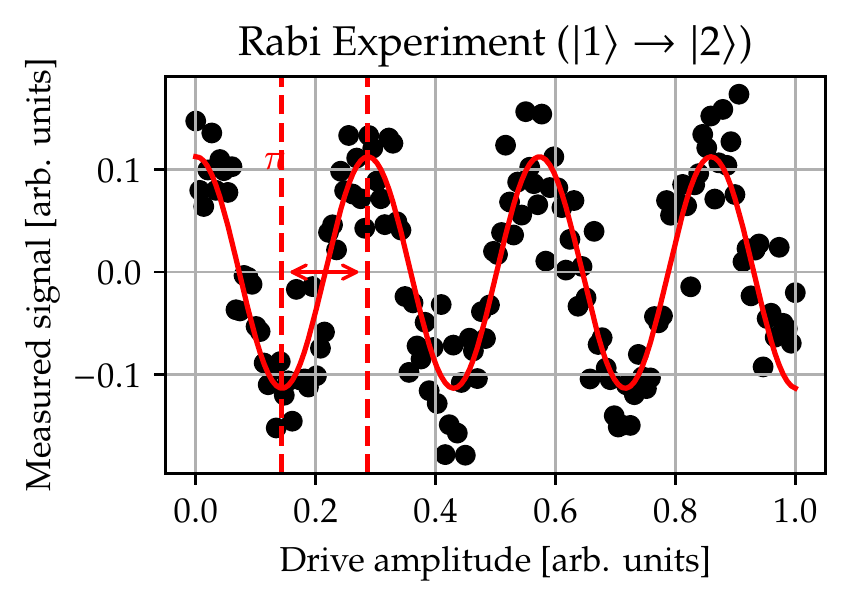}
  }
  \caption{Frequency and amplitude sweeps to define qutrit operations in the $\ket{1} \to \ket{2}$ subspace of the superconducting transmon. }
\end{figure}

For the transmons that implements our qutrit, we first found the transition frequency $f_{12}$. Figure~\ref{fig:freq} shows this sweep in frequency to find the excitation. We then performed a Rabi experiment to obtain the amplitude of the $\pi_{1\to2}$ pulse to define rotations in the $(12)$ sub-space. Figure~\ref{fig:rabi} shows the result of this calibration. 

\begin{figure}[htp]
  \centering
  \includegraphics[width=1\linewidth]{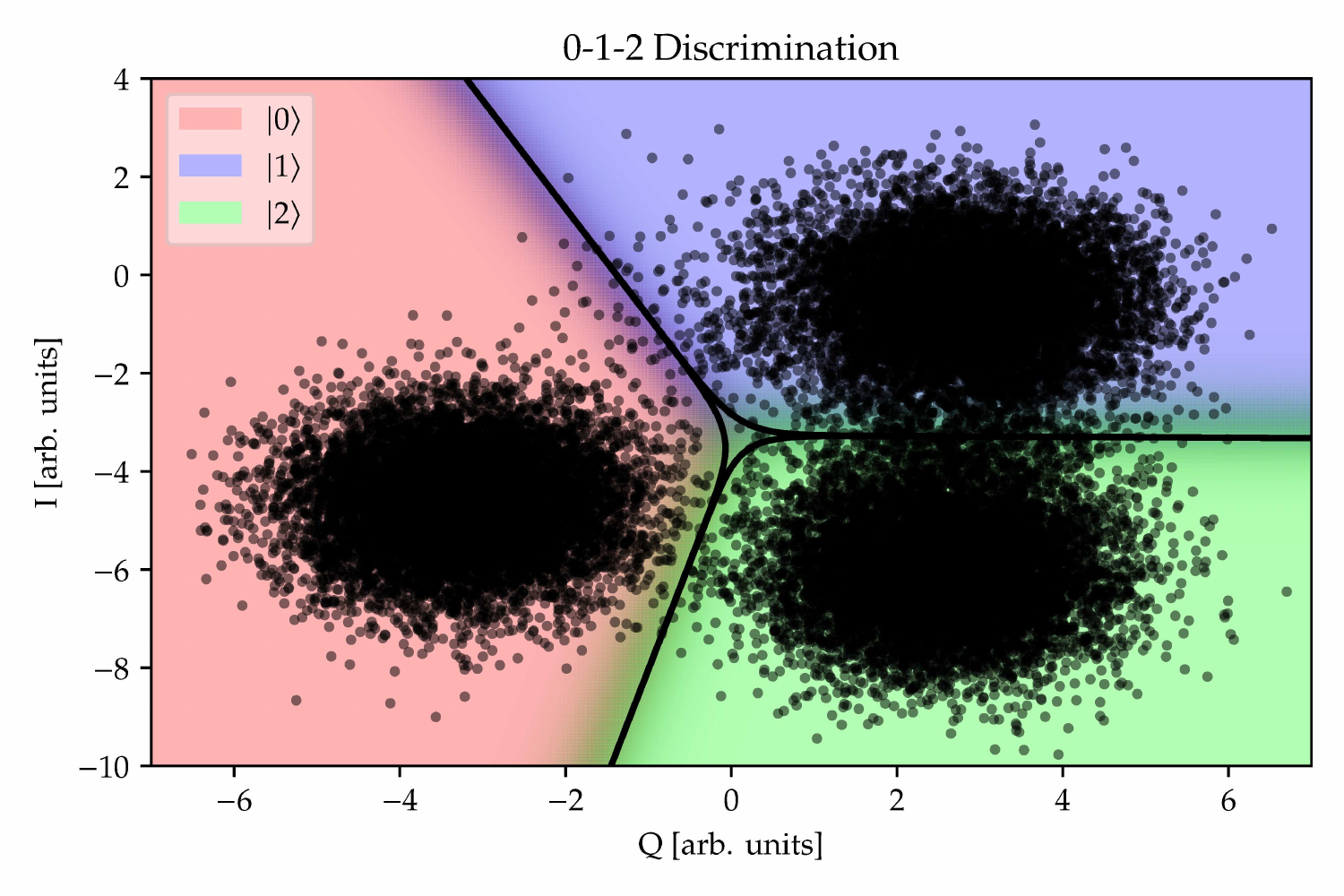}
  \caption{Discriminator to classify the measurement results for a superconducting transmon qutrit. The accuracy of the discriminator is 0.917. \label{fig:qutrit-discriminator}}
  \vspace{-0.1in}
\end{figure}

 The measurement discriminator to classify qutrit states is shown in Figure~\ref{fig:qutrit-discriminator}, with an accuracy of 0.917. To improve the accuracy of the discriminator, measurement error mitigation was performed by correcting the average counts via a correction matrix.
The matrix was generated by preparing 6 basis input states ($\ket{00}, \ket{01}, \ket{02}, \ket{10}, \ket{11}, \ket{12}$) and computing the corresponding probabilities of measuring counts in other basis states.

\section{Qutrit Quantum State Tomography}\label{sec:qutrit-tomography}

In the qubit case, an arbitrary qubit density state
\begin{equation}
  \rho = \frac{1}{2}(I + a_x\sigma_x + a_y\sigma_y + a_z\sigma_z),
\end{equation}
with real parameters $a_j$, can be recovered by computing the expectation values of the Pauli matrices.
For example the expectation value of the $\sigma_x$ Pauli matrix
    \begin{align}
      \expectationvalue{\sigma_x} = \frac{1}{2} (a_x \cdot 2) = a_x
    \end{align}
yields the coefficient $a_x$ of our density state.
We utilized the fact that Pauli matrices follow the identity $\mathrm{Tr}(\sigma_\alpha \sigma_\beta) = 2\delta_{\alpha,\beta}$.
The expectation value of the Pauli matrices has a direct relation with the density matrix of the state. Therefore by computing the expectation values of an appropriate set of observables we can compute the density state.

In general, given an observable $M$, we may diagonalize it by a unitary matrix $U$ and a diagonal matrix with real entries $\Lambda$ corresponding to the eigenvalues
\begin{align}
  \bra{\psi} M \ket{\psi} &= \bra{\psi} U^\dagger \Lambda U \ket{\psi} = \bra{\psi^\prime} \Lambda \ket{\psi^\prime} \nonumber \\
  &= \sum_i \braket{\psi^\prime}{i}\bra{i}\Lambda\ket{i}\braket{i}{\psi^\prime} = \sum_i \lambda_i |\bra{i}\ket{\psi^\prime}|^2.
\end{align}
Therefore, we let the quantum computer perform the operation $U$, and then measure in a standard basis $\ket{i}$.
The expectation value of the observable is then recovered by multiplying the outcomes by the eigenvalues $\lambda_i$.

In our qutrit case, a general density state has the form
\begin{equation}
  \rho = \frac{1}{3} \mathbb{I}_3 + \vec{n} \cdot \vec{\lambda}
\end{equation}
where $\vec{n} = (n_1, n_2, \hdots, n_8)$ are 8 real parameters and $\vec{\lambda} = (\lambda_1, \lambda_2, \hdots \lambda_8)$ are $3\times3$ Gell-Mann matrices. Similar to the Pauli matrices, the Gell-Mann matrices satisfy the identity $\mathrm{Tr}(\lambda_\alpha \lambda_\beta) = 2\delta_{\alpha,\beta}$.
Therefore, computing the expectation value $\expectationvalue{\lambda_i}$ of a qutrit density state will uncover the coefficient $n_i$.


\bibliography{phy-steer}

\begin{thebibliography}{68}%
\makeatletter
\providecommand \@ifxundefined [1]{%
 \@ifx{#1\undefined}
}%
\providecommand \@ifnum [1]{%
 \ifnum #1\expandafter \@firstoftwo
 \else \expandafter \@secondoftwo
 \fi
}%
\providecommand \@ifx [1]{%
 \ifx #1\expandafter \@firstoftwo
 \else \expandafter \@secondoftwo
 \fi
}%
\providecommand \natexlab [1]{#1}%
\providecommand \enquote  [1]{``#1''}%
\providecommand \bibnamefont  [1]{#1}%
\providecommand \bibfnamefont [1]{#1}%
\providecommand \citenamefont [1]{#1}%
\providecommand \href@noop [0]{\@secondoftwo}%
\providecommand \href [0]{\begingroup \@sanitize@url \@href}%
\providecommand \@href[1]{\@@startlink{#1}\@@href}%
\providecommand \@@href[1]{\endgroup#1\@@endlink}%
\providecommand \@sanitize@url [0]{\catcode `\\12\catcode `\$12\catcode `\&12\catcode `\#12\catcode `\^12\catcode `\_12\catcode `\%12\relax}%
\providecommand \@@startlink[1]{}%
\providecommand \@@endlink[0]{}%
\providecommand \url  [0]{\begingroup\@sanitize@url \@url }%
\providecommand \@url [1]{\endgroup\@href {#1}{\urlprefix }}%
\providecommand \urlprefix  [0]{URL }%
\providecommand \Eprint [0]{\href }%
\providecommand \doibase [0]{https://doi.org/}%
\providecommand \selectlanguage [0]{\@gobble}%
\providecommand \bibinfo  [0]{\@secondoftwo}%
\providecommand \bibfield  [0]{\@secondoftwo}%
\providecommand \translation [1]{[#1]}%
\providecommand \BibitemOpen [0]{}%
\providecommand \bibitemStop [0]{}%
\providecommand \bibitemNoStop [0]{.\EOS\space}%
\providecommand \EOS [0]{\spacefactor3000\relax}%
\providecommand \BibitemShut  [1]{\csname bibitem#1\endcsname}%
\let\auto@bib@innerbib\@empty
\bibitem [{\citenamefont {Kak}(1999)}]{kakInitializationProblemQuantum1999}%
  \BibitemOpen
  \bibfield  {author} {\bibinfo {author} {\bibfnamefont {S.}~\bibnamefont {Kak}},\ }\bibfield  {title} {\bibinfo {title} {The {{Initialization Problem}} in {{Quantum Computing}}},\ }\href {https://doi.org/10.1023/A:1018877706849} {\bibfield  {journal} {\bibinfo  {journal} {Foundations of Physics}\ }\textbf {\bibinfo {volume} {29}},\ \bibinfo {pages} {267} (\bibinfo {year} {1999})}\BibitemShut {NoStop}%
\bibitem [{\citenamefont {DiVincenzo}(2000)}]{divincenzoPhysicalImplementationQuantum2000a}%
  \BibitemOpen
  \bibfield  {author} {\bibinfo {author} {\bibfnamefont {D.~P.}\ \bibnamefont {DiVincenzo}},\ }\bibfield  {title} {\bibinfo {title} {The {{Physical Implementation}} of {{Quantum Computation}}},\ }\href {https://doi.org/10.1002/1521-3978(200009)48:9/11<771::AID-PROP771>3.0.CO;2-E} {\bibfield  {journal} {\bibinfo  {journal} {Fortschritte der Physik}\ }\textbf {\bibinfo {volume} {48}},\ \bibinfo {pages} {771} (\bibinfo {year} {2000})}\BibitemShut {NoStop}%
\bibitem [{\citenamefont {Kitaev}(1997)}]{kitaevQuantumComputationsAlgorithms1997}%
  \BibitemOpen
  \bibfield  {author} {\bibinfo {author} {\bibfnamefont {A.~Y.}\ \bibnamefont {Kitaev}},\ }\bibfield  {title} {\bibinfo {title} {Quantum computations: Algorithms and error correction},\ }\href {https://doi.org/10.1070/RM1997v052n06ABEH002155} {\bibfield  {journal} {\bibinfo  {journal} {Russ. Math. Surv.}\ }\textbf {\bibinfo {volume} {52}},\ \bibinfo {pages} {1191} (\bibinfo {year} {1997})}\BibitemShut {NoStop}%
\bibitem [{\citenamefont {Rigetti}\ \emph {et~al.}(2012)\citenamefont {Rigetti}, \citenamefont {Gambetta}, \citenamefont {Poletto}, \citenamefont {Plourde}, \citenamefont {Chow}, \citenamefont {C{\'o}rcoles}, \citenamefont {Smolin}, \citenamefont {Merkel}, \citenamefont {Rozen}, \citenamefont {Keefe}, \citenamefont {Rothwell}, \citenamefont {Ketchen},\ and\ \citenamefont {Steffen}}]{rigettiSuperconductingQubitWaveguide2012a}%
  \BibitemOpen
  \bibfield  {author} {\bibinfo {author} {\bibfnamefont {C.}~\bibnamefont {Rigetti}}, \bibinfo {author} {\bibfnamefont {J.~M.}\ \bibnamefont {Gambetta}}, \bibinfo {author} {\bibfnamefont {S.}~\bibnamefont {Poletto}}, \bibinfo {author} {\bibfnamefont {B.~L.~T.}\ \bibnamefont {Plourde}}, \bibinfo {author} {\bibfnamefont {J.~M.}\ \bibnamefont {Chow}}, \bibinfo {author} {\bibfnamefont {A.~D.}\ \bibnamefont {C{\'o}rcoles}}, \bibinfo {author} {\bibfnamefont {J.~A.}\ \bibnamefont {Smolin}}, \bibinfo {author} {\bibfnamefont {S.~T.}\ \bibnamefont {Merkel}}, \bibinfo {author} {\bibfnamefont {J.~R.}\ \bibnamefont {Rozen}}, \bibinfo {author} {\bibfnamefont {G.~A.}\ \bibnamefont {Keefe}}, \bibinfo {author} {\bibfnamefont {M.~B.}\ \bibnamefont {Rothwell}}, \bibinfo {author} {\bibfnamefont {M.~B.}\ \bibnamefont {Ketchen}},\ and\ \bibinfo {author} {\bibfnamefont {M.}~\bibnamefont {Steffen}},\ }\bibfield  {title} {\bibinfo {title} {Superconducting qubit in a waveguide cavity with a coherence time approaching 0.1 ms},\ }\href {https://doi.org/10.1103/PhysRevB.86.100506} {\bibfield  {journal} {\bibinfo  {journal} {Phys. Rev. B}\ }\textbf {\bibinfo {volume} {86}},\ \bibinfo {pages} {100506} (\bibinfo {year} {2012})}\BibitemShut {NoStop}%
\bibitem [{\citenamefont {Harty}\ \emph {et~al.}(2014)\citenamefont {Harty}, \citenamefont {Allcock}, \citenamefont {Ballance}, \citenamefont {Guidoni}, \citenamefont {Janacek}, \citenamefont {Linke}, \citenamefont {Stacey},\ and\ \citenamefont {Lucas}}]{hartyHighFidelityPreparationGates2014a}%
  \BibitemOpen
  \bibfield  {author} {\bibinfo {author} {\bibfnamefont {T.~P.}\ \bibnamefont {Harty}}, \bibinfo {author} {\bibfnamefont {D.~T.~C.}\ \bibnamefont {Allcock}}, \bibinfo {author} {\bibfnamefont {C.~J.}\ \bibnamefont {Ballance}}, \bibinfo {author} {\bibfnamefont {L.}~\bibnamefont {Guidoni}}, \bibinfo {author} {\bibfnamefont {H.~A.}\ \bibnamefont {Janacek}}, \bibinfo {author} {\bibfnamefont {N.~M.}\ \bibnamefont {Linke}}, \bibinfo {author} {\bibfnamefont {D.~N.}\ \bibnamefont {Stacey}},\ and\ \bibinfo {author} {\bibfnamefont {D.~M.}\ \bibnamefont {Lucas}},\ }\bibfield  {title} {\bibinfo {title} {High-{{Fidelity Preparation}}, {{Gates}}, {{Memory}}, and {{Readout}} of a {{Trapped-Ion Quantum Bit}}},\ }\href {https://doi.org/10.1103/PhysRevLett.113.220501} {\bibfield  {journal} {\bibinfo  {journal} {Phys. Rev. Lett.}\ }\textbf {\bibinfo {volume} {113}},\ \bibinfo {pages} {220501} (\bibinfo {year} {2014})}\BibitemShut {NoStop}%
\bibitem [{\citenamefont {Basilewitsch}\ \emph {et~al.}(2021)\citenamefont {Basilewitsch}, \citenamefont {Fischer}, \citenamefont {Reich}, \citenamefont {Sugny},\ and\ \citenamefont {Koch}}]{basilewitschFundamentalBoundsQubit2021a}%
  \BibitemOpen
  \bibfield  {author} {\bibinfo {author} {\bibfnamefont {D.}~\bibnamefont {Basilewitsch}}, \bibinfo {author} {\bibfnamefont {J.}~\bibnamefont {Fischer}}, \bibinfo {author} {\bibfnamefont {D.~M.}\ \bibnamefont {Reich}}, \bibinfo {author} {\bibfnamefont {D.}~\bibnamefont {Sugny}},\ and\ \bibinfo {author} {\bibfnamefont {C.~P.}\ \bibnamefont {Koch}},\ }\bibfield  {title} {\bibinfo {title} {Fundamental bounds on qubit reset},\ }\href {https://doi.org/10.1103/PhysRevResearch.3.013110} {\bibfield  {journal} {\bibinfo  {journal} {Phys. Rev. Res.}\ }\textbf {\bibinfo {volume} {3}},\ \bibinfo {pages} {013110} (\bibinfo {year} {2021})}\BibitemShut {NoStop}%
\bibitem [{\citenamefont {Tornow}\ \emph {et~al.}(2022)\citenamefont {Tornow}, \citenamefont {Kanazawa}, \citenamefont {Shanks},\ and\ \citenamefont {Egger}}]{tornowMinimumQuantumRunTime2022a}%
  \BibitemOpen
  \bibfield  {author} {\bibinfo {author} {\bibfnamefont {C.}~\bibnamefont {Tornow}}, \bibinfo {author} {\bibfnamefont {N.}~\bibnamefont {Kanazawa}}, \bibinfo {author} {\bibfnamefont {W.~E.}\ \bibnamefont {Shanks}},\ and\ \bibinfo {author} {\bibfnamefont {D.~J.}\ \bibnamefont {Egger}},\ }\bibfield  {title} {\bibinfo {title} {Minimum {{Quantum Run-Time Characterization}} and {{Calibration}} via {{Restless Measurements}} with {{Dynamic Repetition Rates}}},\ }\href {https://doi.org/10.1103/PhysRevApplied.17.064061} {\bibfield  {journal} {\bibinfo  {journal} {Phys. Rev. Appl.}\ }\textbf {\bibinfo {volume} {17}},\ \bibinfo {pages} {064061} (\bibinfo {year} {2022})}\BibitemShut {NoStop}%
\bibitem [{\citenamefont {Johnson}\ \emph {et~al.}(2012)\citenamefont {Johnson}, \citenamefont {Macklin}, \citenamefont {Slichter}, \citenamefont {Vijay}, \citenamefont {Weingarten}, \citenamefont {Clarke},\ and\ \citenamefont {Siddiqi}}]{johnsonHeraldedStatePreparation2012}%
  \BibitemOpen
  \bibfield  {author} {\bibinfo {author} {\bibfnamefont {J.~E.}\ \bibnamefont {Johnson}}, \bibinfo {author} {\bibfnamefont {C.}~\bibnamefont {Macklin}}, \bibinfo {author} {\bibfnamefont {D.~H.}\ \bibnamefont {Slichter}}, \bibinfo {author} {\bibfnamefont {R.}~\bibnamefont {Vijay}}, \bibinfo {author} {\bibfnamefont {E.~B.}\ \bibnamefont {Weingarten}}, \bibinfo {author} {\bibfnamefont {J.}~\bibnamefont {Clarke}},\ and\ \bibinfo {author} {\bibfnamefont {I.}~\bibnamefont {Siddiqi}},\ }\bibfield  {title} {\bibinfo {title} {Heralded {{State Preparation}} in a {{Superconducting Qubit}}},\ }\href {https://doi.org/10.1103/PhysRevLett.109.050506} {\bibfield  {journal} {\bibinfo  {journal} {Phys. Rev. Lett.}\ }\textbf {\bibinfo {volume} {109}},\ \bibinfo {pages} {050506} (\bibinfo {year} {2012})}\BibitemShut {NoStop}%
\bibitem [{\citenamefont {Rist{\`e}}\ \emph {et~al.}(2012)\citenamefont {Rist{\`e}}, \citenamefont {{van Leeuwen}}, \citenamefont {Ku}, \citenamefont {Lehnert},\ and\ \citenamefont {DiCarlo}}]{risteInitializationMeasurementSuperconducting2012}%
  \BibitemOpen
  \bibfield  {author} {\bibinfo {author} {\bibfnamefont {D.}~\bibnamefont {Rist{\`e}}}, \bibinfo {author} {\bibfnamefont {J.~G.}\ \bibnamefont {{van Leeuwen}}}, \bibinfo {author} {\bibfnamefont {H.-S.}\ \bibnamefont {Ku}}, \bibinfo {author} {\bibfnamefont {K.~W.}\ \bibnamefont {Lehnert}},\ and\ \bibinfo {author} {\bibfnamefont {L.}~\bibnamefont {DiCarlo}},\ }\bibfield  {title} {\bibinfo {title} {Initialization by {{Measurement}} of a {{Superconducting Quantum Bit Circuit}}},\ }\href {https://doi.org/10.1103/PhysRevLett.109.050507} {\bibfield  {journal} {\bibinfo  {journal} {Phys. Rev. Lett.}\ }\textbf {\bibinfo {volume} {109}},\ \bibinfo {pages} {050507} (\bibinfo {year} {2012})}\BibitemShut {NoStop}%
\bibitem [{\citenamefont {Kraus}\ \emph {et~al.}(2008)\citenamefont {Kraus}, \citenamefont {B{\"u}chler}, \citenamefont {Diehl}, \citenamefont {Kantian}, \citenamefont {Micheli},\ and\ \citenamefont {Zoller}}]{krausPreparationEntangledStates2008}%
  \BibitemOpen
  \bibfield  {author} {\bibinfo {author} {\bibfnamefont {B.}~\bibnamefont {Kraus}}, \bibinfo {author} {\bibfnamefont {H.~P.}\ \bibnamefont {B{\"u}chler}}, \bibinfo {author} {\bibfnamefont {S.}~\bibnamefont {Diehl}}, \bibinfo {author} {\bibfnamefont {A.}~\bibnamefont {Kantian}}, \bibinfo {author} {\bibfnamefont {A.}~\bibnamefont {Micheli}},\ and\ \bibinfo {author} {\bibfnamefont {P.}~\bibnamefont {Zoller}},\ }\bibfield  {title} {\bibinfo {title} {Preparation of entangled states by quantum {{Markov}} processes},\ }\href {https://doi.org/10.1103/PhysRevA.78.042307} {\bibfield  {journal} {\bibinfo  {journal} {Phys. Rev. A}\ }\textbf {\bibinfo {volume} {78}},\ \bibinfo {pages} {042307} (\bibinfo {year} {2008})}\BibitemShut {NoStop}%
\bibitem [{\citenamefont {Verstraete}\ \emph {et~al.}(2009)\citenamefont {Verstraete}, \citenamefont {Wolf},\ and\ \citenamefont {Ignacio~Cirac}}]{verstraeteQuantumComputationQuantumstate2009}%
  \BibitemOpen
  \bibfield  {author} {\bibinfo {author} {\bibfnamefont {F.}~\bibnamefont {Verstraete}}, \bibinfo {author} {\bibfnamefont {M.~M.}\ \bibnamefont {Wolf}},\ and\ \bibinfo {author} {\bibfnamefont {J.}~\bibnamefont {Ignacio~Cirac}},\ }\bibfield  {title} {\bibinfo {title} {Quantum computation and quantum-state engineering driven by dissipation},\ }\href {https://doi.org/10.1038/nphys1342} {\bibfield  {journal} {\bibinfo  {journal} {Nature Phys}\ }\textbf {\bibinfo {volume} {5}},\ \bibinfo {pages} {633} (\bibinfo {year} {2009})}\BibitemShut {NoStop}%
\bibitem [{\citenamefont {Weimer}\ \emph {et~al.}(2010)\citenamefont {Weimer}, \citenamefont {M{\"u}ller}, \citenamefont {Lesanovsky}, \citenamefont {Zoller},\ and\ \citenamefont {B{\"u}chler}}]{weimerRydbergQuantumSimulator2010}%
  \BibitemOpen
  \bibfield  {author} {\bibinfo {author} {\bibfnamefont {H.}~\bibnamefont {Weimer}}, \bibinfo {author} {\bibfnamefont {M.}~\bibnamefont {M{\"u}ller}}, \bibinfo {author} {\bibfnamefont {I.}~\bibnamefont {Lesanovsky}}, \bibinfo {author} {\bibfnamefont {P.}~\bibnamefont {Zoller}},\ and\ \bibinfo {author} {\bibfnamefont {H.~P.}\ \bibnamefont {B{\"u}chler}},\ }\bibfield  {title} {\bibinfo {title} {A {{Rydberg}} quantum simulator},\ }\href {https://doi.org/10.1038/nphys1614} {\bibfield  {journal} {\bibinfo  {journal} {Nature Phys}\ }\textbf {\bibinfo {volume} {6}},\ \bibinfo {pages} {382} (\bibinfo {year} {2010})}\BibitemShut {NoStop}%
\bibitem [{\citenamefont {Boykin}\ \emph {et~al.}(2002)\citenamefont {Boykin}, \citenamefont {Mor}, \citenamefont {Roychowdhury}, \citenamefont {Vatan},\ and\ \citenamefont {Vrijen}}]{boykinAlgorithmicCoolingScalable2002}%
  \BibitemOpen
  \bibfield  {author} {\bibinfo {author} {\bibfnamefont {P.~O.}\ \bibnamefont {Boykin}}, \bibinfo {author} {\bibfnamefont {T.}~\bibnamefont {Mor}}, \bibinfo {author} {\bibfnamefont {V.}~\bibnamefont {Roychowdhury}}, \bibinfo {author} {\bibfnamefont {F.}~\bibnamefont {Vatan}},\ and\ \bibinfo {author} {\bibfnamefont {R.}~\bibnamefont {Vrijen}},\ }\bibfield  {title} {\bibinfo {title} {Algorithmic cooling and scalable {{NMR}} quantum computers},\ }\href {https://doi.org/10.1073/pnas.241641898} {\bibfield  {journal} {\bibinfo  {journal} {Proceedings of the National Academy of Sciences}\ }\textbf {\bibinfo {volume} {99}},\ \bibinfo {pages} {3388} (\bibinfo {year} {2002})}\BibitemShut {NoStop}%
\bibitem [{\citenamefont {Fernandez}\ \emph {et~al.}(2004)\citenamefont {Fernandez}, \citenamefont {Lloyd}, \citenamefont {Mor},\ and\ \citenamefont {Roychowdhury}}]{fernandezAlgorithmicCoolingSpins2004}%
  \BibitemOpen
  \bibfield  {author} {\bibinfo {author} {\bibfnamefont {J.~M.}\ \bibnamefont {Fernandez}}, \bibinfo {author} {\bibfnamefont {S.}~\bibnamefont {Lloyd}}, \bibinfo {author} {\bibfnamefont {T.}~\bibnamefont {Mor}},\ and\ \bibinfo {author} {\bibfnamefont {V.}~\bibnamefont {Roychowdhury}},\ }\href {https://doi.org/10.48550/arXiv.quant-ph/0401135} {\bibinfo {title} {Algorithmic {{Cooling}} of {{Spins}}: {{A Practicable Method}} for {{Increasing Polarization}}}} (\bibinfo {year} {2004}),\ \Eprint {https://arxiv.org/abs/quant-ph/0401135} {arxiv:quant-ph/0401135} \BibitemShut {NoStop}%
\bibitem [{\citenamefont {Brassard}\ \emph {et~al.}(2014)\citenamefont {Brassard}, \citenamefont {Elias}, \citenamefont {Mor},\ and\ \citenamefont {Weinstein}}]{brassardProspectsLimitationsAlgorithmic2014}%
  \BibitemOpen
  \bibfield  {author} {\bibinfo {author} {\bibfnamefont {G.}~\bibnamefont {Brassard}}, \bibinfo {author} {\bibfnamefont {Y.}~\bibnamefont {Elias}}, \bibinfo {author} {\bibfnamefont {T.}~\bibnamefont {Mor}},\ and\ \bibinfo {author} {\bibfnamefont {Y.}~\bibnamefont {Weinstein}},\ }\bibfield  {title} {\bibinfo {title} {Prospects and limitations of algorithmic cooling},\ }\href {https://doi.org/10.1140/epjp/i2014-14258-0} {\bibfield  {journal} {\bibinfo  {journal} {Eur. Phys. J. Plus}\ }\textbf {\bibinfo {volume} {129}},\ \bibinfo {pages} {258} (\bibinfo {year} {2014})}\BibitemShut {NoStop}%
\bibitem [{\citenamefont {{Rodr{\'i}guez-Briones}}\ \emph {et~al.}(2017)\citenamefont {{Rodr{\'i}guez-Briones}}, \citenamefont {Li}, \citenamefont {Peng}, \citenamefont {Mor}, \citenamefont {Weinstein},\ and\ \citenamefont {Laflamme}}]{rodriguez-brionesHeatbathAlgorithmicCooling2017}%
  \BibitemOpen
  \bibfield  {author} {\bibinfo {author} {\bibfnamefont {N.~A.}\ \bibnamefont {{Rodr{\'i}guez-Briones}}}, \bibinfo {author} {\bibfnamefont {J.}~\bibnamefont {Li}}, \bibinfo {author} {\bibfnamefont {X.}~\bibnamefont {Peng}}, \bibinfo {author} {\bibfnamefont {T.}~\bibnamefont {Mor}}, \bibinfo {author} {\bibfnamefont {Y.}~\bibnamefont {Weinstein}},\ and\ \bibinfo {author} {\bibfnamefont {R.}~\bibnamefont {Laflamme}},\ }\bibfield  {title} {\bibinfo {title} {Heat-bath algorithmic cooling with correlated qubit-environment interactions},\ }\href {https://doi.org/10.1088/1367-2630/aa8fe0} {\bibfield  {journal} {\bibinfo  {journal} {New J. Phys.}\ }\textbf {\bibinfo {volume} {19}},\ \bibinfo {pages} {113047} (\bibinfo {year} {2017})}\BibitemShut {NoStop}%
\bibitem [{\citenamefont {Park}\ \emph {et~al.}(2015)\citenamefont {Park}, \citenamefont {{Rodriguez-Briones}}, \citenamefont {Feng}, \citenamefont {Darabad}, \citenamefont {Baugh},\ and\ \citenamefont {Laflamme}}]{parkHeatBathAlgorithmic2015}%
  \BibitemOpen
  \bibfield  {author} {\bibinfo {author} {\bibfnamefont {D.~K.}\ \bibnamefont {Park}}, \bibinfo {author} {\bibfnamefont {N.~A.}\ \bibnamefont {{Rodriguez-Briones}}}, \bibinfo {author} {\bibfnamefont {G.}~\bibnamefont {Feng}}, \bibinfo {author} {\bibfnamefont {R.~R.}\ \bibnamefont {Darabad}}, \bibinfo {author} {\bibfnamefont {J.}~\bibnamefont {Baugh}},\ and\ \bibinfo {author} {\bibfnamefont {R.}~\bibnamefont {Laflamme}},\ }\href {https://doi.org/10.48550/arXiv.1501.00952} {\bibinfo {title} {Heat {{Bath Algorithmic Cooling}} with {{Spins}}: {{Review}} and {{Prospects}}}} (\bibinfo {year} {2015}),\ \Eprint {https://arxiv.org/abs/1501.00952} {arxiv:1501.00952 [quant-ph]} \BibitemShut {NoStop}%
\bibitem [{\citenamefont {Breuer}\ \emph {et~al.}(2016)\citenamefont {Breuer}, \citenamefont {Laine}, \citenamefont {Piilo},\ and\ \citenamefont {Vacchini}}]{breuerColloquiumNonMarkovianDynamics2016}%
  \BibitemOpen
  \bibfield  {author} {\bibinfo {author} {\bibfnamefont {H.-P.}\ \bibnamefont {Breuer}}, \bibinfo {author} {\bibfnamefont {E.-M.}\ \bibnamefont {Laine}}, \bibinfo {author} {\bibfnamefont {J.}~\bibnamefont {Piilo}},\ and\ \bibinfo {author} {\bibfnamefont {B.}~\bibnamefont {Vacchini}},\ }\bibfield  {title} {\bibinfo {title} {Colloquium: {{Non-Markovian}} dynamics in open quantum systems},\ }\href {https://doi.org/10.1103/RevModPhys.88.021002} {\bibfield  {journal} {\bibinfo  {journal} {Rev. Mod. Phys.}\ }\textbf {\bibinfo {volume} {88}},\ \bibinfo {pages} {021002} (\bibinfo {year} {2016})}\BibitemShut {NoStop}%
\bibitem [{\citenamefont {White}\ \emph {et~al.}(2020)\citenamefont {White}, \citenamefont {Hill}, \citenamefont {Pollock}, \citenamefont {Hollenberg},\ and\ \citenamefont {Modi}}]{whiteDemonstrationNonMarkovianProcess2020a}%
  \BibitemOpen
  \bibfield  {author} {\bibinfo {author} {\bibfnamefont {G.~A.~L.}\ \bibnamefont {White}}, \bibinfo {author} {\bibfnamefont {C.~D.}\ \bibnamefont {Hill}}, \bibinfo {author} {\bibfnamefont {F.~A.}\ \bibnamefont {Pollock}}, \bibinfo {author} {\bibfnamefont {L.~C.~L.}\ \bibnamefont {Hollenberg}},\ and\ \bibinfo {author} {\bibfnamefont {K.}~\bibnamefont {Modi}},\ }\bibfield  {title} {\bibinfo {title} {Demonstration of non-{{Markovian}} process characterisation and control on a quantum processor},\ }\href {https://doi.org/10.1038/s41467-020-20113-3} {\bibfield  {journal} {\bibinfo  {journal} {Nat Commun}\ }\textbf {\bibinfo {volume} {11}},\ \bibinfo {pages} {6301} (\bibinfo {year} {2020})}\BibitemShut {NoStop}%
\bibitem [{\citenamefont {Reed}\ \emph {et~al.}(2010)\citenamefont {Reed}, \citenamefont {Johnson}, \citenamefont {Houck}, \citenamefont {DiCarlo}, \citenamefont {Chow}, \citenamefont {Schuster}, \citenamefont {Frunzio},\ and\ \citenamefont {Schoelkopf}}]{reedFastResetSuppressing2010}%
  \BibitemOpen
  \bibfield  {author} {\bibinfo {author} {\bibfnamefont {M.~D.}\ \bibnamefont {Reed}}, \bibinfo {author} {\bibfnamefont {B.~R.}\ \bibnamefont {Johnson}}, \bibinfo {author} {\bibfnamefont {A.~A.}\ \bibnamefont {Houck}}, \bibinfo {author} {\bibfnamefont {L.}~\bibnamefont {DiCarlo}}, \bibinfo {author} {\bibfnamefont {J.~M.}\ \bibnamefont {Chow}}, \bibinfo {author} {\bibfnamefont {D.~I.}\ \bibnamefont {Schuster}}, \bibinfo {author} {\bibfnamefont {L.}~\bibnamefont {Frunzio}},\ and\ \bibinfo {author} {\bibfnamefont {R.~J.}\ \bibnamefont {Schoelkopf}},\ }\bibfield  {title} {\bibinfo {title} {Fast reset and suppressing spontaneous emission of a superconducting qubit},\ }\href {https://doi.org/10.1063/1.3435463} {\bibfield  {journal} {\bibinfo  {journal} {Appl. Phys. Lett.}\ }\textbf {\bibinfo {volume} {96}},\ \bibinfo {pages} {203110} (\bibinfo {year} {2010})}\BibitemShut {NoStop}%
\bibitem [{\citenamefont {Geerlings}\ \emph {et~al.}(2013)\citenamefont {Geerlings}, \citenamefont {Leghtas}, \citenamefont {Pop}, \citenamefont {Shankar}, \citenamefont {Frunzio}, \citenamefont {Schoelkopf}, \citenamefont {Mirrahimi},\ and\ \citenamefont {Devoret}}]{geerlingsDemonstratingDrivenReset2013}%
  \BibitemOpen
  \bibfield  {author} {\bibinfo {author} {\bibfnamefont {K.}~\bibnamefont {Geerlings}}, \bibinfo {author} {\bibfnamefont {Z.}~\bibnamefont {Leghtas}}, \bibinfo {author} {\bibfnamefont {I.~M.}\ \bibnamefont {Pop}}, \bibinfo {author} {\bibfnamefont {S.}~\bibnamefont {Shankar}}, \bibinfo {author} {\bibfnamefont {L.}~\bibnamefont {Frunzio}}, \bibinfo {author} {\bibfnamefont {R.~J.}\ \bibnamefont {Schoelkopf}}, \bibinfo {author} {\bibfnamefont {M.}~\bibnamefont {Mirrahimi}},\ and\ \bibinfo {author} {\bibfnamefont {M.~H.}\ \bibnamefont {Devoret}},\ }\bibfield  {title} {\bibinfo {title} {Demonstrating a {{Driven Reset Protocol}} for a {{Superconducting Qubit}}},\ }\href {https://doi.org/10.1103/PhysRevLett.110.120501} {\bibfield  {journal} {\bibinfo  {journal} {Phys. Rev. Lett.}\ }\textbf {\bibinfo {volume} {110}},\ \bibinfo {pages} {120501} (\bibinfo {year} {2013})}\BibitemShut {NoStop}%
\bibitem [{\citenamefont {Basilewitsch}\ \emph {et~al.}(2017)\citenamefont {Basilewitsch}, \citenamefont {Schmidt}, \citenamefont {Sugny}, \citenamefont {Maniscalco},\ and\ \citenamefont {Koch}}]{basilewitschBeatingLimitsInitial2017}%
  \BibitemOpen
  \bibfield  {author} {\bibinfo {author} {\bibfnamefont {D.}~\bibnamefont {Basilewitsch}}, \bibinfo {author} {\bibfnamefont {R.}~\bibnamefont {Schmidt}}, \bibinfo {author} {\bibfnamefont {D.}~\bibnamefont {Sugny}}, \bibinfo {author} {\bibfnamefont {S.}~\bibnamefont {Maniscalco}},\ and\ \bibinfo {author} {\bibfnamefont {C.~P.}\ \bibnamefont {Koch}},\ }\bibfield  {title} {\bibinfo {title} {Beating the limits with initial correlations},\ }\href {https://doi.org/10.1088/1367-2630/aa96f8} {\bibfield  {journal} {\bibinfo  {journal} {New J. Phys.}\ }\textbf {\bibinfo {volume} {19}},\ \bibinfo {pages} {113042} (\bibinfo {year} {2017})}\BibitemShut {NoStop}%
\bibitem [{\citenamefont {Peng}\ \emph {et~al.}(2020)\citenamefont {Peng}, \citenamefont {Harrow}, \citenamefont {Ozols},\ and\ \citenamefont {Wu}}]{pengSimulatingLargeQuantum2020}%
  \BibitemOpen
  \bibfield  {author} {\bibinfo {author} {\bibfnamefont {T.}~\bibnamefont {Peng}}, \bibinfo {author} {\bibfnamefont {A.~W.}\ \bibnamefont {Harrow}}, \bibinfo {author} {\bibfnamefont {M.}~\bibnamefont {Ozols}},\ and\ \bibinfo {author} {\bibfnamefont {X.}~\bibnamefont {Wu}},\ }\bibfield  {title} {\bibinfo {title} {Simulating {{Large Quantum Circuits}} on a {{Small Quantum Computer}}},\ }\href {https://doi.org/10.1103/PhysRevLett.125.150504} {\bibfield  {journal} {\bibinfo  {journal} {Phys. Rev. Lett.}\ }\textbf {\bibinfo {volume} {125}},\ \bibinfo {pages} {150504} (\bibinfo {year} {2020})}\BibitemShut {NoStop}%
\bibitem [{\citenamefont {Lowe}\ \emph {et~al.}(2023)\citenamefont {Lowe}, \citenamefont {Medvidovi{\'c}}, \citenamefont {Hayes}, \citenamefont {O'Riordan}, \citenamefont {Bromley}, \citenamefont {Arrazola},\ and\ \citenamefont {Killoran}}]{loweFastQuantumCircuit2023}%
  \BibitemOpen
  \bibfield  {author} {\bibinfo {author} {\bibfnamefont {A.}~\bibnamefont {Lowe}}, \bibinfo {author} {\bibfnamefont {M.}~\bibnamefont {Medvidovi{\'c}}}, \bibinfo {author} {\bibfnamefont {A.}~\bibnamefont {Hayes}}, \bibinfo {author} {\bibfnamefont {L.~J.}\ \bibnamefont {O'Riordan}}, \bibinfo {author} {\bibfnamefont {T.~R.}\ \bibnamefont {Bromley}}, \bibinfo {author} {\bibfnamefont {J.~M.}\ \bibnamefont {Arrazola}},\ and\ \bibinfo {author} {\bibfnamefont {N.}~\bibnamefont {Killoran}},\ }\bibfield  {title} {\bibinfo {title} {Fast quantum circuit cutting with randomized measurements},\ }\href {https://doi.org/10.22331/q-2023-03-02-934} {\bibfield  {journal} {\bibinfo  {journal} {Quantum}\ }\textbf {\bibinfo {volume} {7}},\ \bibinfo {pages} {934} (\bibinfo {year} {2023})}\BibitemShut {NoStop}%
\bibitem [{\citenamefont {Temme}\ \emph {et~al.}(2017)\citenamefont {Temme}, \citenamefont {Bravyi},\ and\ \citenamefont {Gambetta}}]{temmeErrorMitigationShortDepth2017a}%
  \BibitemOpen
  \bibfield  {author} {\bibinfo {author} {\bibfnamefont {K.}~\bibnamefont {Temme}}, \bibinfo {author} {\bibfnamefont {S.}~\bibnamefont {Bravyi}},\ and\ \bibinfo {author} {\bibfnamefont {J.~M.}\ \bibnamefont {Gambetta}},\ }\bibfield  {title} {\bibinfo {title} {Error {{Mitigation}} for {{Short-Depth Quantum Circuits}}},\ }\href {https://doi.org/10.1103/PhysRevLett.119.180509} {\bibfield  {journal} {\bibinfo  {journal} {Phys. Rev. Lett.}\ }\textbf {\bibinfo {volume} {119}},\ \bibinfo {pages} {180509} (\bibinfo {year} {2017})}\BibitemShut {NoStop}%
\bibitem [{\citenamefont {Li}\ and\ \citenamefont {Benjamin}(2017)}]{liEfficientVariationalQuantum2017a}%
  \BibitemOpen
  \bibfield  {author} {\bibinfo {author} {\bibfnamefont {Y.}~\bibnamefont {Li}}\ and\ \bibinfo {author} {\bibfnamefont {S.~C.}\ \bibnamefont {Benjamin}},\ }\bibfield  {title} {\bibinfo {title} {Efficient {{Variational Quantum Simulator Incorporating Active Error Minimization}}},\ }\href {https://doi.org/10.1103/PhysRevX.7.021050} {\bibfield  {journal} {\bibinfo  {journal} {Phys. Rev. X}\ }\textbf {\bibinfo {volume} {7}},\ \bibinfo {pages} {021050} (\bibinfo {year} {2017})}\BibitemShut {NoStop}%
\bibitem [{\citenamefont {Satzinger}\ \emph {et~al.}(2021)\citenamefont {Satzinger}, \citenamefont {Liu}, \citenamefont {Smith}, \citenamefont {Knapp}, \citenamefont {Newman}, \citenamefont {Jones}, \citenamefont {Chen}, \citenamefont {Quintana}, \citenamefont {Mi}, \citenamefont {Dunsworth}, \citenamefont {Gidney}, \citenamefont {Aleiner}, \citenamefont {Arute}, \citenamefont {Arya}, \citenamefont {Atalaya}, \citenamefont {Babbush}, \citenamefont {Bardin}, \citenamefont {Barends}, \citenamefont {Basso}, \citenamefont {Bengtsson}, \citenamefont {Bilmes}, \citenamefont {Broughton}, \citenamefont {Buckley}, \citenamefont {Buell}, \citenamefont {Burkett}, \citenamefont {Bushnell}, \citenamefont {Chiaro}, \citenamefont {Collins}, \citenamefont {Courtney}, \citenamefont {Demura}, \citenamefont {Derk}, \citenamefont {Eppens}, \citenamefont {Erickson}, \citenamefont {Faoro}, \citenamefont {Farhi}, \citenamefont {Fowler}, \citenamefont {Foxen}, \citenamefont {Giustina}, \citenamefont {Greene}, \citenamefont {Gross}, \citenamefont {Harrigan}, \citenamefont {Harrington}, \citenamefont {Hilton}, \citenamefont {Hong}, \citenamefont {Huang}, \citenamefont {Huggins}, \citenamefont {Ioffe}, \citenamefont {Isakov}, \citenamefont {Jeffrey}, \citenamefont {Jiang}, \citenamefont {Kafri}, \citenamefont {Kechedzhi}, \citenamefont {Khattar}, \citenamefont {Kim}, \citenamefont {Klimov}, \citenamefont {Korotkov}, \citenamefont {Kostritsa}, \citenamefont {Landhuis}, \citenamefont {Laptev}, \citenamefont {Locharla}, \citenamefont {Lucero}, \citenamefont {Martin}, \citenamefont {McClean}, \citenamefont {McEwen}, \citenamefont {Miao}, \citenamefont {Mohseni}, \citenamefont {Montazeri}, \citenamefont {Mruczkiewicz}, \citenamefont {Mutus}, \citenamefont {Naaman}, \citenamefont {Neeley}, \citenamefont {Neill}, \citenamefont {Niu}, \citenamefont {O'Brien}, \citenamefont {Opremcak}, \citenamefont {Pat{\'o}}, \citenamefont {Petukhov}, \citenamefont {Rubin}, \citenamefont {Sank}, \citenamefont {Shvarts}, \citenamefont {Strain}, \citenamefont {Szalay}, \citenamefont {Villalonga}, \citenamefont {White}, \citenamefont {Yao}, \citenamefont {Yeh}, \citenamefont {Yoo}, \citenamefont {Zalcman}, \citenamefont {Neven}, \citenamefont {Boixo}, \citenamefont {Megrant}, \citenamefont {Chen}, \citenamefont {Kelly}, \citenamefont {Smelyanskiy}, \citenamefont {Kitaev}, \citenamefont {Knap}, \citenamefont {Pollmann},\ and\ \citenamefont {Roushan}}]{satzingerRealizingTopologicallyOrdered2021a}%
  \BibitemOpen
  \bibfield  {author} {\bibinfo {author} {\bibfnamefont {K.~J.}\ \bibnamefont {Satzinger}}, \bibinfo {author} {\bibfnamefont {Y.-J.}\ \bibnamefont {Liu}}, \bibinfo {author} {\bibfnamefont {A.}~\bibnamefont {Smith}}, \bibinfo {author} {\bibfnamefont {C.}~\bibnamefont {Knapp}}, \bibinfo {author} {\bibfnamefont {M.}~\bibnamefont {Newman}}, \bibinfo {author} {\bibfnamefont {C.}~\bibnamefont {Jones}}, \bibinfo {author} {\bibfnamefont {Z.}~\bibnamefont {Chen}}, \bibinfo {author} {\bibfnamefont {C.}~\bibnamefont {Quintana}}, \bibinfo {author} {\bibfnamefont {X.}~\bibnamefont {Mi}}, \bibinfo {author} {\bibfnamefont {A.}~\bibnamefont {Dunsworth}}, \bibinfo {author} {\bibfnamefont {C.}~\bibnamefont {Gidney}}, \bibinfo {author} {\bibfnamefont {I.}~\bibnamefont {Aleiner}}, \bibinfo {author} {\bibfnamefont {F.}~\bibnamefont {Arute}}, \bibinfo {author} {\bibfnamefont {K.}~\bibnamefont {Arya}}, \bibinfo {author} {\bibfnamefont {J.}~\bibnamefont {Atalaya}}, \bibinfo {author} {\bibfnamefont {R.}~\bibnamefont {Babbush}}, \bibinfo {author} {\bibfnamefont {J.~C.}\ \bibnamefont {Bardin}}, \bibinfo {author} {\bibfnamefont {R.}~\bibnamefont {Barends}}, \bibinfo {author} {\bibfnamefont {J.}~\bibnamefont {Basso}}, \bibinfo {author} {\bibfnamefont {A.}~\bibnamefont {Bengtsson}}, \bibinfo {author} {\bibfnamefont {A.}~\bibnamefont {Bilmes}}, \bibinfo {author} {\bibfnamefont {M.}~\bibnamefont {Broughton}}, \bibinfo {author} {\bibfnamefont {B.~B.}\ \bibnamefont {Buckley}}, \bibinfo {author} {\bibfnamefont {D.~A.}\ \bibnamefont {Buell}}, \bibinfo {author} {\bibfnamefont {B.}~\bibnamefont {Burkett}}, \bibinfo {author} {\bibfnamefont {N.}~\bibnamefont {Bushnell}}, \bibinfo {author} {\bibfnamefont {B.}~\bibnamefont {Chiaro}}, \bibinfo {author} {\bibfnamefont {R.}~\bibnamefont {Collins}}, \bibinfo {author} {\bibfnamefont {W.}~\bibnamefont {Courtney}}, \bibinfo {author} {\bibfnamefont {S.}~\bibnamefont {Demura}}, \bibinfo {author} {\bibfnamefont {A.~R.}\ \bibnamefont {Derk}}, \bibinfo {author} {\bibfnamefont {D.}~\bibnamefont {Eppens}}, \bibinfo {author} {\bibfnamefont {C.}~\bibnamefont {Erickson}}, \bibinfo {author} {\bibfnamefont {L.}~\bibnamefont {Faoro}}, \bibinfo {author} {\bibfnamefont {E.}~\bibnamefont {Farhi}}, \bibinfo {author} {\bibfnamefont {A.~G.}\ \bibnamefont {Fowler}}, \bibinfo {author} {\bibfnamefont {B.}~\bibnamefont {Foxen}}, \bibinfo {author} {\bibfnamefont {M.}~\bibnamefont {Giustina}}, \bibinfo {author} {\bibfnamefont {A.}~\bibnamefont {Greene}}, \bibinfo {author} {\bibfnamefont {J.~A.}\ \bibnamefont {Gross}}, \bibinfo {author} {\bibfnamefont {M.~P.}\ \bibnamefont {Harrigan}}, \bibinfo {author} {\bibfnamefont {S.~D.}\ \bibnamefont {Harrington}}, \bibinfo {author} {\bibfnamefont {J.}~\bibnamefont {Hilton}}, \bibinfo {author} {\bibfnamefont {S.}~\bibnamefont {Hong}}, \bibinfo {author} {\bibfnamefont {T.}~\bibnamefont {Huang}}, \bibinfo {author} {\bibfnamefont {W.~J.}\ \bibnamefont {Huggins}}, \bibinfo {author} {\bibfnamefont {L.~B.}\ \bibnamefont {Ioffe}}, \bibinfo {author} {\bibfnamefont {S.~V.}\ \bibnamefont {Isakov}}, \bibinfo {author} {\bibfnamefont {E.}~\bibnamefont {Jeffrey}}, \bibinfo {author} {\bibfnamefont {Z.}~\bibnamefont {Jiang}}, \bibinfo {author} {\bibfnamefont {D.}~\bibnamefont {Kafri}}, \bibinfo {author} {\bibfnamefont {K.}~\bibnamefont {Kechedzhi}}, \bibinfo {author} {\bibfnamefont {T.}~\bibnamefont {Khattar}}, \bibinfo {author} {\bibfnamefont {S.}~\bibnamefont {Kim}}, \bibinfo {author} {\bibfnamefont {P.~V.}\ \bibnamefont {Klimov}}, \bibinfo {author} {\bibfnamefont {A.~N.}\ \bibnamefont {Korotkov}}, \bibinfo {author} {\bibfnamefont {F.}~\bibnamefont {Kostritsa}}, \bibinfo {author} {\bibfnamefont {D.}~\bibnamefont {Landhuis}}, \bibinfo {author} {\bibfnamefont {P.}~\bibnamefont {Laptev}}, \bibinfo {author} {\bibfnamefont {A.}~\bibnamefont {Locharla}}, \bibinfo {author} {\bibfnamefont {E.}~\bibnamefont {Lucero}}, \bibinfo {author} {\bibfnamefont {O.}~\bibnamefont {Martin}}, \bibinfo {author} {\bibfnamefont {J.~R.}\ \bibnamefont {McClean}}, \bibinfo {author} {\bibfnamefont {M.}~\bibnamefont {McEwen}}, \bibinfo {author} {\bibfnamefont {K.~C.}\ \bibnamefont {Miao}}, \bibinfo {author} {\bibfnamefont {M.}~\bibnamefont {Mohseni}}, \bibinfo {author} {\bibfnamefont {S.}~\bibnamefont {Montazeri}}, \bibinfo {author} {\bibfnamefont {W.}~\bibnamefont {Mruczkiewicz}}, \bibinfo {author} {\bibfnamefont {J.}~\bibnamefont {Mutus}}, \bibinfo {author} {\bibfnamefont {O.}~\bibnamefont {Naaman}}, \bibinfo {author} {\bibfnamefont {M.}~\bibnamefont {Neeley}}, \bibinfo {author} {\bibfnamefont {C.}~\bibnamefont {Neill}}, \bibinfo {author} {\bibfnamefont {M.~Y.}\ \bibnamefont {Niu}}, \bibinfo {author} {\bibfnamefont {T.~E.}\ \bibnamefont {O'Brien}}, \bibinfo {author} {\bibfnamefont {A.}~\bibnamefont {Opremcak}}, \bibinfo {author} {\bibfnamefont {B.}~\bibnamefont {Pat{\'o}}}, \bibinfo {author} {\bibfnamefont {A.}~\bibnamefont {Petukhov}}, \bibinfo {author} {\bibfnamefont {N.~C.}\ \bibnamefont {Rubin}}, \bibinfo {author} {\bibfnamefont {D.}~\bibnamefont {Sank}}, \bibinfo {author} {\bibfnamefont {V.}~\bibnamefont {Shvarts}}, \bibinfo {author} {\bibfnamefont {D.}~\bibnamefont {Strain}}, \bibinfo {author} {\bibfnamefont {M.}~\bibnamefont {Szalay}}, \bibinfo {author} {\bibfnamefont {B.}~\bibnamefont {Villalonga}}, \bibinfo {author} {\bibfnamefont {T.~C.}\ \bibnamefont {White}}, \bibinfo {author} {\bibfnamefont {Z.}~\bibnamefont {Yao}}, \bibinfo {author} {\bibfnamefont {P.}~\bibnamefont {Yeh}}, \bibinfo {author} {\bibfnamefont {J.}~\bibnamefont {Yoo}}, \bibinfo {author} {\bibfnamefont {A.}~\bibnamefont {Zalcman}}, \bibinfo {author} {\bibfnamefont {H.}~\bibnamefont {Neven}}, \bibinfo {author} {\bibfnamefont {S.}~\bibnamefont {Boixo}}, \bibinfo {author} {\bibfnamefont {A.}~\bibnamefont {Megrant}}, \bibinfo {author} {\bibfnamefont {Y.}~\bibnamefont {Chen}}, \bibinfo {author} {\bibfnamefont {J.}~\bibnamefont {Kelly}}, \bibinfo {author} {\bibfnamefont {V.}~\bibnamefont {Smelyanskiy}}, \bibinfo {author} {\bibfnamefont {A.}~\bibnamefont {Kitaev}}, \bibinfo {author} {\bibfnamefont {M.}~\bibnamefont {Knap}}, \bibinfo {author} {\bibfnamefont {F.}~\bibnamefont {Pollmann}},\ and\ \bibinfo {author} {\bibfnamefont {P.}~\bibnamefont {Roushan}},\ }\bibfield  {title} {\bibinfo {title} {Realizing topologically ordered states on a quantum processor},\ }\href {https://doi.org/10.1126/science.abi8378} {\bibfield  {journal} {\bibinfo  {journal} {Science}\ }\textbf {\bibinfo {volume} {374}},\ \bibinfo {pages} {1237} (\bibinfo {year} {2021})}\BibitemShut {NoStop}%
\bibitem [{\citenamefont {Acharya}\ \emph {et~al.}(2023)\citenamefont {Acharya}, \citenamefont {Aleiner}, \citenamefont {Allen}, \citenamefont {Andersen}, \citenamefont {Ansmann}, \citenamefont {Arute}, \citenamefont {Arya}, \citenamefont {Asfaw}, \citenamefont {Atalaya}, \citenamefont {Babbush}, \citenamefont {Bacon}, \citenamefont {Bardin}, \citenamefont {Basso}, \citenamefont {Bengtsson}, \citenamefont {Boixo}, \citenamefont {Bortoli}, \citenamefont {Bourassa}, \citenamefont {Bovaird}, \citenamefont {Brill}, \citenamefont {Broughton}, \citenamefont {Buckley}, \citenamefont {Buell}, \citenamefont {Burger}, \citenamefont {Burkett}, \citenamefont {Bushnell}, \citenamefont {Chen}, \citenamefont {Chen}, \citenamefont {Chiaro}, \citenamefont {Cogan}, \citenamefont {Collins}, \citenamefont {Conner}, \citenamefont {Courtney}, \citenamefont {Crook}, \citenamefont {Curtin}, \citenamefont {Debroy}, \citenamefont {Del Toro~Barba}, \citenamefont {Demura}, \citenamefont {Dunsworth}, \citenamefont {Eppens}, \citenamefont {Erickson}, \citenamefont {Faoro}, \citenamefont {Farhi}, \citenamefont {Fatemi}, \citenamefont {Flores~Burgos}, \citenamefont {Forati}, \citenamefont {Fowler}, \citenamefont {Foxen}, \citenamefont {Giang}, \citenamefont {Gidney}, \citenamefont {Gilboa}, \citenamefont {Giustina}, \citenamefont {Grajales~Dau}, \citenamefont {Gross}, \citenamefont {Habegger}, \citenamefont {Hamilton}, \citenamefont {Harrigan}, \citenamefont {Harrington}, \citenamefont {Higgott}, \citenamefont {Hilton}, \citenamefont {Hoffmann}, \citenamefont {Hong}, \citenamefont {Huang}, \citenamefont {Huff}, \citenamefont {Huggins}, \citenamefont {Ioffe}, \citenamefont {Isakov}, \citenamefont {Iveland}, \citenamefont {Jeffrey}, \citenamefont {Jiang}, \citenamefont {Jones}, \citenamefont {Juhas}, \citenamefont {Kafri}, \citenamefont {Kechedzhi}, \citenamefont {Kelly}, \citenamefont {Khattar}, \citenamefont {Khezri}, \citenamefont {Kieferov{\'a}}, \citenamefont {Kim}, \citenamefont {Kitaev}, \citenamefont {Klimov}, \citenamefont {Klots}, \citenamefont {Korotkov}, \citenamefont {Kostritsa}, \citenamefont {Kreikebaum}, \citenamefont {Landhuis}, \citenamefont {Laptev}, \citenamefont {Lau}, \citenamefont {Laws}, \citenamefont {Lee}, \citenamefont {Lee}, \citenamefont {Lester}, \citenamefont {Lill}, \citenamefont {Liu}, \citenamefont {Locharla}, \citenamefont {Lucero}, \citenamefont {Malone}, \citenamefont {Marshall}, \citenamefont {Martin}, \citenamefont {McClean}, \citenamefont {McCourt}, \citenamefont {McEwen}, \citenamefont {Megrant}, \citenamefont {Meurer~Costa}, \citenamefont {Mi}, \citenamefont {Miao}, \citenamefont {Mohseni}, \citenamefont {Montazeri}, \citenamefont {Morvan}, \citenamefont {Mount}, \citenamefont {Mruczkiewicz}, \citenamefont {Naaman}, \citenamefont {Neeley}, \citenamefont {Neill}, \citenamefont {Nersisyan}, \citenamefont {Neven}, \citenamefont {Newman}, \citenamefont {Ng}, \citenamefont {Nguyen}, \citenamefont {Nguyen}, \citenamefont {Niu}, \citenamefont {O'Brien}, \citenamefont {Opremcak}, \citenamefont {Platt}, \citenamefont {Petukhov}, \citenamefont {Potter}, \citenamefont {Pryadko}, \citenamefont {Quintana}, \citenamefont {Roushan}, \citenamefont {Rubin}, \citenamefont {Saei}, \citenamefont {Sank}, \citenamefont {Sankaragomathi}, \citenamefont {Satzinger}, \citenamefont {Schurkus}, \citenamefont {Schuster}, \citenamefont {Shearn}, \citenamefont {Shorter}, \citenamefont {Shvarts}, \citenamefont {Skruzny}, \citenamefont {Smelyanskiy}, \citenamefont {Smith}, \citenamefont {Sterling}, \citenamefont {Strain}, \citenamefont {Szalay}, \citenamefont {Torres}, \citenamefont {Vidal}, \citenamefont {Villalonga}, \citenamefont {Vollgraff~Heidweiller}, \citenamefont {White}, \citenamefont {Xing}, \citenamefont {Yao}, \citenamefont {Yeh}, \citenamefont {Yoo}, \citenamefont {Young}, \citenamefont {Zalcman}, \citenamefont {Zhang}, \citenamefont {Zhu},\ and\ \citenamefont {{Google Quantum AI}}}]{acharyaSuppressingQuantumErrors2023}%
  \BibitemOpen
  \bibfield  {author} {\bibinfo {author} {\bibfnamefont {R.}~\bibnamefont {Acharya}}, \bibinfo {author} {\bibfnamefont {I.}~\bibnamefont {Aleiner}}, \bibinfo {author} {\bibfnamefont {R.}~\bibnamefont {Allen}}, \bibinfo {author} {\bibfnamefont {T.~I.}\ \bibnamefont {Andersen}}, \bibinfo {author} {\bibfnamefont {M.}~\bibnamefont {Ansmann}}, \bibinfo {author} {\bibfnamefont {F.}~\bibnamefont {Arute}}, \bibinfo {author} {\bibfnamefont {K.}~\bibnamefont {Arya}}, \bibinfo {author} {\bibfnamefont {A.}~\bibnamefont {Asfaw}}, \bibinfo {author} {\bibfnamefont {J.}~\bibnamefont {Atalaya}}, \bibinfo {author} {\bibfnamefont {R.}~\bibnamefont {Babbush}}, \bibinfo {author} {\bibfnamefont {D.}~\bibnamefont {Bacon}}, \bibinfo {author} {\bibfnamefont {J.~C.}\ \bibnamefont {Bardin}}, \bibinfo {author} {\bibfnamefont {J.}~\bibnamefont {Basso}}, \bibinfo {author} {\bibfnamefont {A.}~\bibnamefont {Bengtsson}}, \bibinfo {author} {\bibfnamefont {S.}~\bibnamefont {Boixo}}, \bibinfo {author} {\bibfnamefont {G.}~\bibnamefont {Bortoli}}, \bibinfo {author} {\bibfnamefont {A.}~\bibnamefont {Bourassa}}, \bibinfo {author} {\bibfnamefont {J.}~\bibnamefont {Bovaird}}, \bibinfo {author} {\bibfnamefont {L.}~\bibnamefont {Brill}}, \bibinfo {author} {\bibfnamefont {M.}~\bibnamefont {Broughton}}, \bibinfo {author} {\bibfnamefont {B.~B.}\ \bibnamefont {Buckley}}, \bibinfo {author} {\bibfnamefont {D.~A.}\ \bibnamefont {Buell}}, \bibinfo {author} {\bibfnamefont {T.}~\bibnamefont {Burger}}, \bibinfo {author} {\bibfnamefont {B.}~\bibnamefont {Burkett}}, \bibinfo {author} {\bibfnamefont {N.}~\bibnamefont {Bushnell}}, \bibinfo {author} {\bibfnamefont {Y.}~\bibnamefont {Chen}}, \bibinfo {author} {\bibfnamefont {Z.}~\bibnamefont {Chen}}, \bibinfo {author} {\bibfnamefont {B.}~\bibnamefont {Chiaro}}, \bibinfo {author} {\bibfnamefont {J.}~\bibnamefont {Cogan}}, \bibinfo {author} {\bibfnamefont {R.}~\bibnamefont {Collins}}, \bibinfo {author} {\bibfnamefont {P.}~\bibnamefont {Conner}}, \bibinfo {author} {\bibfnamefont {W.}~\bibnamefont {Courtney}}, \bibinfo {author} {\bibfnamefont {A.~L.}\ \bibnamefont {Crook}}, \bibinfo {author} {\bibfnamefont {B.}~\bibnamefont {Curtin}}, \bibinfo {author} {\bibfnamefont {D.~M.}\ \bibnamefont {Debroy}}, \bibinfo {author} {\bibfnamefont {A.}~\bibnamefont {Del Toro~Barba}}, \bibinfo {author} {\bibfnamefont {S.}~\bibnamefont {Demura}}, \bibinfo {author} {\bibfnamefont {A.}~\bibnamefont {Dunsworth}}, \bibinfo {author} {\bibfnamefont {D.}~\bibnamefont {Eppens}}, \bibinfo {author} {\bibfnamefont {C.}~\bibnamefont {Erickson}}, \bibinfo {author} {\bibfnamefont {L.}~\bibnamefont {Faoro}}, \bibinfo {author} {\bibfnamefont {E.}~\bibnamefont {Farhi}}, \bibinfo {author} {\bibfnamefont {R.}~\bibnamefont {Fatemi}}, \bibinfo {author} {\bibfnamefont {L.}~\bibnamefont {Flores~Burgos}}, \bibinfo {author} {\bibfnamefont {E.}~\bibnamefont {Forati}}, \bibinfo {author} {\bibfnamefont {A.~G.}\ \bibnamefont {Fowler}}, \bibinfo {author} {\bibfnamefont {B.}~\bibnamefont {Foxen}}, \bibinfo {author} {\bibfnamefont {W.}~\bibnamefont {Giang}}, \bibinfo {author} {\bibfnamefont {C.}~\bibnamefont {Gidney}}, \bibinfo {author} {\bibfnamefont {D.}~\bibnamefont {Gilboa}}, \bibinfo {author} {\bibfnamefont {M.}~\bibnamefont {Giustina}}, \bibinfo {author} {\bibfnamefont {A.}~\bibnamefont {Grajales~Dau}}, \bibinfo {author} {\bibfnamefont {J.~A.}\ \bibnamefont {Gross}}, \bibinfo {author} {\bibfnamefont {S.}~\bibnamefont {Habegger}}, \bibinfo {author} {\bibfnamefont {M.~C.}\ \bibnamefont {Hamilton}}, \bibinfo {author} {\bibfnamefont {M.~P.}\ \bibnamefont {Harrigan}}, \bibinfo {author} {\bibfnamefont {S.~D.}\ \bibnamefont {Harrington}}, \bibinfo {author} {\bibfnamefont {O.}~\bibnamefont {Higgott}}, \bibinfo {author} {\bibfnamefont {J.}~\bibnamefont {Hilton}}, \bibinfo {author} {\bibfnamefont {M.}~\bibnamefont {Hoffmann}}, \bibinfo {author} {\bibfnamefont {S.}~\bibnamefont {Hong}}, \bibinfo {author} {\bibfnamefont {T.}~\bibnamefont {Huang}}, \bibinfo {author} {\bibfnamefont {A.}~\bibnamefont {Huff}}, \bibinfo {author} {\bibfnamefont {W.~J.}\ \bibnamefont {Huggins}}, \bibinfo {author} {\bibfnamefont {L.~B.}\ \bibnamefont {Ioffe}}, \bibinfo {author} {\bibfnamefont {S.~V.}\ \bibnamefont {Isakov}}, \bibinfo {author} {\bibfnamefont {J.}~\bibnamefont {Iveland}}, \bibinfo {author} {\bibfnamefont {E.}~\bibnamefont {Jeffrey}}, \bibinfo {author} {\bibfnamefont {Z.}~\bibnamefont {Jiang}}, \bibinfo {author} {\bibfnamefont {C.}~\bibnamefont {Jones}}, \bibinfo {author} {\bibfnamefont {P.}~\bibnamefont {Juhas}}, \bibinfo {author} {\bibfnamefont {D.}~\bibnamefont {Kafri}}, \bibinfo {author} {\bibfnamefont {K.}~\bibnamefont {Kechedzhi}}, \bibinfo {author} {\bibfnamefont {J.}~\bibnamefont {Kelly}}, \bibinfo {author} {\bibfnamefont {T.}~\bibnamefont {Khattar}}, \bibinfo {author} {\bibfnamefont {M.}~\bibnamefont {Khezri}}, \bibinfo {author} {\bibfnamefont {M.}~\bibnamefont {Kieferov{\'a}}}, \bibinfo {author} {\bibfnamefont {S.}~\bibnamefont {Kim}}, \bibinfo {author} {\bibfnamefont {A.}~\bibnamefont {Kitaev}}, \bibinfo {author} {\bibfnamefont {P.~V.}\ \bibnamefont {Klimov}}, \bibinfo {author} {\bibfnamefont {A.~R.}\ \bibnamefont {Klots}}, \bibinfo {author} {\bibfnamefont {A.~N.}\ \bibnamefont {Korotkov}}, \bibinfo {author} {\bibfnamefont {F.}~\bibnamefont {Kostritsa}}, \bibinfo {author} {\bibfnamefont {J.~M.}\ \bibnamefont {Kreikebaum}}, \bibinfo {author} {\bibfnamefont {D.}~\bibnamefont {Landhuis}}, \bibinfo {author} {\bibfnamefont {P.}~\bibnamefont {Laptev}}, \bibinfo {author} {\bibfnamefont {K.-M.}\ \bibnamefont {Lau}}, \bibinfo {author} {\bibfnamefont {L.}~\bibnamefont {Laws}}, \bibinfo {author} {\bibfnamefont {J.}~\bibnamefont {Lee}}, \bibinfo {author} {\bibfnamefont {K.}~\bibnamefont {Lee}}, \bibinfo {author} {\bibfnamefont {B.~J.}\ \bibnamefont {Lester}}, \bibinfo {author} {\bibfnamefont {A.}~\bibnamefont {Lill}}, \bibinfo {author} {\bibfnamefont {W.}~\bibnamefont {Liu}}, \bibinfo {author} {\bibfnamefont {A.}~\bibnamefont {Locharla}}, \bibinfo {author} {\bibfnamefont {E.}~\bibnamefont {Lucero}}, \bibinfo {author} {\bibfnamefont {F.~D.}\ \bibnamefont {Malone}}, \bibinfo {author} {\bibfnamefont {J.}~\bibnamefont {Marshall}}, \bibinfo {author} {\bibfnamefont {O.}~\bibnamefont {Martin}}, \bibinfo {author} {\bibfnamefont {J.~R.}\ \bibnamefont {McClean}}, \bibinfo {author} {\bibfnamefont {T.}~\bibnamefont {McCourt}}, \bibinfo {author} {\bibfnamefont {M.}~\bibnamefont {McEwen}}, \bibinfo {author} {\bibfnamefont {A.}~\bibnamefont {Megrant}}, \bibinfo {author} {\bibfnamefont {B.}~\bibnamefont {Meurer~Costa}}, \bibinfo {author} {\bibfnamefont {X.}~\bibnamefont {Mi}}, \bibinfo {author} {\bibfnamefont {K.~C.}\ \bibnamefont {Miao}}, \bibinfo {author} {\bibfnamefont {M.}~\bibnamefont {Mohseni}}, \bibinfo {author} {\bibfnamefont {S.}~\bibnamefont {Montazeri}}, \bibinfo {author} {\bibfnamefont {A.}~\bibnamefont {Morvan}}, \bibinfo {author} {\bibfnamefont {E.}~\bibnamefont {Mount}}, \bibinfo {author} {\bibfnamefont {W.}~\bibnamefont {Mruczkiewicz}}, \bibinfo {author} {\bibfnamefont {O.}~\bibnamefont {Naaman}}, \bibinfo {author} {\bibfnamefont {M.}~\bibnamefont {Neeley}}, \bibinfo {author} {\bibfnamefont {C.}~\bibnamefont {Neill}}, \bibinfo {author} {\bibfnamefont {A.}~\bibnamefont {Nersisyan}}, \bibinfo {author} {\bibfnamefont {H.}~\bibnamefont {Neven}}, \bibinfo {author} {\bibfnamefont {M.}~\bibnamefont {Newman}}, \bibinfo {author} {\bibfnamefont {J.~H.}\ \bibnamefont {Ng}}, \bibinfo {author} {\bibfnamefont {A.}~\bibnamefont {Nguyen}}, \bibinfo {author} {\bibfnamefont {M.}~\bibnamefont {Nguyen}}, \bibinfo {author} {\bibfnamefont {M.~Y.}\ \bibnamefont {Niu}}, \bibinfo {author} {\bibfnamefont {T.~E.}\ \bibnamefont {O'Brien}}, \bibinfo {author} {\bibfnamefont {A.}~\bibnamefont {Opremcak}}, \bibinfo {author} {\bibfnamefont {J.}~\bibnamefont {Platt}}, \bibinfo {author} {\bibfnamefont {A.}~\bibnamefont {Petukhov}}, \bibinfo {author} {\bibfnamefont {R.}~\bibnamefont {Potter}}, \bibinfo {author} {\bibfnamefont {L.~P.}\ \bibnamefont {Pryadko}}, \bibinfo {author} {\bibfnamefont {C.}~\bibnamefont {Quintana}}, \bibinfo {author} {\bibfnamefont {P.}~\bibnamefont {Roushan}}, \bibinfo {author} {\bibfnamefont {N.~C.}\ \bibnamefont {Rubin}}, \bibinfo {author} {\bibfnamefont {N.}~\bibnamefont {Saei}}, \bibinfo {author} {\bibfnamefont {D.}~\bibnamefont {Sank}}, \bibinfo {author} {\bibfnamefont {K.}~\bibnamefont {Sankaragomathi}}, \bibinfo {author} {\bibfnamefont {K.~J.}\ \bibnamefont {Satzinger}}, \bibinfo {author} {\bibfnamefont {H.~F.}\ \bibnamefont {Schurkus}}, \bibinfo {author} {\bibfnamefont {C.}~\bibnamefont {Schuster}}, \bibinfo {author} {\bibfnamefont {M.~J.}\ \bibnamefont {Shearn}}, \bibinfo {author} {\bibfnamefont {A.}~\bibnamefont {Shorter}}, \bibinfo {author} {\bibfnamefont {V.}~\bibnamefont {Shvarts}}, \bibinfo {author} {\bibfnamefont {J.}~\bibnamefont {Skruzny}}, \bibinfo {author} {\bibfnamefont {V.}~\bibnamefont {Smelyanskiy}}, \bibinfo {author} {\bibfnamefont {W.~C.}\ \bibnamefont {Smith}}, \bibinfo {author} {\bibfnamefont {G.}~\bibnamefont {Sterling}}, \bibinfo {author} {\bibfnamefont {D.}~\bibnamefont {Strain}}, \bibinfo {author} {\bibfnamefont {M.}~\bibnamefont {Szalay}}, \bibinfo {author} {\bibfnamefont {A.}~\bibnamefont {Torres}}, \bibinfo {author} {\bibfnamefont {G.}~\bibnamefont {Vidal}}, \bibinfo {author} {\bibfnamefont {B.}~\bibnamefont {Villalonga}}, \bibinfo {author} {\bibfnamefont {C.}~\bibnamefont {Vollgraff~Heidweiller}}, \bibinfo {author} {\bibfnamefont {T.}~\bibnamefont {White}}, \bibinfo {author} {\bibfnamefont {C.}~\bibnamefont {Xing}}, \bibinfo {author} {\bibfnamefont {Z.~J.}\ \bibnamefont {Yao}}, \bibinfo {author} {\bibfnamefont {P.}~\bibnamefont {Yeh}}, \bibinfo {author} {\bibfnamefont {J.}~\bibnamefont {Yoo}}, \bibinfo {author} {\bibfnamefont {G.}~\bibnamefont {Young}}, \bibinfo {author} {\bibfnamefont {A.}~\bibnamefont {Zalcman}}, \bibinfo {author} {\bibfnamefont {Y.}~\bibnamefont {Zhang}}, \bibinfo {author} {\bibfnamefont {N.}~\bibnamefont {Zhu}},\ and\ \bibinfo {author}
  {\bibnamefont {{Google Quantum AI}}},\ }\bibfield  {title} {\bibinfo {title} {Suppressing quantum errors by scaling a surface code logical qubit},\ }\href {https://doi.org/10.1038/s41586-022-05434-1} {\bibfield  {journal} {\bibinfo  {journal} {Nature}\ }\textbf {\bibinfo {volume} {614}},\ \bibinfo {pages} {676} (\bibinfo {year} {2023})}\BibitemShut {NoStop}%
\bibitem [{\citenamefont {Schr{\"o}dinger}(1929)}]{schroedingerErfassungQuantengesetzeDurch1929}%
  \BibitemOpen
  \bibfield  {author} {\bibinfo {author} {\bibfnamefont {E.}~\bibnamefont {Schr{\"o}dinger}},\ }\bibfield  {title} {\bibinfo {title} {{Die Erfassung der Quantengesetze durch kontinuierliche Funktionen}},\ }\href {https://doi.org/10.1007/BF01505681} {\bibfield  {journal} {\bibinfo  {journal} {Naturwissenschaften}\ }\textbf {\bibinfo {volume} {17}},\ \bibinfo {pages} {486} (\bibinfo {year} {1929})}\BibitemShut {NoStop}%
\bibitem [{\citenamefont {Schr{\"o}dinger}(1935)}]{schrodingerDiscussionProbabilityRelations1935}%
  \BibitemOpen
  \bibfield  {author} {\bibinfo {author} {\bibfnamefont {E.}~\bibnamefont {Schr{\"o}dinger}},\ }\bibfield  {title} {\bibinfo {title} {Discussion of {{Probability Relations}} between {{Separated Systems}}},\ }\href {https://doi.org/10.1017/S0305004100013554} {\bibfield  {journal} {\bibinfo  {journal} {Mathematical Proceedings of the Cambridge Philosophical Society}\ }\textbf {\bibinfo {volume} {31}},\ \bibinfo {pages} {555} (\bibinfo {year} {1935})}\BibitemShut {NoStop}%
\bibitem [{\citenamefont {Roy}\ \emph {et~al.}(2020)\citenamefont {Roy}, \citenamefont {Chalker}, \citenamefont {Gornyi},\ and\ \citenamefont {Gefen}}]{royMeasurementinducedSteeringQuantum2020a}%
  \BibitemOpen
  \bibfield  {author} {\bibinfo {author} {\bibfnamefont {S.}~\bibnamefont {Roy}}, \bibinfo {author} {\bibfnamefont {J.~T.}\ \bibnamefont {Chalker}}, \bibinfo {author} {\bibfnamefont {I.~V.}\ \bibnamefont {Gornyi}},\ and\ \bibinfo {author} {\bibfnamefont {Y.}~\bibnamefont {Gefen}},\ }\bibfield  {title} {\bibinfo {title} {Measurement-induced steering of quantum systems},\ }\href {https://doi.org/10.1103/PhysRevResearch.2.033347} {\bibfield  {journal} {\bibinfo  {journal} {Phys. Rev. Res.}\ }\textbf {\bibinfo {volume} {2}},\ \bibinfo {pages} {033347} (\bibinfo {year} {2020})}\BibitemShut {NoStop}%
\bibitem [{\citenamefont {Kumar}\ \emph {et~al.}(2020)\citenamefont {Kumar}, \citenamefont {Snizhko},\ and\ \citenamefont {Gefen}}]{kumarEngineeringTwoqubitMixed2020}%
  \BibitemOpen
  \bibfield  {author} {\bibinfo {author} {\bibfnamefont {P.}~\bibnamefont {Kumar}}, \bibinfo {author} {\bibfnamefont {K.}~\bibnamefont {Snizhko}},\ and\ \bibinfo {author} {\bibfnamefont {Y.}~\bibnamefont {Gefen}},\ }\bibfield  {title} {\bibinfo {title} {Engineering two-qubit mixed states with weak measurements},\ }\href {https://doi.org/10.1103/PhysRevResearch.2.042014} {\bibfield  {journal} {\bibinfo  {journal} {Phys. Rev. Res.}\ }\textbf {\bibinfo {volume} {2}},\ \bibinfo {pages} {042014} (\bibinfo {year} {2020})}\BibitemShut {NoStop}%
\bibitem [{\citenamefont {Kumar}\ \emph {et~al.}(2022)\citenamefont {Kumar}, \citenamefont {Snizhko}, \citenamefont {Gefen},\ and\ \citenamefont {Rosenow}}]{kumarOptimizedSteeringQuantum2022}%
  \BibitemOpen
  \bibfield  {author} {\bibinfo {author} {\bibfnamefont {P.}~\bibnamefont {Kumar}}, \bibinfo {author} {\bibfnamefont {K.}~\bibnamefont {Snizhko}}, \bibinfo {author} {\bibfnamefont {Y.}~\bibnamefont {Gefen}},\ and\ \bibinfo {author} {\bibfnamefont {B.}~\bibnamefont {Rosenow}},\ }\bibfield  {title} {\bibinfo {title} {Optimized steering: {{Quantum}} state engineering and exceptional points},\ }\href {https://doi.org/10.1103/PhysRevA.105.L010203} {\bibfield  {journal} {\bibinfo  {journal} {Phys. Rev. A}\ }\textbf {\bibinfo {volume} {105}},\ \bibinfo {pages} {L010203} (\bibinfo {year} {2022})}\BibitemShut {NoStop}%
\bibitem [{\citenamefont {Herasymenko}\ \emph {et~al.}(2022)\citenamefont {Herasymenko}, \citenamefont {Gornyi},\ and\ \citenamefont {Gefen}}]{herasymenkoMeasurementdrivenNavigationManybody2022a}%
  \BibitemOpen
  \bibfield  {author} {\bibinfo {author} {\bibfnamefont {Y.}~\bibnamefont {Herasymenko}}, \bibinfo {author} {\bibfnamefont {I.}~\bibnamefont {Gornyi}},\ and\ \bibinfo {author} {\bibfnamefont {Y.}~\bibnamefont {Gefen}},\ }\href {https://doi.org/10.48550/arXiv.2111.09306} {\bibinfo {title} {Measurement-driven navigation in many-body {{Hilbert}} space: {{Active-decision}} steering}} (\bibinfo {year} {2022}),\ \Eprint {https://arxiv.org/abs/2111.09306} {arxiv:2111.09306 [cond-mat, physics:quant-ph]} \BibitemShut {NoStop}%
\bibitem [{\citenamefont {D'Alessandro}\ and\ \citenamefont {Romano}(2006)}]{dalessandroDecompositionsUnitaryEvolutions2006}%
  \BibitemOpen
  \bibfield  {author} {\bibinfo {author} {\bibfnamefont {D.}~\bibnamefont {D'Alessandro}}\ and\ \bibinfo {author} {\bibfnamefont {R.}~\bibnamefont {Romano}},\ }\bibfield  {title} {\bibinfo {title} {Decompositions of unitary evolutions and entanglement dynamics of bipartite quantum systems},\ }\href {https://doi.org/10.1063/1.2245205} {\bibfield  {journal} {\bibinfo  {journal} {J. Math. Phys.}\ }\textbf {\bibinfo {volume} {47}},\ \bibinfo {pages} {082109} (\bibinfo {year} {2006})}\BibitemShut {NoStop}%
\bibitem [{\citenamefont {D'Alessandro}(2021)}]{dalessandroIntroductionQuantumControl2021}%
  \BibitemOpen
  \bibfield  {author} {\bibinfo {author} {\bibfnamefont {D.}~\bibnamefont {D'Alessandro}},\ }\href@noop {} {\emph {\bibinfo {title} {Introduction to {{Quantum Control}} and {{Dynamics}}}}}\ (\bibinfo  {publisher} {{CRC Press}},\ \bibinfo {year} {2021})\BibitemShut {NoStop}%
\bibitem [{\citenamefont {Low}\ and\ \citenamefont {Chuang}(2019)}]{lowHamiltonianSimulationQubitization2019a}%
  \BibitemOpen
  \bibfield  {author} {\bibinfo {author} {\bibfnamefont {G.~H.}\ \bibnamefont {Low}}\ and\ \bibinfo {author} {\bibfnamefont {I.~L.}\ \bibnamefont {Chuang}},\ }\bibfield  {title} {\bibinfo {title} {Hamiltonian {{Simulation}} by {{Qubitization}}},\ }\href {https://doi.org/10.22331/q-2019-07-12-163} {\bibfield  {journal} {\bibinfo  {journal} {Quantum}\ }\textbf {\bibinfo {volume} {3}},\ \bibinfo {pages} {163} (\bibinfo {year} {2019})}\BibitemShut {NoStop}%
\bibitem [{\citenamefont {K{\"o}kc{\"u}}\ \emph {et~al.}(2022)\citenamefont {K{\"o}kc{\"u}}, \citenamefont {Steckmann}, \citenamefont {Wang}, \citenamefont {Freericks}, \citenamefont {Dumitrescu},\ and\ \citenamefont {Kemper}}]{kokcuFixedDepthHamiltonian2022}%
  \BibitemOpen
  \bibfield  {author} {\bibinfo {author} {\bibfnamefont {E.}~\bibnamefont {K{\"o}kc{\"u}}}, \bibinfo {author} {\bibfnamefont {T.}~\bibnamefont {Steckmann}}, \bibinfo {author} {\bibfnamefont {Y.}~\bibnamefont {Wang}}, \bibinfo {author} {\bibfnamefont {J.~K.}\ \bibnamefont {Freericks}}, \bibinfo {author} {\bibfnamefont {E.~F.}\ \bibnamefont {Dumitrescu}},\ and\ \bibinfo {author} {\bibfnamefont {A.~F.}\ \bibnamefont {Kemper}},\ }\bibfield  {title} {\bibinfo {title} {Fixed {{Depth Hamiltonian Simulation}} via {{Cartan Decomposition}}},\ }\href {https://doi.org/10.1103/PhysRevLett.129.070501} {\bibfield  {journal} {\bibinfo  {journal} {Phys. Rev. Lett.}\ }\textbf {\bibinfo {volume} {129}},\ \bibinfo {pages} {070501} (\bibinfo {year} {2022})}\BibitemShut {NoStop}%
\bibitem [{\citenamefont {Zhang}\ \emph {et~al.}(2003)\citenamefont {Zhang}, \citenamefont {Vala}, \citenamefont {Sastry},\ and\ \citenamefont {Whaley}}]{zhangGeometricTheoryNonlocal2003}%
  \BibitemOpen
  \bibfield  {author} {\bibinfo {author} {\bibfnamefont {J.}~\bibnamefont {Zhang}}, \bibinfo {author} {\bibfnamefont {J.}~\bibnamefont {Vala}}, \bibinfo {author} {\bibfnamefont {S.}~\bibnamefont {Sastry}},\ and\ \bibinfo {author} {\bibfnamefont {K.~B.}\ \bibnamefont {Whaley}},\ }\bibfield  {title} {\bibinfo {title} {Geometric theory of nonlocal two-qubit operations},\ }\href {https://doi.org/10.1103/PhysRevA.67.042313} {\bibfield  {journal} {\bibinfo  {journal} {Phys. Rev. A}\ }\textbf {\bibinfo {volume} {67}},\ \bibinfo {pages} {042313} (\bibinfo {year} {2003})}\BibitemShut {NoStop}%
\bibitem [{IBM()}]{IBMQuantum}%
  \BibitemOpen
  \href@noop {} {\bibinfo {title} {{{IBM Quantum}}}},\ \bibinfo {howpublished} {https://quantum-computing.ibm.com/}\BibitemShut {NoStop}%
\bibitem [{\citenamefont {Alexander}\ \emph {et~al.}(2020)\citenamefont {Alexander}, \citenamefont {Kanazawa}, \citenamefont {Egger}, \citenamefont {Capelluto}, \citenamefont {Wood}, \citenamefont {{Javadi-Abhari}},\ and\ \citenamefont {McKay}}]{alexanderQiskitPulseProgramming2020a}%
  \BibitemOpen
  \bibfield  {author} {\bibinfo {author} {\bibfnamefont {T.}~\bibnamefont {Alexander}}, \bibinfo {author} {\bibfnamefont {N.}~\bibnamefont {Kanazawa}}, \bibinfo {author} {\bibfnamefont {D.~J.}\ \bibnamefont {Egger}}, \bibinfo {author} {\bibfnamefont {L.}~\bibnamefont {Capelluto}}, \bibinfo {author} {\bibfnamefont {C.~J.}\ \bibnamefont {Wood}}, \bibinfo {author} {\bibfnamefont {A.}~\bibnamefont {{Javadi-Abhari}}},\ and\ \bibinfo {author} {\bibfnamefont {D.~C.}\ \bibnamefont {McKay}},\ }\bibfield  {title} {\bibinfo {title} {Qiskit pulse: Programming quantum computers through the cloud with pulses},\ }\href {https://doi.org/10.1088/2058-9565/aba404} {\bibfield  {journal} {\bibinfo  {journal} {Quantum Sci. Technol.}\ }\textbf {\bibinfo {volume} {5}},\ \bibinfo {pages} {044006} (\bibinfo {year} {2020})}\BibitemShut {NoStop}%
\bibitem [{\citenamefont {Krantz}\ \emph {et~al.}(2019)\citenamefont {Krantz}, \citenamefont {Kjaergaard}, \citenamefont {Yan}, \citenamefont {Orlando}, \citenamefont {Gustavsson},\ and\ \citenamefont {Oliver}}]{krantzQuantumEngineerGuide2019a}%
  \BibitemOpen
  \bibfield  {author} {\bibinfo {author} {\bibfnamefont {P.}~\bibnamefont {Krantz}}, \bibinfo {author} {\bibfnamefont {M.}~\bibnamefont {Kjaergaard}}, \bibinfo {author} {\bibfnamefont {F.}~\bibnamefont {Yan}}, \bibinfo {author} {\bibfnamefont {T.~P.}\ \bibnamefont {Orlando}}, \bibinfo {author} {\bibfnamefont {S.}~\bibnamefont {Gustavsson}},\ and\ \bibinfo {author} {\bibfnamefont {W.~D.}\ \bibnamefont {Oliver}},\ }\bibfield  {title} {\bibinfo {title} {A quantum engineer's guide to superconducting qubits},\ }\href {https://doi.org/10.1063/1.5089550} {\bibfield  {journal} {\bibinfo  {journal} {Applied Physics Reviews}\ }\textbf {\bibinfo {volume} {6}},\ \bibinfo {pages} {021318} (\bibinfo {year} {2019})}\BibitemShut {NoStop}%
\bibitem [{\citenamefont {Galda}\ \emph {et~al.}(2021)\citenamefont {Galda}, \citenamefont {Cubeddu}, \citenamefont {Kanazawa}, \citenamefont {Narang},\ and\ \citenamefont {{Earnest-Noble}}}]{galdaImplementingTernaryDecomposition2021}%
  \BibitemOpen
  \bibfield  {author} {\bibinfo {author} {\bibfnamefont {A.}~\bibnamefont {Galda}}, \bibinfo {author} {\bibfnamefont {M.}~\bibnamefont {Cubeddu}}, \bibinfo {author} {\bibfnamefont {N.}~\bibnamefont {Kanazawa}}, \bibinfo {author} {\bibfnamefont {P.}~\bibnamefont {Narang}},\ and\ \bibinfo {author} {\bibfnamefont {N.}~\bibnamefont {{Earnest-Noble}}},\ }\href {https://doi.org/10.48550/arXiv.2109.00558} {\bibinfo {title} {Implementing a {{Ternary Decomposition}} of the {{Toffoli Gate}} on {{Fixed-FrequencyTransmon Qutrits}}}} (\bibinfo {year} {2021}),\ \Eprint {https://arxiv.org/abs/2109.00558} {arxiv:2109.00558 [quant-ph]} \BibitemShut {NoStop}%
\bibitem [{\citenamefont {Yurtalan}\ \emph {et~al.}(2020)\citenamefont {Yurtalan}, \citenamefont {Shi}, \citenamefont {Kononenko}, \citenamefont {Lupascu},\ and\ \citenamefont {Ashhab}}]{yurtalanImplementationWalshHadamardGate2020}%
  \BibitemOpen
  \bibfield  {author} {\bibinfo {author} {\bibfnamefont {M.~A.}\ \bibnamefont {Yurtalan}}, \bibinfo {author} {\bibfnamefont {J.}~\bibnamefont {Shi}}, \bibinfo {author} {\bibfnamefont {M.}~\bibnamefont {Kononenko}}, \bibinfo {author} {\bibfnamefont {A.}~\bibnamefont {Lupascu}},\ and\ \bibinfo {author} {\bibfnamefont {S.}~\bibnamefont {Ashhab}},\ }\bibfield  {title} {\bibinfo {title} {Implementation of a {{Walsh-Hadamard Gate}} in a {{Superconducting Qutrit}}},\ }\href {https://doi.org/10.1103/PhysRevLett.125.180504} {\bibfield  {journal} {\bibinfo  {journal} {Phys. Rev. Lett.}\ }\textbf {\bibinfo {volume} {125}},\ \bibinfo {pages} {180504} (\bibinfo {year} {2020})}\BibitemShut {NoStop}%
\bibitem [{\citenamefont {Smolin}\ \emph {et~al.}(2012)\citenamefont {Smolin}, \citenamefont {Gambetta},\ and\ \citenamefont {Smith}}]{smolinEfficientMethodComputing2012}%
  \BibitemOpen
  \bibfield  {author} {\bibinfo {author} {\bibfnamefont {J.~A.}\ \bibnamefont {Smolin}}, \bibinfo {author} {\bibfnamefont {J.~M.}\ \bibnamefont {Gambetta}},\ and\ \bibinfo {author} {\bibfnamefont {G.}~\bibnamefont {Smith}},\ }\bibfield  {title} {\bibinfo {title} {Efficient {{Method}} for {{Computing}} the {{Maximum-Likelihood Quantum State}} from {{Measurements}} with {{Additive Gaussian Noise}}},\ }\href {https://doi.org/10.1103/PhysRevLett.108.070502} {\bibfield  {journal} {\bibinfo  {journal} {Phys. Rev. Lett.}\ }\textbf {\bibinfo {volume} {108}},\ \bibinfo {pages} {070502} (\bibinfo {year} {2012})}\BibitemShut {NoStop}%
\bibitem [{\citenamefont {Sheldon}\ \emph {et~al.}(2016)\citenamefont {Sheldon}, \citenamefont {Magesan}, \citenamefont {Chow},\ and\ \citenamefont {Gambetta}}]{sheldonProcedureSystematicallyTuning2016}%
  \BibitemOpen
  \bibfield  {author} {\bibinfo {author} {\bibfnamefont {S.}~\bibnamefont {Sheldon}}, \bibinfo {author} {\bibfnamefont {E.}~\bibnamefont {Magesan}}, \bibinfo {author} {\bibfnamefont {J.~M.}\ \bibnamefont {Chow}},\ and\ \bibinfo {author} {\bibfnamefont {J.~M.}\ \bibnamefont {Gambetta}},\ }\bibfield  {title} {\bibinfo {title} {Procedure for systematically tuning up cross-talk in the cross-resonance gate},\ }\href {https://doi.org/10.1103/PhysRevA.93.060302} {\bibfield  {journal} {\bibinfo  {journal} {Phys. Rev. A}\ }\textbf {\bibinfo {volume} {93}},\ \bibinfo {pages} {060302} (\bibinfo {year} {2016})}\BibitemShut {NoStop}%
\bibitem [{\citenamefont {Wood}(2020)}]{woodSpecialSessionNoise2020b}%
  \BibitemOpen
  \bibfield  {author} {\bibinfo {author} {\bibfnamefont {C.~J.}\ \bibnamefont {Wood}},\ }\bibfield  {title} {\bibinfo {title} {Special {{Session}}: {{Noise Characterization}} and {{Error Mitigation}} in {{Near-Term Quantum Computers}}},\ }in\ \href {https://doi.org/10.1109/ICCD50377.2020.00016} {\emph {\bibinfo {booktitle} {2020 {{IEEE}} 38th {{International Conference}} on {{Computer Design}} ({{ICCD}})}}}\ (\bibinfo {year} {2020})\ pp.\ \bibinfo {pages} {13--16}\BibitemShut {NoStop}%
\bibitem [{\citenamefont {Martinis}\ \emph {et~al.}(2002)\citenamefont {Martinis}, \citenamefont {Nam}, \citenamefont {Aumentado},\ and\ \citenamefont {Urbina}}]{martinisRabiOscillationsLarge2002}%
  \BibitemOpen
  \bibfield  {author} {\bibinfo {author} {\bibfnamefont {J.~M.}\ \bibnamefont {Martinis}}, \bibinfo {author} {\bibfnamefont {S.}~\bibnamefont {Nam}}, \bibinfo {author} {\bibfnamefont {J.}~\bibnamefont {Aumentado}},\ and\ \bibinfo {author} {\bibfnamefont {C.}~\bibnamefont {Urbina}},\ }\bibfield  {title} {\bibinfo {title} {Rabi {{Oscillations}} in a {{Large Josephson-Junction Qubit}}},\ }\href {https://doi.org/10.1103/PhysRevLett.89.117901} {\bibfield  {journal} {\bibinfo  {journal} {Phys. Rev. Lett.}\ }\textbf {\bibinfo {volume} {89}},\ \bibinfo {pages} {117901} (\bibinfo {year} {2002})}\BibitemShut {NoStop}%
\bibitem [{\citenamefont {Cooper}\ \emph {et~al.}(2004)\citenamefont {Cooper}, \citenamefont {Steffen}, \citenamefont {McDermott}, \citenamefont {Simmonds}, \citenamefont {Oh}, \citenamefont {Hite}, \citenamefont {Pappas},\ and\ \citenamefont {Martinis}}]{cooperObservationQuantumOscillations2004}%
  \BibitemOpen
  \bibfield  {author} {\bibinfo {author} {\bibfnamefont {K.~B.}\ \bibnamefont {Cooper}}, \bibinfo {author} {\bibfnamefont {M.}~\bibnamefont {Steffen}}, \bibinfo {author} {\bibfnamefont {R.}~\bibnamefont {McDermott}}, \bibinfo {author} {\bibfnamefont {R.~W.}\ \bibnamefont {Simmonds}}, \bibinfo {author} {\bibfnamefont {S.}~\bibnamefont {Oh}}, \bibinfo {author} {\bibfnamefont {D.~A.}\ \bibnamefont {Hite}}, \bibinfo {author} {\bibfnamefont {D.~P.}\ \bibnamefont {Pappas}},\ and\ \bibinfo {author} {\bibfnamefont {J.~M.}\ \bibnamefont {Martinis}},\ }\bibfield  {title} {\bibinfo {title} {Observation of {{Quantum Oscillations}} between a {{Josephson Phase Qubit}} and a {{Microscopic Resonator Using Fast Readout}}},\ }\href {https://doi.org/10.1103/PhysRevLett.93.180401} {\bibfield  {journal} {\bibinfo  {journal} {Phys. Rev. Lett.}\ }\textbf {\bibinfo {volume} {93}},\ \bibinfo {pages} {180401} (\bibinfo {year} {2004})}\BibitemShut {NoStop}%
\bibitem [{\citenamefont {Lucero}\ \emph {et~al.}(2008)\citenamefont {Lucero}, \citenamefont {Hofheinz}, \citenamefont {Ansmann}, \citenamefont {Bialczak}, \citenamefont {Katz}, \citenamefont {Neeley}, \citenamefont {O'Connell}, \citenamefont {Wang}, \citenamefont {Cleland},\ and\ \citenamefont {Martinis}}]{luceroHighFidelityGatesSingle2008}%
  \BibitemOpen
  \bibfield  {author} {\bibinfo {author} {\bibfnamefont {E.}~\bibnamefont {Lucero}}, \bibinfo {author} {\bibfnamefont {M.}~\bibnamefont {Hofheinz}}, \bibinfo {author} {\bibfnamefont {M.}~\bibnamefont {Ansmann}}, \bibinfo {author} {\bibfnamefont {R.~C.}\ \bibnamefont {Bialczak}}, \bibinfo {author} {\bibfnamefont {N.}~\bibnamefont {Katz}}, \bibinfo {author} {\bibfnamefont {M.}~\bibnamefont {Neeley}}, \bibinfo {author} {\bibfnamefont {A.~D.}\ \bibnamefont {O'Connell}}, \bibinfo {author} {\bibfnamefont {H.}~\bibnamefont {Wang}}, \bibinfo {author} {\bibfnamefont {A.~N.}\ \bibnamefont {Cleland}},\ and\ \bibinfo {author} {\bibfnamefont {J.~M.}\ \bibnamefont {Martinis}},\ }\bibfield  {title} {\bibinfo {title} {High-{{Fidelity Gates}} in a {{Single Josephson Qubit}}},\ }\href {https://doi.org/10.1103/PhysRevLett.100.247001} {\bibfield  {journal} {\bibinfo  {journal} {Phys. Rev. Lett.}\ }\textbf {\bibinfo {volume} {100}},\ \bibinfo {pages} {247001} (\bibinfo {year} {2008})}\BibitemShut {NoStop}%
\bibitem [{\citenamefont {Valenzuela}\ \emph {et~al.}(2006)\citenamefont {Valenzuela}, \citenamefont {Oliver}, \citenamefont {Berns}, \citenamefont {Berggren}, \citenamefont {Levitov},\ and\ \citenamefont {Orlando}}]{valenzuelaMicrowaveInducedCoolingSuperconducting2006}%
  \BibitemOpen
  \bibfield  {author} {\bibinfo {author} {\bibfnamefont {S.~O.}\ \bibnamefont {Valenzuela}}, \bibinfo {author} {\bibfnamefont {W.~D.}\ \bibnamefont {Oliver}}, \bibinfo {author} {\bibfnamefont {D.~M.}\ \bibnamefont {Berns}}, \bibinfo {author} {\bibfnamefont {K.~K.}\ \bibnamefont {Berggren}}, \bibinfo {author} {\bibfnamefont {L.~S.}\ \bibnamefont {Levitov}},\ and\ \bibinfo {author} {\bibfnamefont {T.~P.}\ \bibnamefont {Orlando}},\ }\bibfield  {title} {\bibinfo {title} {Microwave-{{Induced Cooling}} of a {{Superconducting Qubit}}},\ }\href {https://doi.org/10.1126/science.1134008} {\bibfield  {journal} {\bibinfo  {journal} {Science}\ }\textbf {\bibinfo {volume} {314}},\ \bibinfo {pages} {1589} (\bibinfo {year} {2006})}\BibitemShut {NoStop}%
\bibitem [{\citenamefont {Neeley}\ \emph {et~al.}(2009)\citenamefont {Neeley}, \citenamefont {Ansmann}, \citenamefont {Bialczak}, \citenamefont {Hofheinz}, \citenamefont {Lucero}, \citenamefont {O'Connell}, \citenamefont {Sank}, \citenamefont {Wang}, \citenamefont {Wenner}, \citenamefont {Cleland}, \citenamefont {Geller},\ and\ \citenamefont {Martinis}}]{neeleyEmulationQuantumSpin2009}%
  \BibitemOpen
  \bibfield  {author} {\bibinfo {author} {\bibfnamefont {M.}~\bibnamefont {Neeley}}, \bibinfo {author} {\bibfnamefont {M.}~\bibnamefont {Ansmann}}, \bibinfo {author} {\bibfnamefont {R.~C.}\ \bibnamefont {Bialczak}}, \bibinfo {author} {\bibfnamefont {M.}~\bibnamefont {Hofheinz}}, \bibinfo {author} {\bibfnamefont {E.}~\bibnamefont {Lucero}}, \bibinfo {author} {\bibfnamefont {A.~D.}\ \bibnamefont {O'Connell}}, \bibinfo {author} {\bibfnamefont {D.}~\bibnamefont {Sank}}, \bibinfo {author} {\bibfnamefont {H.}~\bibnamefont {Wang}}, \bibinfo {author} {\bibfnamefont {J.}~\bibnamefont {Wenner}}, \bibinfo {author} {\bibfnamefont {A.~N.}\ \bibnamefont {Cleland}}, \bibinfo {author} {\bibfnamefont {M.~R.}\ \bibnamefont {Geller}},\ and\ \bibinfo {author} {\bibfnamefont {J.~M.}\ \bibnamefont {Martinis}},\ }\bibfield  {title} {\bibinfo {title} {Emulation of a {{Quantum Spin}} with a {{Superconducting Phase Qudit}}},\ }\href {https://doi.org/10.1126/science.1173440} {\bibfield  {journal} {\bibinfo  {journal} {Science}\ }\textbf {\bibinfo {volume} {325}},\ \bibinfo {pages} {722} (\bibinfo {year} {2009})}\BibitemShut {NoStop}%
\bibitem [{\citenamefont {Bianchetti}\ \emph {et~al.}(2010)\citenamefont {Bianchetti}, \citenamefont {Filipp}, \citenamefont {Baur}, \citenamefont {Fink}, \citenamefont {Lang}, \citenamefont {Steffen}, \citenamefont {Boissonneault}, \citenamefont {Blais},\ and\ \citenamefont {Wallraff}}]{bianchettiControlTomographyThree2010}%
  \BibitemOpen
  \bibfield  {author} {\bibinfo {author} {\bibfnamefont {R.}~\bibnamefont {Bianchetti}}, \bibinfo {author} {\bibfnamefont {S.}~\bibnamefont {Filipp}}, \bibinfo {author} {\bibfnamefont {M.}~\bibnamefont {Baur}}, \bibinfo {author} {\bibfnamefont {J.~M.}\ \bibnamefont {Fink}}, \bibinfo {author} {\bibfnamefont {C.}~\bibnamefont {Lang}}, \bibinfo {author} {\bibfnamefont {L.}~\bibnamefont {Steffen}}, \bibinfo {author} {\bibfnamefont {M.}~\bibnamefont {Boissonneault}}, \bibinfo {author} {\bibfnamefont {A.}~\bibnamefont {Blais}},\ and\ \bibinfo {author} {\bibfnamefont {A.}~\bibnamefont {Wallraff}},\ }\bibfield  {title} {\bibinfo {title} {Control and {{Tomography}} of a {{Three Level Superconducting Artificial Atom}}},\ }\href {https://doi.org/10.1103/PhysRevLett.105.223601} {\bibfield  {journal} {\bibinfo  {journal} {Phys. Rev. Lett.}\ }\textbf {\bibinfo {volume} {105}},\ \bibinfo {pages} {223601} (\bibinfo {year} {2010})}\BibitemShut {NoStop}%
\bibitem [{\citenamefont {Abdumalikov}\ \emph {et~al.}(2010)\citenamefont {Abdumalikov}, \citenamefont {Astafiev}, \citenamefont {Zagoskin}, \citenamefont {Pashkin}, \citenamefont {Nakamura},\ and\ \citenamefont {Tsai}}]{abdumalikovElectromagneticallyInducedTransparency2010}%
  \BibitemOpen
  \bibfield  {author} {\bibinfo {author} {\bibfnamefont {A.~A.}\ \bibnamefont {Abdumalikov}}, \bibinfo {author} {\bibfnamefont {O.}~\bibnamefont {Astafiev}}, \bibinfo {author} {\bibfnamefont {A.~M.}\ \bibnamefont {Zagoskin}}, \bibinfo {author} {\bibfnamefont {{\relax Yu}.~A.}\ \bibnamefont {Pashkin}}, \bibinfo {author} {\bibfnamefont {Y.}~\bibnamefont {Nakamura}},\ and\ \bibinfo {author} {\bibfnamefont {J.~S.}\ \bibnamefont {Tsai}},\ }\bibfield  {title} {\bibinfo {title} {Electromagnetically {{Induced Transparency}} on a {{Single Artificial Atom}}},\ }\href {https://doi.org/10.1103/PhysRevLett.104.193601} {\bibfield  {journal} {\bibinfo  {journal} {Phys. Rev. Lett.}\ }\textbf {\bibinfo {volume} {104}},\ \bibinfo {pages} {193601} (\bibinfo {year} {2010})}\BibitemShut {NoStop}%
\bibitem [{\citenamefont {Abdumalikov~Jr}\ \emph {et~al.}(2013)\citenamefont {Abdumalikov~Jr}, \citenamefont {Fink}, \citenamefont {Juliusson}, \citenamefont {Pechal}, \citenamefont {Berger}, \citenamefont {Wallraff},\ and\ \citenamefont {Filipp}}]{abdumalikovjrExperimentalRealizationNonAbelian2013a}%
  \BibitemOpen
  \bibfield  {author} {\bibinfo {author} {\bibfnamefont {A.~A.}\ \bibnamefont {Abdumalikov~Jr}}, \bibinfo {author} {\bibfnamefont {J.~M.}\ \bibnamefont {Fink}}, \bibinfo {author} {\bibfnamefont {K.}~\bibnamefont {Juliusson}}, \bibinfo {author} {\bibfnamefont {M.}~\bibnamefont {Pechal}}, \bibinfo {author} {\bibfnamefont {S.}~\bibnamefont {Berger}}, \bibinfo {author} {\bibfnamefont {A.}~\bibnamefont {Wallraff}},\ and\ \bibinfo {author} {\bibfnamefont {S.}~\bibnamefont {Filipp}},\ }\bibfield  {title} {\bibinfo {title} {Experimental realization of non-{{Abelian}} non-adiabatic geometric gates},\ }\href {https://doi.org/10.1038/nature12010} {\bibfield  {journal} {\bibinfo  {journal} {Nature}\ }\textbf {\bibinfo {volume} {496}},\ \bibinfo {pages} {482} (\bibinfo {year} {2013})}\BibitemShut {NoStop}%
\bibitem [{\citenamefont {Jerger}\ \emph {et~al.}(2016)\citenamefont {Jerger}, \citenamefont {Reshitnyk}, \citenamefont {Oppliger}, \citenamefont {Poto{\v c}nik}, \citenamefont {Mondal}, \citenamefont {Wallraff}, \citenamefont {Goodenough}, \citenamefont {Wehner}, \citenamefont {Juliusson}, \citenamefont {Langford},\ and\ \citenamefont {Fedorov}}]{jergerContextualityNonlocalitySuperconducting2016}%
  \BibitemOpen
  \bibfield  {author} {\bibinfo {author} {\bibfnamefont {M.}~\bibnamefont {Jerger}}, \bibinfo {author} {\bibfnamefont {Y.}~\bibnamefont {Reshitnyk}}, \bibinfo {author} {\bibfnamefont {M.}~\bibnamefont {Oppliger}}, \bibinfo {author} {\bibfnamefont {A.}~\bibnamefont {Poto{\v c}nik}}, \bibinfo {author} {\bibfnamefont {M.}~\bibnamefont {Mondal}}, \bibinfo {author} {\bibfnamefont {A.}~\bibnamefont {Wallraff}}, \bibinfo {author} {\bibfnamefont {K.}~\bibnamefont {Goodenough}}, \bibinfo {author} {\bibfnamefont {S.}~\bibnamefont {Wehner}}, \bibinfo {author} {\bibfnamefont {K.}~\bibnamefont {Juliusson}}, \bibinfo {author} {\bibfnamefont {N.~K.}\ \bibnamefont {Langford}},\ and\ \bibinfo {author} {\bibfnamefont {A.}~\bibnamefont {Fedorov}},\ }\bibfield  {title} {\bibinfo {title} {Contextuality without nonlocality in a superconducting quantum system},\ }\href {https://doi.org/10.1038/ncomms12930} {\bibfield  {journal} {\bibinfo  {journal} {Nat Commun}\ }\textbf {\bibinfo {volume} {7}},\ \bibinfo {pages} {12930} (\bibinfo {year} {2016})}\BibitemShut {NoStop}%
\bibitem [{\citenamefont {Tan}\ \emph {et~al.}(2018)\citenamefont {Tan}, \citenamefont {Zhang}, \citenamefont {Liu}, \citenamefont {Xue}, \citenamefont {Yu}, \citenamefont {Zhu}, \citenamefont {Yan}, \citenamefont {Zhu},\ and\ \citenamefont {Yu}}]{tanTopologicalMaxwellMetal2018}%
  \BibitemOpen
  \bibfield  {author} {\bibinfo {author} {\bibfnamefont {X.}~\bibnamefont {Tan}}, \bibinfo {author} {\bibfnamefont {D.-W.}\ \bibnamefont {Zhang}}, \bibinfo {author} {\bibfnamefont {Q.}~\bibnamefont {Liu}}, \bibinfo {author} {\bibfnamefont {G.}~\bibnamefont {Xue}}, \bibinfo {author} {\bibfnamefont {H.-F.}\ \bibnamefont {Yu}}, \bibinfo {author} {\bibfnamefont {Y.-Q.}\ \bibnamefont {Zhu}}, \bibinfo {author} {\bibfnamefont {H.}~\bibnamefont {Yan}}, \bibinfo {author} {\bibfnamefont {S.-L.}\ \bibnamefont {Zhu}},\ and\ \bibinfo {author} {\bibfnamefont {Y.}~\bibnamefont {Yu}},\ }\bibfield  {title} {\bibinfo {title} {Topological {{Maxwell Metal Bands}} in a {{Superconducting Qutrit}}},\ }\href {https://doi.org/10.1103/PhysRevLett.120.130503} {\bibfield  {journal} {\bibinfo  {journal} {Phys. Rev. Lett.}\ }\textbf {\bibinfo {volume} {120}},\ \bibinfo {pages} {130503} (\bibinfo {year} {2018})}\BibitemShut {NoStop}%
\bibitem [{\citenamefont {{H{\"o}nigl-Decrinis}}\ \emph {et~al.}(2018)\citenamefont {{H{\"o}nigl-Decrinis}}, \citenamefont {Antonov}, \citenamefont {Shaikhaidarov}, \citenamefont {Antonov}, \citenamefont {Dmitriev},\ and\ \citenamefont {Astafiev}}]{honigl-decrinisMixingCoherentWaves2018}%
  \BibitemOpen
  \bibfield  {author} {\bibinfo {author} {\bibfnamefont {T.}~\bibnamefont {{H{\"o}nigl-Decrinis}}}, \bibinfo {author} {\bibfnamefont {I.~V.}\ \bibnamefont {Antonov}}, \bibinfo {author} {\bibfnamefont {R.}~\bibnamefont {Shaikhaidarov}}, \bibinfo {author} {\bibfnamefont {V.~N.}\ \bibnamefont {Antonov}}, \bibinfo {author} {\bibfnamefont {A.~{\relax Yu}.}\ \bibnamefont {Dmitriev}},\ and\ \bibinfo {author} {\bibfnamefont {O.~V.}\ \bibnamefont {Astafiev}},\ }\bibfield  {title} {\bibinfo {title} {Mixing of coherent waves in a single three-level artificial atom},\ }\href {https://doi.org/10.1103/PhysRevA.98.041801} {\bibfield  {journal} {\bibinfo  {journal} {Phys. Rev. A}\ }\textbf {\bibinfo {volume} {98}},\ \bibinfo {pages} {041801} (\bibinfo {year} {2018})}\BibitemShut {NoStop}%
\bibitem [{\citenamefont {Veps{\"a}l{\"a}inen}\ and\ \citenamefont {Paraoanu}(2020)}]{vepsalainenSimulatingSpinChains2020}%
  \BibitemOpen
  \bibfield  {author} {\bibinfo {author} {\bibfnamefont {A.}~\bibnamefont {Veps{\"a}l{\"a}inen}}\ and\ \bibinfo {author} {\bibfnamefont {G.~S.}\ \bibnamefont {Paraoanu}},\ }\bibfield  {title} {\bibinfo {title} {Simulating {{Spin Chains Using}} a {{Superconducting Circuit}}: {{Gauge Invariance}}, {{Superadiabatic Transport}}, and {{Broken Time-Reversal Symmetry}}},\ }\href {https://doi.org/10.1002/qute.201900121} {\bibfield  {journal} {\bibinfo  {journal} {Advanced Quantum Technologies}\ }\textbf {\bibinfo {volume} {3}},\ \bibinfo {pages} {1900121} (\bibinfo {year} {2020})}\BibitemShut {NoStop}%
\bibitem [{\citenamefont {Fedorov}\ \emph {et~al.}(2012)\citenamefont {Fedorov}, \citenamefont {Steffen}, \citenamefont {Baur}, \citenamefont {{da Silva}},\ and\ \citenamefont {Wallraff}}]{fedorovImplementationToffoliGate2012}%
  \BibitemOpen
  \bibfield  {author} {\bibinfo {author} {\bibfnamefont {A.}~\bibnamefont {Fedorov}}, \bibinfo {author} {\bibfnamefont {L.}~\bibnamefont {Steffen}}, \bibinfo {author} {\bibfnamefont {M.}~\bibnamefont {Baur}}, \bibinfo {author} {\bibfnamefont {M.~P.}\ \bibnamefont {{da Silva}}},\ and\ \bibinfo {author} {\bibfnamefont {A.}~\bibnamefont {Wallraff}},\ }\bibfield  {title} {\bibinfo {title} {Implementation of a {{Toffoli}} gate with superconducting circuits},\ }\href {https://doi.org/10.1038/nature10713} {\bibfield  {journal} {\bibinfo  {journal} {Nature}\ }\textbf {\bibinfo {volume} {481}},\ \bibinfo {pages} {170} (\bibinfo {year} {2012})}\BibitemShut {NoStop}%
\bibitem [{\citenamefont {Morvan}\ \emph {et~al.}(2021)\citenamefont {Morvan}, \citenamefont {Ramasesh}, \citenamefont {Blok}, \citenamefont {Kreikebaum}, \citenamefont {O'Brien}, \citenamefont {Chen}, \citenamefont {Mitchell}, \citenamefont {Naik}, \citenamefont {Santiago},\ and\ \citenamefont {Siddiqi}}]{morvanQutritRandomizedBenchmarking2021}%
  \BibitemOpen
  \bibfield  {author} {\bibinfo {author} {\bibfnamefont {A.}~\bibnamefont {Morvan}}, \bibinfo {author} {\bibfnamefont {V.~V.}\ \bibnamefont {Ramasesh}}, \bibinfo {author} {\bibfnamefont {M.~S.}\ \bibnamefont {Blok}}, \bibinfo {author} {\bibfnamefont {J.~M.}\ \bibnamefont {Kreikebaum}}, \bibinfo {author} {\bibfnamefont {K.}~\bibnamefont {O'Brien}}, \bibinfo {author} {\bibfnamefont {L.}~\bibnamefont {Chen}}, \bibinfo {author} {\bibfnamefont {B.~K.}\ \bibnamefont {Mitchell}}, \bibinfo {author} {\bibfnamefont {R.~K.}\ \bibnamefont {Naik}}, \bibinfo {author} {\bibfnamefont {D.~I.}\ \bibnamefont {Santiago}},\ and\ \bibinfo {author} {\bibfnamefont {I.}~\bibnamefont {Siddiqi}},\ }\bibfield  {title} {\bibinfo {title} {Qutrit {{Randomized Benchmarking}}},\ }\href {https://doi.org/10.1103/PhysRevLett.126.210504} {\bibfield  {journal} {\bibinfo  {journal} {Phys. Rev. Lett.}\ }\textbf {\bibinfo {volume} {126}},\ \bibinfo {pages} {210504} (\bibinfo {year} {2021})}\BibitemShut {NoStop}%
\bibitem [{\citenamefont {Gaebler}\ \emph {et~al.}(2021)\citenamefont {Gaebler}, \citenamefont {Baldwin}, \citenamefont {Moses}, \citenamefont {Dreiling}, \citenamefont {Figgatt}, \citenamefont {{Foss-Feig}}, \citenamefont {Hayes},\ and\ \citenamefont {Pino}}]{gaeblerSuppressionMidcircuitMeasurement2021}%
  \BibitemOpen
  \bibfield  {author} {\bibinfo {author} {\bibfnamefont {J.~P.}\ \bibnamefont {Gaebler}}, \bibinfo {author} {\bibfnamefont {C.~H.}\ \bibnamefont {Baldwin}}, \bibinfo {author} {\bibfnamefont {S.~A.}\ \bibnamefont {Moses}}, \bibinfo {author} {\bibfnamefont {J.~M.}\ \bibnamefont {Dreiling}}, \bibinfo {author} {\bibfnamefont {C.}~\bibnamefont {Figgatt}}, \bibinfo {author} {\bibfnamefont {M.}~\bibnamefont {{Foss-Feig}}}, \bibinfo {author} {\bibfnamefont {D.}~\bibnamefont {Hayes}},\ and\ \bibinfo {author} {\bibfnamefont {J.~M.}\ \bibnamefont {Pino}},\ }\bibfield  {title} {\bibinfo {title} {Suppression of midcircuit measurement crosstalk errors with micromotion},\ }\href {https://doi.org/10.1103/PhysRevA.104.062440} {\bibfield  {journal} {\bibinfo  {journal} {Phys. Rev. A}\ }\textbf {\bibinfo {volume} {104}},\ \bibinfo {pages} {062440} (\bibinfo {year} {2021})}\BibitemShut {NoStop}%
\bibitem [{\citenamefont {McClure}\ \emph {et~al.}(2016)\citenamefont {McClure}, \citenamefont {Paik}, \citenamefont {Bishop}, \citenamefont {Steffen}, \citenamefont {Chow},\ and\ \citenamefont {Gambetta}}]{mcclureRapidDrivenReset2016}%
  \BibitemOpen
  \bibfield  {author} {\bibinfo {author} {\bibfnamefont {D.~T.}\ \bibnamefont {McClure}}, \bibinfo {author} {\bibfnamefont {H.}~\bibnamefont {Paik}}, \bibinfo {author} {\bibfnamefont {L.~S.}\ \bibnamefont {Bishop}}, \bibinfo {author} {\bibfnamefont {M.}~\bibnamefont {Steffen}}, \bibinfo {author} {\bibfnamefont {J.~M.}\ \bibnamefont {Chow}},\ and\ \bibinfo {author} {\bibfnamefont {J.~M.}\ \bibnamefont {Gambetta}},\ }\bibfield  {title} {\bibinfo {title} {Rapid {{Driven Reset}} of a {{Qubit Readout Resonator}}},\ }\href {https://doi.org/10.1103/PhysRevApplied.5.011001} {\bibfield  {journal} {\bibinfo  {journal} {Phys. Rev. Appl.}\ }\textbf {\bibinfo {volume} {5}},\ \bibinfo {pages} {011001} (\bibinfo {year} {2016})}\BibitemShut {NoStop}%
\bibitem [{\citenamefont {Peruzzo}\ \emph {et~al.}(2014)\citenamefont {Peruzzo}, \citenamefont {McClean}, \citenamefont {Shadbolt}, \citenamefont {Yung}, \citenamefont {Zhou}, \citenamefont {Love}, \citenamefont {{Aspuru-Guzik}},\ and\ \citenamefont {O'Brien}}]{peruzzoVariationalEigenvalueSolver2014a}%
  \BibitemOpen
  \bibfield  {author} {\bibinfo {author} {\bibfnamefont {A.}~\bibnamefont {Peruzzo}}, \bibinfo {author} {\bibfnamefont {J.}~\bibnamefont {McClean}}, \bibinfo {author} {\bibfnamefont {P.}~\bibnamefont {Shadbolt}}, \bibinfo {author} {\bibfnamefont {M.-H.}\ \bibnamefont {Yung}}, \bibinfo {author} {\bibfnamefont {X.-Q.}\ \bibnamefont {Zhou}}, \bibinfo {author} {\bibfnamefont {P.~J.}\ \bibnamefont {Love}}, \bibinfo {author} {\bibfnamefont {A.}~\bibnamefont {{Aspuru-Guzik}}},\ and\ \bibinfo {author} {\bibfnamefont {J.~L.}\ \bibnamefont {O'Brien}},\ }\bibfield  {title} {\bibinfo {title} {A variational eigenvalue solver on a photonic quantum processor},\ }\href {https://doi.org/10.1038/ncomms5213} {\bibfield  {journal} {\bibinfo  {journal} {Nat Commun}\ }\textbf {\bibinfo {volume} {5}},\ \bibinfo {pages} {4213} (\bibinfo {year} {2014})}\BibitemShut {NoStop}%
\bibitem [{\citenamefont {Grant}\ \emph {et~al.}(2019)\citenamefont {Grant}, \citenamefont {Wossnig}, \citenamefont {Ostaszewski},\ and\ \citenamefont {Benedetti}}]{grantInitializationStrategyAddressing2019a}%
  \BibitemOpen
  \bibfield  {author} {\bibinfo {author} {\bibfnamefont {E.}~\bibnamefont {Grant}}, \bibinfo {author} {\bibfnamefont {L.}~\bibnamefont {Wossnig}}, \bibinfo {author} {\bibfnamefont {M.}~\bibnamefont {Ostaszewski}},\ and\ \bibinfo {author} {\bibfnamefont {M.}~\bibnamefont {Benedetti}},\ }\bibfield  {title} {\bibinfo {title} {An initialization strategy for addressing barren plateaus in parametrized quantum circuits},\ }\href {https://doi.org/10.22331/q-2019-12-09-214} {\bibfield  {journal} {\bibinfo  {journal} {Quantum}\ }\textbf {\bibinfo {volume} {3}},\ \bibinfo {pages} {214} (\bibinfo {year} {2019})}\BibitemShut {NoStop}%
\bibitem [{\citenamefont {Wilczek}\ and\ \citenamefont {Zee}(1984)}]{wilczekAppearanceGaugeStructure1984a}%
  \BibitemOpen
  \bibfield  {author} {\bibinfo {author} {\bibfnamefont {F.}~\bibnamefont {Wilczek}}\ and\ \bibinfo {author} {\bibfnamefont {A.}~\bibnamefont {Zee}},\ }\bibfield  {title} {\bibinfo {title} {Appearance of {{Gauge Structure}} in {{Simple Dynamical Systems}}},\ }\href {https://doi.org/10.1103/PhysRevLett.52.2111} {\bibfield  {journal} {\bibinfo  {journal} {Phys. Rev. Lett.}\ }\textbf {\bibinfo {volume} {52}},\ \bibinfo {pages} {2111} (\bibinfo {year} {1984})}\BibitemShut {NoStop}%
\bibitem [{\citenamefont {Snizhko}\ \emph {et~al.}(2019)\citenamefont {Snizhko}, \citenamefont {Egger},\ and\ \citenamefont {Gefen}}]{snizhkoNonAbelianGeometricDephasing2019a}%
  \BibitemOpen
  \bibfield  {author} {\bibinfo {author} {\bibfnamefont {K.}~\bibnamefont {Snizhko}}, \bibinfo {author} {\bibfnamefont {R.}~\bibnamefont {Egger}},\ and\ \bibinfo {author} {\bibfnamefont {Y.}~\bibnamefont {Gefen}},\ }\bibfield  {title} {\bibinfo {title} {Non-{{Abelian Geometric Dephasing}}},\ }\href {https://doi.org/10.1103/PhysRevLett.123.060405} {\bibfield  {journal} {\bibinfo  {journal} {Phys. Rev. Lett.}\ }\textbf {\bibinfo {volume} {123}},\ \bibinfo {pages} {060405} (\bibinfo {year} {2019})}\BibitemShut {NoStop}%
\bibitem [{\citenamefont {Zanardi}\ and\ \citenamefont {Rasetti}(1999)}]{zanardiHolonomicQuantumComputation1999b}%
  \BibitemOpen
  \bibfield  {author} {\bibinfo {author} {\bibfnamefont {P.}~\bibnamefont {Zanardi}}\ and\ \bibinfo {author} {\bibfnamefont {M.}~\bibnamefont {Rasetti}},\ }\bibfield  {title} {\bibinfo {title} {Holonomic {{Quantum Computation}}},\ }\href {https://doi.org/10.1016/S0375-9601(99)00803-8} {\bibfield  {journal} {\bibinfo  {journal} {Physics Letters A}\ }\textbf {\bibinfo {volume} {264}},\ \bibinfo {pages} {94} (\bibinfo {year} {1999})},\ \Eprint {https://arxiv.org/abs/quant-ph/9904011} {arxiv:quant-ph/9904011} \BibitemShut {NoStop}%
\end{thebibliography}%

\end{document}